\documentclass[11pt]{article}
\usepackage{jheppub}

\usepackage{epsfig}
\usepackage{amssymb}
\usepackage{amsmath}
\usepackage{multirow}

\usepackage{graphicx, subfigure, array, placeins, float}
\usepackage{epstopdf}
\usepackage{extarrows}

\usepackage{dsfont,bbm}

\makeatletter \@addtoreset{equation}{section} \makeatother

\newcommand{\be}{\begin{equation}}
\newcommand{\ee}{\end{equation}}
\newcommand{\bea}{\begin{eqnarray}}
\newcommand{\eea}{\end{eqnarray}}
\newcommand{\la}{\langle}
\newcommand{\ra}{\rangle}

\newcommand{\mmm}{-}
\newcommand{\mmp}{+}

\allowdisplaybreaks[4]
\graphicspath{{./figs/}}

\preprint{
USTC-ICTS/PCFT-20-37}
\title{Two-Loop anomalous dimensions of QCD operators up to dimension-sixteen and Higgs EFT amplitudes}

\author[a]{Qingjun Jin,}
\emailAdd{qjin@gscaep.ac.cn}
\author[b]{Ke Ren,}
\emailAdd{renke@itp.ac.cn}
\author[b,c,d,e]{and Gang Yang}
\emailAdd{yangg@itp.ac.cn}
\affiliation[a]{Graduate School of China Academy of Engineering Physics, No.~10 Xibeiwang East Road, Haidian District, Beijing, 100193, China}
\affiliation[b]{CAS Key Laboratory of Theoretical Physics, Institute of Theoretical Physics, \\Chinese Academy of Sciences, Beijing 100190, China}
\affiliation[c]{School of Fundamental Physics and Mathematical Sciences, Hangzhou Institute for Advanced Study, UCAS, Hangzhou 310024, China}
\affiliation[d]{International Centre for Theoretical Physics Asia-Pacific, Beijing/Hangzhou, China}
\affiliation[e]{Peng Huanwu Center for Fundamental Theory, Hefei, Anhui 230026, China}

\abstract{
We consider two-loop renormalization of high-dimensional Lorentz scalar operators in the gluonic sector of QCD. These operators appear also in the Higgs effective theory obtained by integrating out the top quark loop in the gluon fusion process. We first discuss the classification of operators and how to construct a good set of basis using both off-shell field theory method and on-shell form factor formalism. To study loop corrections, we apply efficient unitarity-IBP strategy and compute the two-loop minimal form factors of length-3 operators up to dimension sixteen. From the UV divergences of form factor results, we extract the renormalization matrices and analyze the operator mixing behavior in detail. The form factors we compute are also equivalent to Higgs plus three-gluon amplitudes that capture high-order top mass corrections in Higgs EFT. We obtain the analytic finite remainder functions which exhibit several universal transcendentality structures.
}

\begin{document}

\maketitle

\setcounter{footnote}{0}

\section{Introduction}

Understanding hadron spectrum is one central problem in QCD.
The hadron states can be understood as gauge invariant composite operators which are composed of gluon and quark fields.
The problem of computing the hadron spectrum is then reduced to calculating the (anomalous) dimension of composite operators.
Due the non-perturbative nature of confinement, an analytic derivation of anomalous dimensions remains a dream;\footnote{In the simplified toy model of planar ${\cal N}=4$ super Yang-Mills (SYM), this goal is in certain sense achieved, thanks to the infinite number of hidden symmetries in the theory, see \cite{Beisert:2010jr} for a review.}
on the other hand,
at high energy scale the asymptotic freedom ensures that a perturbative expansion still applies.
A good knowledge of such perturbative information is helpful to understand the RG flow of the spectrum, and should also provide an important probe to the full spectrum.
One goal of this paper is to provide a working framework that can be efficiently used to compute the anomalous dimension of high-dimensional operators as well as at high loop orders.
To be concrete, we will focus on gauge invariant and Lorentz invariant local operators ${\cal O}(x)$ where all elementary fields are located at a common point in spacetime.

As another motivation, the local operators we consider are also related to the Higgs effective action, which describes the Higgs production via gluon fusion process at LHC.
The Higgs particle has no direct interaction with gluons but through Yukawa coupling with quarks.
The coupling is proportional to the mass of quarks, which is dominated by the heaviest top quark \cite{Ellis:1975ap, Georgi:1977gs}.
To simplify the computation, a useful approximation is to use an effective field theory (EFT) which describes the interaction between Higgs and gluons by integrating out heavy top quark \cite{Wilczek:1977zn, Shifman:1979eb, Dawson:1990zj, Djouadi:1991tka, Kniehl:1995tn, Chetyrkin:1997sg, Chetyrkin:1997un}.
The EFT Lagrangian can be schematically given as:
\begin{equation}
{\cal L}_{\rm eff} = \hat{C}_0H \mathcal{O}_{4;0} + \sum_{k=1}^\infty {1\over m_{\rm t}^{2k}} \sum_{i} \hat{C}_i H \mathcal{O}_{4+2k;i} \,,
\label{eq:HiggsEFT}
\end{equation}
where $\hat{C}_i$ is the Wilson coefficient,  $H$ is the Higgs field, and ${\cal O}_{\Delta_0;i}$ are the effective operators of canonical dimension $\Delta_0$.
For the Higgs plus one jet production, the contribution of higher dimension operators can be important when the Higgs transverse momentum is comparable to the top mass.
The two-loop Higgs plus three-parton amplitudes with the leading operator ${\cal O}_{4;0} = {\rm Tr}(F_{\mu\nu}F^{\mu\nu})$ were computed in \cite{Gehrmann:2011aa},
and similar two-loop amplitudes with dimension-six operators were computed in \cite{Jin:2018fak, Jin:2019opr}.
The two-loop amplitudes with higher dimension operators may be used to improve the precision for the cross section of Higgs plus a jet production at N$^2$LO, which is so far known in the infinite top mass limit \cite{Boughezal:2013uia,Chen:2014gva,Boughezal:2015aha,Boughezal:2015dra, Harlander:2016hcx, Anastasiou:2016hlm, Chen:2016zka}.
At NLO QCD accuracy, the full top mass effect can be taken into account by integrating the top quark loop directly \cite{Lindert:2018iug,Jones:2018hbb,Neumann:2018bsx}.
See also \cite{Heinrich:2020ybq} for a recent extensive review about related studies on Higgs amplitudes and their phenomenological applications.

To study the operator spectrum and the corresponding Higgs amplitudes, we consider the form factor which is defined as a matrix element between an operator ${\cal O}(x)$ and $n$ on-shell states (see \emph{e.g.} \cite{Yang:2019vag} for an introduction):
\begin{equation}
{\cal F}_{{\cal O},n} = \int d^4 x \, e^{-i q\cdot x} \langle p_1, \ldots, p_n | {\cal O}(x) |0 \rangle \,.
\end{equation}
Such form factor is equivalent to a Higgs plus $n$-parton amplitude in the Higgs EFT \eqref{eq:HiggsEFT}, where $q^2 = m_H^2$.
In this following, we will often refer Higgs amplitudes as form factors.
The concrete goal of this paper is to classify the operators of general high dimensions and to compute their anomalous dimensions and related Higgs amplitudes up to two loops.

Since we discuss multiple high dimension operators, the first problem to solve is to classify the operators and construct a convenient operator basis (at classical level).
The operators at a given canonical dimension are generally not independent with each other, since they may be related to each other through equations of motion or Bianchi identities.
We will apply both (off-shell) field theory method and (on-shell) minimal form factor method to construct the operator basis.
We point out that, besides counting the dimension of the basis, an important aspect for this work is also how to construct a \emph{good} set of basis operators explicitly that will be convenient for the high loop computations.

To compute the loop form factors for the basis operators,
we apply the unitarity-IBP strategy that combines unitarity cut \cite{Bern:1994zx, Bern:1994cg, Britto:2004nc} and integration by parts (IBP) methods \cite{Chetyrkin:1981qh, Tkachov:1981wb}.
This strategy has been applied to computing form factors (and Higgs amplitudes) in \cite{Jin:2018fak, Jin:2019ile, Jin:2019opr} and for pure gluon amplitudes in \cite{Boels:2018nrr, Jin:2019nya}.
Similar strategy has also been used in the numerical unitarity approach \cite{Abreu:2017xsl, Abreu:2017hqn}, in which unitarity and IBP are used together with numerical variables to avoid large intermediate expressions.
The idea of using cuts to simplify IBP has also been used in \emph{e.g.}~\cite{Kosower:2011ty, Larsen:2015ped, Ita:2015tya, Georgoudis:2016wff}.

Given the form factor results, one can extract the ultraviolet (UV) divergences by subtracting the universal infrared (IR) divergences.
The basis operators of same classical dimensions can mix with each other through renormalization at quantum level (see \emph{e.g.}~\cite{Collins:1984xc}).
This is captured by the renormalization matrix $Z$ defined in ${\cal O}_{R,i} = Z_i ^{~j} {\cal O}_{B,j}$.
With the explicit two-loop results, we can study various aspects of the operator mixing in detail.
By diagonalizing the renormalization matrix, one can also compute anomalous dimensions.

As mentioned above, the form factors we consider also correspond to the Higgs amplitudes. In particular, high dimension operators corresponds to the high-order top mass corrections in \eqref{eq:HiggsEFT}. The central quantity of the Higgs amplitudes is the finite remainder function.
We obtain the analytic expressions, and find that that maximal transcendental part is universal, namely, independent of the operators. This provides further support of the maximal transcendentality principle introduced in study of anomalous dimensions in \cite{Kotikov:2002ab, Kotikov:2004er} and then observed for the form factor of ${\rm tr}(F^2)$ in \cite{Brandhuber:2012vm} and later also for other operators \cite{Brandhuber:2014ica, Loebbert:2015ova, Brandhuber:2016fni, Loebbert:2016xkw, Banerjee:2016kri, Brandhuber:2017bkg,  Banerjee:2017faz, Jin:2018fak, Brandhuber:2018xzk, Brandhuber:2018kqb, Jin:2019ile, Jin:2019opr}. Besides, the lower transcendentality degree-3 and degree-2 parts also contain universal building blocks, which was also observed in the previous results with dimension-six operators \cite{Jin:2019opr}.

This paper is structured as follows.
In Section \ref{sec:operatorbasis}, we construct the operator basis using both the field theory method and on-shell minimal form factor methods.
Section \ref{sec:computation} describes the strategy of computing form factors based on the unitarity-IBP method.
In Section \ref{sec:anomalous-dim}, the renormalization of form factors are discussed in detail, from which we extract the renormalization matrix of the operators and compute anomalous dimensions.
In Section \ref{sec:finite}, we consider the full Higgs amplitudes and in particular we focus on the finite remainder functions and study their analytic properties.
Section \ref{sec:conclusion} provides a summary and outlook.
Several technical details are included in the Appendix \ref{app:miniprimOper}-\ref{app:O81remainder}.
Finally, for the convenience of the reader, we summarize the structure of the paper as following graph:
%
%
$$
\begin{matrix}
\begin{tabular}{| c |}
\hline  QCD /  \\
Higgs EFT  \\ \hline
\end{tabular}
 \end{matrix}
\xlongrightarrow[\mbox{on-shell spinor}]{\mbox{off-shell field}}
\begin{tabular}{| c |}
\hline  Operator \\ basis  \\ \hline
\end{tabular}
\xlongrightarrow[\mbox{IBP}]{\mbox{Unitarity}}
\begin{tabular}{| c |}
\hline  Loop \\ form factors  \\ \hline
\end{tabular}
\Rightarrow \Bigg\{
\begin{matrix}
\begin{tabular}{| c |}
\hline  Anomalous \\ dimensions  \\ \hline
\end{tabular}
\\  \\
\begin{tabular}{| c |}
\hline  Remainders  \\ \hline
\end{tabular}
 \end{matrix}
$$

\section{Constructing operator basis}
\label{sec:operatorbasis}

In this section we consider the construction of operator basis.  We will first consider the field theory method
and then apply the on-shell form factor method.
Besides counting the number of basis, a central goal is to explain how to construct a convenient set of basis operators that will facilitate the high loop computations.
We will provide explicit basis for length-3 operators up to dimension 16, and in later sections
we will compute their anomalous dimension and related Higgs EFT amplitudes.

\subsection{Operator setup}

We consider local gauge invariant scalar operators in pure Yang-Mills theory composed of field strength $F_{\mu\nu}$ and covariant derivatives $D_\mu$. The field strength carries an adjoint color index as $F_{\mu\nu} = F^a_{\mu\nu} T^a$, where $T^a$ are the adjoint generators of gauge group and satisfy
\begin{equation}
[T^a, T^b] = i f^{abc} T^c \,.
\end{equation}
The covariant derivative acts in the standard way as
\begin{equation}
D_\mu \  \star = \partial_\mu + i g [A_\mu, \star]\,, \qquad [D_\mu, D_\nu] \ \star = i g [F_{\mu\nu}, \star] \,.
\end{equation}
A gauge invariant scalar operator takes the following form
\begin{equation}
c(a_1,...,a_n)\big(D_{\mu_{11}}...D_{\mu_{1m_1}}F_{\nu_1\rho_1} \big)^{a_1}
 \cdots
\big( D_{\mu_{n1}}...D_{\mu_{nm_n}}F_{\nu_n\rho_n}\big)^{a_n}  X(\eta, \epsilon) \,,
\label{eq:def-general-opers}
\end{equation}
where $c(a_1,...,a_n)$ are color factors, such as given in terms of products of ${\rm Tr}(..T^{a_i}.. T^{a_j}..)$. And to form a scalar operator, all Lorentz indices $\{\mu_i, \nu_i, \rho_i\}$ are contracted in pairs by metric $\eta^{\mu \nu}$ or by antisymmetric tensor $\epsilon^{\mu \nu \rho \sigma}$, which are contained in the function $X(\eta, \epsilon)$. In this paper,  for simplicity we will consider the parity even operators where $X$ contains only $\eta$'s.

For convenience of  the upcoming discussions, we define following useful quantities for the operators:
\begin{itemize}
\item \emph{Canonical dimension} of an operator:
\begin{equation}
\text{dim}({\cal O}) = \Delta_0({\cal O})= \text{($\#$ of $D$'s)} + 2 \times \text{($\#$ of $F$'s)} \,.
\end{equation}
Since we consider Lorentz scalar operators,
this dimension is always an even integer number, starting with dim=4.
The canonical dimension typically receives quantum corrections at loop level,
and the correction is called the anomalous dimension $\gamma({\cal O})$.
\item \emph{Length} of an operator:
\begin{equation}
\text{len}({\cal O}) = \text{($\#$ of $F$'s)} \,.
\end{equation}
We will also call an $F$ in an operator together with all the derivatives in front of it, \emph{i.e.} $(D \ldots D F)$, as one \emph{block}. Obviously, the number of blocks is equal to the length of the operator.
\item \emph{Descendants}. If an operator can be written as a total derivative of lower dimensional operator, it is called a descendant. Since we always consider Lorentz invariant and gauge invariant operators,
    the overall  derivatives are just covariant derivatives and always appear in pairs so a descendant
    must take the form like
\begin{equation}
{\cal O} = \partial^2 ({\cal O}')=D^2 ({\cal O}')\,.
\end{equation}
\end{itemize}

Below we also summarize the color factors for length-2 to length-4 operators:
\begin{enumerate}
\item The length-2 case has a unique color factor:
\begin{equation}
\label{eq:colorL2}
c(a,b) = {\rm Tr}(T^a T^b)  = \delta^{a b} \,.
\end{equation}
\item In length-3 case there are two inequivalent color factors:
\begin{equation}
\label{eq:colorL3trace}
c(a, b, c) = {\rm Tr}(T^a T^b T^c) \ \  \text{and} \ \ {\rm Tr}(T^a T^c T^b) \,.
\end{equation}
Equivalently, one can introduce two other color factors as
\begin{equation}
\label{eq:colorL3fd}
f^{abc} = {\rm Tr}(T^a T^b T^c) -  {\rm Tr}(T^a T^c T^b) \,, \qquad d^{abc} = {\rm Tr}(T^a T^b T^c) + {\rm Tr}(T^a T^c T^b) \,,
\end{equation}
which are fully anti-symmetric or symmetric in the color indices, respectively.
\item Length-4 is the first case where double traces appear. The color factors are:
\begin{equation}
{\rm Tr}(T^{a_1} T^{a_{\sigma(2)}} T^{a_{\sigma(3)}} T^{a_{\sigma(4)}}), \quad {\rm Tr}(T^{a_1} T^{a_{\tilde\sigma(2)}}) {\rm Tr}(T^{a_{\tilde\sigma(3)}} T^{a_{\tilde\sigma(4)}}) \,, \qquad \sigma \in S_3,  \tilde\sigma \in Z_3\,.
\nonumber
\end{equation}
\end{enumerate}

Our goal is to find a set of independent operator basis, which requires
no equivalent relation holds among the basis we choose.
Two operators are said to be \emph{equivalent} if their difference is proportional to
the equation of motion (EoM) or  Bianchi identity (BI):
\begin{align}
\label{eq:EoM}
\text{EoM}: \qquad & D_\mu F^{\mu\nu} = 0 \,, \\
\label{eq:Bianchi}
\text{BI}: \qquad & D_\mu F_{\nu\rho}+D_\nu F_{\rho\mu}+D_\rho F_{\mu\nu} = 0 \,.
\end{align}
Below we will use two different ways to do this classification:
(1) field theory method, and (2) on-shell spinor helicity method.

For the convenience of notation, we will often use integer numbers to
represent Lorentz indices and abbreviate product $D_i D_j D_k ....$ to $D_{ijk...}$.
For example,
\begin{equation}
F^{\mu_1 \mu_2}D_{\mu_1} D_{\mu_5} F^{\mu_3\mu_4}D_{\mu_2} D^{\mu_5} F_{\mu_3\mu_4}
\Rightarrow
F_{12}D_{15}F_{34}D_{25}F_{34} \,.
\end{equation}

\subsection{Field theory method}
\label{sec:conventional-method}

The choice of the operator basis is not unique.
To make a preference for the choice, we impose two requirements on basis operators:

\begin{itemize}
\item
We consider operators with smallest length first and then increase the length one by one.
If two operators at length $l$ differ by an operator of high length ($L>l$):
\begin{equation}
\label{eq:equivOP}
O_l  -  O'_l  =  O_{L>l} \,,
\end{equation}
we will say that $O_l$ and $O'_l$ are equivalent at length $l$ level, and only one of them may be kept in the operator basis.
In the following, we will start with the length-2 operators and then consider length-3 operators and finally briefly discuss length-4 operator.
\item
At a given dimension, we prefer to choose descendants as basis operators, so that the total derivatives of lower dimensional operators will contribute automatically as the basis operators in higher dimensional cases.
Since the total derivatives do not change the anomalous dimension of the operator, this choice has an important advantage, in the sense that the renormalization property of a descendant is determined by its lower dimensional counterpart.
We point out  that the descendants can have non-trivial mixing with other operators at the level of renormalization matrices, therefore it is important to keep them in the basis in our computation, see Section~\ref{sec:anomalous-dim} for detailed examples.

\end{itemize}

At a given length and a given dimension, the numbers of $D$ and $F$ are fixed,
and different operators are differed by their ways of Lorentz contraction.
A covariant derivatives $D$ contracts  either with another $D$ or
with an $F$. We call operators with no $DD$ contraction as \emph{primitive operators}.
The full set of operator can be generated by adding pairs of identical $D$s to the primitive operators.

Denote the number of $D$s in an primitive operator as  $\hat{n}_d$.
At a given length $l$, it is easy to see $\hat{n}_d$ ranges from 0 to $2l$ and must be even.
Starting from arbitrary operator, one can get a primitive one by taking off all the identical $DD$ pairs from it, like:
 \begin{equation}
  \mathcal{O}_1={\rm tr}(D_5F_{12}D_{15}F_{34}D_2F_{34}) \
  (\text{non-primitive})
\xlongrightarrow[\mbox{$D_5$ pair}]{\mbox{remove}}
 \mathcal{O}_0={\rm tr}(F_{12}D_1F_{34}D_2F_{34})\
 (\text{primitive}).
\nonumber
\end{equation}
We say the operator $\mathcal{O}_1$ belongs to \emph{primitive class} $\lfloor \mathcal{O}_0 \rfloor$.

\subsubsection*{Length-2 case}

In the length-2 case we illustrate the above classification in detail.
Since there is only one color structure \eqref{eq:colorL2},
we will ignore the color factor in the discussion for simplicity.

First let us classify primitive operators.
Number of $D$s in a primitive operator might take values $\hat{n}_d=0,2,4$,
respectively corresponding to
\begin{equation}
F_{12}F_{12}\,, \quad D_3 F_{12}D_1 F_{23}\,, \quad D_{34} F_{12} D_{12} F_{34} \,.
\end{equation}
Here we have used the EoM constraint \eqref{eq:EoM}, such that $D_i$ should not contract with  $F_{ij}$
that contains the same index.
The above $\hat{n}_d=2$ operator can be reduced to a non-primitive operator using Bianchi identity:
\begin{equation}
\begin{aligned}
D_3 F_{12}D_1 F_{23}&=-D_3F_{12}D_2 F_{31}-D_3F_{12} D_3 F_{12}
=-D_3 F_{12}D_1F_{23}-D_3F_{12} D_3 F_{12}&
\\
&=-\frac{1}{2}D_3F_{12} D_3 F_{12}\in \big\lfloor F_{12}F_{12}\big\rfloor\,.&
\nonumber
\end{aligned}
\end{equation}
The $\hat{n}_d=4$ operator turns out to be a higher length-3 one and therefore can be dropped out
at length-2 level:
\begin{equation}
\begin{aligned}
D_{34}F_{12}D_{12}F_{34}
=[D_{3},\ D_{4}]F_{12}D_{12}F_{34}+D_{43}F_{12}D_{12}F_{34}
=\frac{i g}{2}[F_{34},F_{12}]D_{12}F_{34}\sim (FFF)\,.
\nonumber
\end{aligned}
\end{equation}
In summary, $F_{12}F_{12}$ is the only  primitive operator
for length-2 cases.

Second let us construct non-primitive operators through $DD$ pair insertion.
One can check that a $DD$ pair acting on the same $F$ is always equivalent to a higher length operator:
\begin{align}
D^2F_{12}&=D_3D_3F_{12}=-D_3D_1F_{23}-D_3 D_2 F_{31}
=-[D_3,D_1]F_{23}-[D_3,D_2]F_{31}
\sim (FF)\,.
\end{align}
According to our requirement mentioned around \eqref{eq:equivOP}, at a given length, we only need to consider the cases that
pairs of identical $D_i$s being inserted into different blocks.
Thus the non-primitive length-2 operator obtained from
$DD$ pair insertion must be in the form of $\mathrm{Tr}\big(D_{12...}F_{34}D_{12...}F_{34}\big)$.

Finally we can see such operators  are equivalent to descendants:
\begin{equation}
\begin{aligned}
D_{12...}F_{34}D_{12...}F_{34}
&=\frac{1}{2}D_1^2 \big(D_{4...}F_{23}D_{4...}F_{23}\big)
-  (D^2_1 D_{4...}F_{23})(D_{4...}F_{23})
\\
&=\frac{1}{2}\partial_1^2 \big(D_{4...}F_{23}D_{4...}F_{23}\big)+(\textrm{length-3 operator})\,.
\nonumber
\end{aligned}
\end{equation}
Recursively, an arbitrary length-2 operator, say with dimension  $\Delta_0$, is equivalent to
a total derivative of  $F_{12}F_{12}$ like
\begin{equation}
{(\partial^2)^m} \mathrm{Tr}(F_{12}F_{12}), \qquad
m=\frac{\Delta_0-4}{2}\,.
\end{equation}

\subsubsection*{Length-3 case}

For the length-3 case, the classification of primitive operators is similar to the length-2 case, and we provide the
derivation in Appendix \ref{app:miniprimOper}.
We directly cite the result here: there are two  length-3 primitive
operators
\begin{equation}
\label{length3primitive}
{\cal O}_{\rm P1} =  \mathrm{Tr}(D_1F_{23}D_4F_{23}F_{14}) \,,  \qquad
{\cal O}_{\rm P2} =  \mathrm{Tr}(F_{12}F_{13}F_{23}) \,.
\end{equation}
They can be rewritten as
\begin{equation}
\mathcal{O}_{\mathrm{P1}}=\frac{1}{2}f^{abc} (D_1F_{23})^a (D_4F_{23})^b (F_{14})^c\,,\qquad
\mathcal{O}_{\mathrm{P2}}=\frac{1}{2}f^{abc}(F_{12}^aF_{13}^bF_{23}^c) \,.
\end{equation}

Second, based on that, construction of basis operators at a given dimension $\Delta_0$ is
just to correctly count inequivalent ways of $DD$ pair insertion into blocks of
${\cal O}_{\rm P1}$ and ${\cal O}_{\rm P2}$, which require
 $n=\frac{\Delta_0-8}{2}$ and $m=\frac{\Delta_0-6}{2}$ pairs of identical
 $D$s respectively.
Let us discuss the two cases in more detail.
\begin{itemize}
\item

We first consider inserting pairs of identical $D$s into primitive operator ${\cal O}_{\rm P1}$,
which has blocks \underline{1},\underline{2},\underline{3}, namely $D_1F_{23}$, $D_4F_{23}$, $F_{14}$.
As explained above, $D^2$ acting on a single $F$ vanishes up to higher length, so a  pair
of identical $D$s can only be inserted into two different blocks, which means there are three
choices: \underline{1},\underline{2}  or \underline{1},\underline{3} or \underline{2},\underline{3}.

Taking $n=2$ pairs of $D$s as an example, all the six possible insertions are listed below:
\begin{equation}
\begin{aligned}
\label{eg-partition1}
(D_{i_1}D_{i_2},1,D_{i_1}D_{i_2}),
(D_{i_1}D_{i_2},D_{i_1}D_{i_2},1),
(1,D_{i_1}D_{i_2},D_{i_1}D_{i_2}),
\\
(D_{i_1}D_{i_2},D_{i_2},D_{i_1}),
(D_{i_1},D_{i_2},D_{i_1}D_{i_2}),
(D_{i_1},D_{i_1}D_{i_2},D_{i_2})\,.
\end{aligned}
\end{equation}
where the
three slots represent the positions of block $D_1F_{23}$, $D_4F_{23}$, $F_{14}$.
Alternatively, they can be obtained from the three-part partitions of number $n=2$:
\begin{equation}
\begin{aligned}
0+2+0,\ 2+0+0,\ 0+0+2,\
1+1+0,\ 0+1+1,\ 1+0+1\,.
\end{aligned}
\end{equation}
The three parts in a sum
correspond to numbers of $D$-pairs inserted into three block pairs (\underline{1},\underline{2}) + (\underline{1},\underline{3}) + (\underline{2},\underline{3}).
For a general $n$, the number of inequivalent insertions
are given by  partition counting $C_{n+2}^2=\frac{(\Delta_0-4)(\Delta_0-6)}{8}$.

\item
The ${\cal O}_{\rm P2}$ case is a little bit different, since the three blocks
$F_{12}$, $F_{13}$, $F_{23}$ have $Z_3$ symmetry under index renaming.
As a result, three insertions given by the first line of  (\ref{eg-partition1}) actually
produce the same operator. The same is true for the second line, so there are only two inequivalent
ways to insert $m=2$ pairs of identical $D$s into three blocks of ${\cal O}_{\rm P2}$.

Generally, we should equalize different $D$-insertions
if they are related by $Z_3$ cyclic symmetry.
A special insertion appears when number $m$ is multiple of 3, because
the trisection partition produces a $D$-insertion which is invariant under $Z_3$ symmetry.
In summary,  for $m\neq 0\ (\mbox{mod}\ 3)$, number of inequivalent $D$-insertions
 equals to one third of the partition number, i.e.,
$\frac{1}{3}C_{m+2}^2=\frac{(\Delta_0-2)(\Delta_0-4)}{24}$.
For $m\equiv 0\ (\mbox{mod}\ 3)$, it equals to
$\frac{1}{3}(C_{m+2}^2-1)+1=\frac{\Delta_0(\Delta_0-6)}{24}+1$.
\end{itemize}

Summing over  above two cases, we obtain the
 number of independent length-3 operators with dimension $\Delta_0$, denoted as
$N(\Delta_0)$:
\begin{equation}
\begin{aligned}
N(\Delta_0)=
\left
\{
\begin{array}{rcl}
\frac{(\Delta_0-4)(\Delta_0-5)}{6},\quad \Delta_0 (\mbox{mod}\ 6) \neq0,    \\
\frac{\Delta_0^2-9\Delta_0+24}{6},\quad \Delta_0 (\mbox{mod}\ 6) \equiv0 .
\end{array}
\right.
\end{aligned}
\end{equation}
Expressions of these $N(\Delta_0)$ basis operators for $\Delta_0=6,...,16$ are listed in Appendix \ref{app:oldbasis}.

Third, operators listed in Appendix \ref{app:oldbasis} are not the final basis choice. Since we prefer to choose descendants as the basis elements,
we need to further solve descendant relations and replace some operators by descendants.

Finally, it is convenient to symmetrize
the operators according to the color factors.
As discussed in \eqref{eq:colorL3trace} - \eqref{eq:colorL3fd},
for length-3 operators,
one can take either single trace products or $\{f^{abc},d^{abc}\}$
as color basis. In this paper we choose the later.
To get an operator with color factor $d^{abc}$ or $f^{abc}$ from a single
traced one, we need to (anti)symmetrize the kinematic part.

To illustrate the above procedure, let us take the dimension 10 case as a concrete example.
To start with, there are five length-3 operators obtained by add DD pairs in the two primary classes:
\begin{equation}
\begin{aligned}
\label{eq:original-dim10}
{\cal O}_{\rm P1} + 2 D: \quad &
\mathcal{O}''_{10;1}=\mathrm{Tr}(D_{12}F_{34}D_{15}F_{34}F_{25}),\
\mathcal{O}''_{10;2}=\mathrm{Tr}(D_{12}F_{34}D_5F_{34}D_1F_{25}), \\
&\mathcal{O}''_{10;3}=\mathrm{Tr}(D_2F_{34}D_{15}F_{34}D_1 F_{25}); \\
{\cal O}_{\rm P2} + 4 D: \quad &
\mathcal{O}''_{10;4}=\mathrm{Tr}(D_{12}F_{34}D_1F_{35}D_2F_{45}),\
\mathcal{O}''_{10;5}=\mathrm{Tr}(D_{12}F_{34}D_{12}F_{35}F_{45}) \,.
\end{aligned}
\end{equation}

As mentioned before, we choose length-3 color factor to be $\{f^{abc},d^{abc}\}$
instead of single trace basis $\{\mathrm{Tr}(T^aT^bT^c),\mathrm{Tr}(T^aT^cT^b)\}$.
To achieve that we need to symmetrize or anti-symmetrize the original ones (\ref{eq:original-dim10}),
and the resultant new basis are:
\begin{equation}
\begin{aligned}
\label{eq:original-dim10b}
\mathcal{O}'_{10;1}&=\mathcal{O}''_{10;1}
=\frac{1}{2}f^{abc}(D_{12}F_{34})^a (D_{15}F_{34})^b(F_{25})^c,
\\
\mathcal{O}'_{10;2}&=\mathcal{O}''_{10;2}
+\mathcal{O}''_{10;3}
=f^{abc}(D_{12}F_{34})^a(D_5F_{34})^b(D_1F_{25})^c,
\\
\mathcal{O}'_{10;3}&=\mathcal{O}''_{10;2}
-\mathcal{O}''_{10;3}
=d^{abc}(D_{12}F_{34})^a(D_5F_{34})^b(D_1F_{25})^c,
\\
\mathcal{O}'_{10;4}&=\mathcal{O}''_{10;4}
\sim \frac{1}{2}f^{abc}(D_{12}F_{34})^a(D_1F_{35})^b(D_2F_{45})^c,
\\
\mathcal{O}'_{10;4}&=\mathcal{O}''_{10;5}
=\frac{1}{2}f^{abc}(D_{12}F_{34})^a(D_{12}F_{35})^b(F_{45})^c.
\end{aligned}
\end{equation}
Here equivalence relation $A \sim B$ means the difference between $A$ and $B$ is a higher length operator.

Furthermore, we prefer to choose descendants as basis operators.
At dimension 10, there are  two descendants from total derivatives of lower dimensional operators
${\cal O}_{\rm P1}$ and ${\cal O}_{\rm P2}$,
and they are related to operators in \eqref{eq:original-dim10b} as:
\begin{equation}
\label{linearcombine1}
\frac{1}{2} \partial^2  {\cal O}_{\rm P1}
\sim
\mathcal{O}'_{10;1}+\mathcal{O}'_{10;2} \,,
\qquad
\frac{1}{12}\partial^4  {\cal O}_{\rm P2}
\sim 2\mathcal{O}'_{10;4}+\mathcal{O}'_{10;5}\,.
\end{equation}
Therefore, we can replace one of
$\mathcal{O}'_{10;1}$ and $\mathcal{O}'_{10;2}$ with $\frac{1}{2} \partial^2  {\cal O}_{{\rm P1}}$,
and one of $\mathcal{O}'_{10;4}$ and $\mathcal{O}'_{10;5}$ with
$\frac{1}{12}\partial^4  {\cal O}_{\rm P2}$.
Here we choose to replace $\mathcal{O}'_{10;2}$ and $\mathcal{O}'_{10;4}$, so
the final choice of length-3 basis operators at dimension 10 is
\begin{equation}
\begin{aligned}
\label{eq:original-dim10c}
f\text{-sector}: & \quad \mathcal{O}'_{10;1}, \ \mathcal{O}'_{10;5},
\ \frac{1}{2} \partial^2 {\cal O}_{\rm P1},
\ \frac{1}{12}\partial^4 {\cal O}_{\rm P2} \,, \\
d\text{-sector}: & \quad \mathcal{O}'_{10;3} \,.
\end{aligned}
\end{equation}

\subsubsection*{Length-4 case}

Above construction should be straightforward to generalize to higher length operators. Here we will not go into details but only give a brief account of the length-4 case.
For length-4 case there are seven types of primitive operators as shown in Table \ref{tab:len4prim}.

\begin{table}[!t]
\centering
\caption{Primitive operators in the length-4 case. $a,b,c,d$ are color indices and brackets $(a \ldots d)$ is short for ${\rm Tr}(T^a \ldots T^d)$.}
\label{tab:len4prim}
\vskip 0.4 cm
\begin{tabular}{|c|c|c|c|}
\hline
Block configurations & Single trace &
Double trace
\\
\hline
$(F^a_{12},\ F^b_{23},\ F^c_{34},\ F^d_{14})$ &  $(abcd)$, $(acbd)$ & $(ab)(cd)$, $(ac)(bd)$

\\
\hline
$(F^a_{12},\ F^b_{12},\ F^c_{34},\ F^d_{34})$ & $(abcd)$, $(acbd)$
& $(ab)(cd)$, $(ac)(bd)$

\\
\hline
$(F^a_{13},\ F^b_{23},\ D_{12}F^c_{45},\ F^d_{45})$   & $(abcd),(acbd),(cabd)$
& $(ab)(cd)$, $(ac)(bd)$

\\
\hline
$(F^a_{13},\ F^b_{23},\ D_1F^c_{45},\ D_2 F^d_{45})$ & $(abcd),(bacd),(acbd),(bcad)$
& $(ab)(cd),(ac)(bd),(ad)(bc)$

\\
\hline
$(F^a_{12},\ D_1F^b_{34},\ D_2F^c_{45},\ F^d_{35})$  & $(abcd),(dbca),(abdc)$
& $(ab)(cd),(ad)(bc)$

\\
\hline
$(F^a_{12},\ D_5 F^b_{23},\ D_1 F^c_{34},\ F^d_{45})$  & $(abcd),(dbca),(abdc),(dbac)$
& $(ab)(dc),(db)(ac),(ad)(bc)$

\\
\hline
$(F^a_{12},\ F^b_{34},\ D_{13}F^c_{56},\ D_{24}F^d_{56})$  & $(abcd),(acbd)$
 & $(ab)(cd),(ac)(bd)$

\\
\hline
\end{tabular}
\end{table}

Here we provide the basis of dimension-8 operators as a concrete example.
At dimension 8, all length-4 operators are primitive ones, and
they correspond to the first two rows of Table \ref{tab:len4prim}.
Taking into account different color orders, one can find that there are four single-trace  and four
double-trace operators in total:
\begin{align}
\label{eq:len4dim8-0}
&\Xi'_1=\mathrm{tr}(F_{12} F_{23} F_{34} F_{14}),  \qquad\qquad
\Xi'_2=\mathrm{tr}(F_{12} F_{34} F_{23} F_{14}),
&
\nonumber\\
&
\Xi'_3=\mathrm{tr}(F_{12} F_{12} F_{34} F_{34}), \qquad\qquad
\Xi'_4=\mathrm{tr}(F_{12} F_{34} F_{12} F_{34}),
&
\nonumber\\
&
\Xi'_5=\mathrm{tr}(F_{12} F_{23})\mathrm{tr} (F_{34} F_{14}),
\qquad \
\Xi'_6=\mathrm{tr}(F_{12} F_{34})\mathrm{tr} (F_{23} F_{14}),
&
\nonumber\\
&
\Xi'_7=\mathrm{tr}(F_{12} F_{12})\mathrm{tr} (F_{34} F_{34}),
\qquad \
\Xi'_8=\mathrm{tr}(F_{12} F_{34})\mathrm{tr} (F_{12} F_{34}).
\end{align}
Adding length-2 and length-3 operators, at dimension-8, the operator basis contains $1+2+8 = 11$ operators in total.

Unlike length-2 and length-3 cases however, length-4
operators from different primitive classes might not be independent,
so naively adding pairs of $D$ into the primitive operators
may create an overcomplete basis set.
For example, consider an identity involving three operators:
\begin{equation}
\mathrm{tr}(D_6F_{12}D_1F_{34}D_6F_{45}D_2F_{53})
=\mathrm{tr}(D_6F_{12}D_6F_{34}D_1F_{45}D_2F_{53})
+\mathrm{tr}(D_6F_{12}D_1F_{34}D_2F_{45}D_6F_{53}) .
\nonumber
\end{equation}
These operators belong to three different primitive classes at the fifth row of
Table \ref{tab:len4prim}, i.e.,
$(abdc),(dbca),(abcd)$.
Therefore the linear independence of primitive operators
doesn't guarantee the independence of non-primitive ones.

Before entering next subsection, let us summarize the strategy of field theory classification:

\begin{enumerate}
\item  First we classify primitive operators which contain no $DD$ contraction.

\item After primitive operators being classified,
one can then generate other (non-primitive) operators by enumerating inequivalent ways of $DD$ pair insertion into primitive ones.

\item
While independent operators obtained from inserting $DD$ pairs into
primitive ones already form a set of basis, they are not a good choice since
we require descendants  to be included.
One can apply identities between descendants and above basis
and solve for part of them in terms of descendants.

\item For length-3 case, we also organize the basis into $f^{abc}$ and $d^{abc}$ sectors, to manifest the symmetry properties.

\item Finally, to obtain a full basis at a given dimension, one needs to sum operators of all possible length. For example, for dimension-8 case, operators up to length-4 all contribute.

\end{enumerate}

\subsection{On-shell spinor helicity method}
\label{sec:onshell-method}

An alternative way to do the classification is to make
use of on-shell technique  and read properties of operators
from their form factors. In other words, we follow the
dictionary between operators and their tree-level minimal form factors
and translate all the operator information into spinor helicity formalism
for mathematical clearness.

\subsubsection*{Operator-spinor dictionary}

A  minimal tree form factor means the number of external gluons is equal to the length of the operator.
One can establish a dictionary from an operator to its
tree-level minimal form factor \cite{Beisert:2010jq, Zwiebel:2011bx, Wilhelm:2014qua}:
\begin{equation}
{\cal O}_L  \ \Leftrightarrow \ {\cal F}_{{\cal O}_L, L}(1,  \ldots, L) \,,
\end{equation}
especially each
single $D$ and $F$ contained by the operator are mapped to certain spinor structures,
as shown in Table \ref{tab:dictionary}.

\begin{table}[!t]
\centering
\caption{Dictionary between operators and on-shell spinors}
\label{tab:dictionary}
\vskip 0.4 cm
\begin{tabular}{|c|c|c|c|c|}
\hline
operator & $s_{ij}$ & $D_{\dot{\alpha}\alpha}$  &
$f_{\alpha\beta}$ & $\bar{f}_{\dot{\alpha}\dot{\beta}}$
\\
\hline
spinor & $\la ij\ra[ji]$ & $\tilde{\lambda}_{\dot{\alpha}}\lambda_{\alpha}$
 & $\lambda_\alpha\lambda_\beta$
 & $-\tilde{\lambda}_{\dot{\alpha}}
\tilde{\lambda}_{\dot{\beta}}$\\
\hline
\end{tabular}
\end{table}

The map of $D$ results from spinor represention of momentum
$P^{\dot{\alpha}\alpha}=\tilde{\lambda}_p^{\dot{\alpha}}\lambda_p^\alpha$.
As for field strength $F$, first one takes decomposition
\begin{align}
\label{eq:ff-decompose}
F_{\mu\nu}\rightarrow F_{\alpha\dot{\alpha}\beta\dot{\beta}}=
\epsilon_{\alpha\beta}\bar{f}_{\dot{\alpha}\dot{\beta}}
+\epsilon_{\dot{\alpha}\dot{\beta}}f_{\alpha\beta}
\end{align}
to obtain self-dual and anti-self-dual components
\begin{equation}
\bar{f}_{\dot{\alpha}\dot{\beta}}=\frac{1}{2}\epsilon^{\alpha\beta}
F_{\alpha\dot{\alpha}\beta\dot{\beta}} \,, \qquad
f_{\alpha\beta}=\frac{1}{2}\epsilon^{\dot{\alpha}\dot{\beta}}
F_{\alpha\dot{\alpha}\beta\dot{\beta}} \,.
\end{equation}
Then one  makes use of LSZ reduction formula
\begin{equation}
\langle \vec{p}|F_{\mu\nu}(0)|\Omega\rangle
=(-i)[e_\nu p_\mu-e_\mu p_\nu ]
\end{equation}
to get their final matrix elements
\begin{align}\label{onshell-f}
&\langle \vec{p}|f_{\alpha\beta}(0)|\Omega\rangle
=
\left
\{
\begin{array}{rcl}
0,    &      & h=+  \\
-\frac{i}{\sqrt{2}}\lambda_\alpha\lambda_\beta,    &      & h=-
\end{array}
\right.\,,
\quad
\langle \vec{p}|\bar{f}_{\dot{\alpha}\dot{\beta}}(0)|\Omega\rangle
=
\left
\{
\begin{array}{rcl}
\frac{i}{\sqrt{2}}\tilde{\lambda}_{\dot{\alpha}}
\tilde{\lambda}_{\dot{\beta}},    &      & h=+  \\
0,    &      & h=-
\end{array}
\right.\,.&
\end{align}
We summarize the correspondence between operators and on-shell spinors
in Table \ref{tab:dictionary}, and
the example on reconstructing operators from spinor-helicity formalism
will be given in upcoming context, see (\ref{eg-O101-2}).

The correspondence listed in Table \ref{tab:dictionary}
is not limited within pure Yang-Mills theory,
and the result can be  generalized when fermions enter in.
The above on-shell language has several advantages:
\begin{enumerate}
\item Equivalent relations between operators
take much simpler forms.
Equation of motion holds automatically, and Bianchi identities
are translated into Schouten identities:
\begin{equation}
\begin{aligned}
D_\mu F^{\mu\nu}
&\rightarrow-[\lambda\lambda]
\lambda_\beta \tilde{\lambda}^{\dot{\beta}}
+\langle\lambda\lambda\rangle
\tilde{\lambda}_{\dot{\beta}}\lambda^\beta=0\,,&
\\
D_\mu F_{\nu\rho}+D_\nu F_{\rho\mu}+D_\rho F_{\mu\nu}
&\rightarrow-\tilde{\lambda}_{\dot{\alpha}}\tilde{\lambda}_{\dot{\beta}}\tilde{\lambda}_{\dot{\gamma}}
(\lambda_\alpha\epsilon_{\beta\gamma}+\lambda_\beta\epsilon_{\gamma\alpha}
+\lambda_\gamma\epsilon_{\alpha\beta}) \nonumber\\
&\qquad +\lambda_\alpha\lambda_\beta\lambda_\gamma
(\tilde{\lambda}_{\dot{\alpha}}\epsilon_{\dot{\beta}\dot{\lambda}}
+\tilde{\lambda}_{\dot{\beta}}\epsilon_{\dot{\gamma}\dot{\alpha}}
+\tilde{\lambda}_{\dot{\gamma}}\epsilon_{\dot{\alpha}\dot{\beta}})\,.&
\end{aligned}
\end{equation}

\item Contribution from higher length operators vanish automatically
since they have vanishing form factor now.
Therefore two operators equivalent up to higher length have identical
tree level minimal form factor.
For example, following three operators
are equivalent at the level of length 2:
\begin{equation}
\label{eq:eg1}
\mathrm{Tr}(D_\rho F_{\mu\nu}D^\nu F^{\mu\rho}),
\quad
\frac{1}{2}\mathrm{Tr}(D_\rho F_{\mu\nu}D^\rho F^{\mu\nu}),
\quad
\frac{1}{4}\partial^2\mathrm{Tr} (F_{\mu\nu}F^{\mu\nu})\,,
\end{equation}
and they have the same form factor under arbitrary helicity setting, like
 $s_{12}\la12\ra^2$ for $1^-2^-$ and 0 for $1^-2^+$.

\item In field theory classification we treat
$DF$ contraction and $DD$ contraction differently.
In on-shell language, $DD$ contraction only contributes to
 scalar factor like $s_{ij}$.
For example:
\begin{align}
\frac{1}{2}\mathrm{Tr}(D_\rho F_{\mu\nu}D^\rho F^{\mu\nu})\in
\big\lfloor \mathrm{Tr}(F_{\mu\nu}F^{\mu\nu})\big\rfloor
&
\quad \Rightarrow \quad
s_{12}^2 \la12\ra^2\in
\big\lfloor \la12\ra^2 \big\rfloor .
\nonumber
\end{align}
\end{enumerate}

Below we will discuss the construction of operator basis based on the spinor-helicity formalism.
As we will see, all the standard steps  in
 field theory method have on-shell analogies.
 Primitive operator classifiction becomes enumerating spinor structures,
and counting $DD$ pair insertion becomes enumerating scalar factors.
Besides, the minimal form factor picture also allows a further choice of the \emph{helicity sector} which will further simplify the kinematic structure.

\subsubsection*{Length 2}

For length-2 operators, the tree-level minimal form factor involves two scattered gluons labeled by 1 and 2.
The only possible spinor bracket structures are $\langle 12\rangle$ and $[12]$, since
forms like $\langle 1|P_i|2]$ must vanish because $P_i$ is either $p_1$ or
$p_2$.

For an operator with minimal dimension,
the only possible expressions of the form factor
are $\langle 12\rangle^2$ and $[12]^2$,
corresponding to projected
operator $\mathrm{tr}(f_{\alpha\beta}f^{\alpha\beta})$ and
$\mathrm{tr}(\bar{f}_{\dot{\alpha}\dot{\beta}}
\bar{f}^{\dot{\alpha}\dot{\beta}})$.
Their sum is $\mathrm{tr}(F^2)$,
and difference is $\epsilon^{\mu\nu\rho\sigma}\mathrm{tr}
(F_{\mu\nu}F_{\rho\sigma})$.
In this paper we will not consider operators with odd $P$-parity,
so only the former is kept.

For an even operator with general dimension $\Delta$, its
minimal form factor is $(s_{12})^{\frac{\Delta-4}{2}}
\la 12\ra^2$ under $(-,-)$ and $(s_{12})^{\frac{\Delta-4}{2}}
[12]^2$ under $(+,+)$.
Taking dimension 6 as an example, from $s_{12}\la 12\ra^2$ one can read
two holomorphic operators, which are  self-dual components of the first two
operators in (\ref{eq:eg1}), and they are both equivalent to the descendant
$\frac{1}{4}\partial^2\mathrm{tr} (F^2)$,
which is chosen as the only
independent length-2 operator at dimension 6:
\begin{equation}
s_{12}\la 12\ra^2\rightarrow
\left
\{
\begin{array}{rcl}
\mathrm{tr}(D_{\gamma\dot{\gamma}}f_{\alpha\beta}D^{\beta\dot{\gamma}}f^{\gamma\alpha})
&\rightarrow \mathrm{tr}(D_\rho F_{\mu\nu}D^\nu F^{\mu\rho})
\\
\mathrm{tr}(D_{\gamma\dot{\gamma}}f_{\alpha\beta}D^{\gamma\dot{\gamma}}f^{\alpha\beta})
&\rightarrow \frac{1}{2}\mathrm{tr}(D_\rho F_{\mu\nu}D^\rho F^{\mu\nu})
\end{array}
\right.
\rightarrow
\frac{1}{4}\partial^2\mathrm{tr} (F_{\mu\nu}F^{\mu\nu}) .
\end{equation}

\subsubsection*{Length 3}

As for length-3 cases, we introduce a new concept \emph{helicity sector} to help the discussion.
The map from operator equivalent class to tree-level minimal form factor
 is not injective.
For example, for two inequivalent operators $\mathcal{O}''_{8;1}$ and $\mathcal{O}''_{8;2}$
given in (\ref{old-dim8}):
\begin{equation}
\mathcal{O}''_{8;1}=\mathrm{Tr}(D_1F_{23}D_4F_{23}F_{14}) \,, \qquad
\mathcal{O}''_{8;2}=\mathrm{Tr}(D_1F_{23}D_1F_{24}F_{34}) \,.
\end{equation}
Their spinor structures under $(-,-,-)$ and $(-,-,+)$ are given in Table
\ref{tab:brackets8dim}. One can see these two operators are undistinguishable
under helicity $(-,-,-)$.

\begin{table}[!t]
\centering
\caption{Tree-level minimal form factors of $\mathcal{O}''_{8;1}$ and $\mathcal{O}''_{8;2}$.}
\label{tab:brackets8dim}
\vskip 0.4 cm
\begin{tabular}{|c|c|c|}
\hline
& $(-,-,+)$ & $(-,-,-)$
\\
\hline
$\mathrm{Tr}(D_1F_{23}D_4F_{23}F_{14})$
& $\la12\ra^3[13][23]$ & $s_{123}\la12\ra\la13\ra\la23\ra$
\\
\hline
$\mathrm{Tr}(D_1F_{23}D_1F_{24}F_{34})$
& 0 & $s_{123}\la12\ra\la13\ra\la23\ra$
\\
\hline
\end{tabular}
\end{table}

A natural solution is to create two new operators using $\mathcal{O}''_{8;1}$
and $\mathcal{O}''_{8;2}$ so that
one has nonzero tree-level minimal form factor only
under $(-,-,+)$ and its h.c.,
and the other is non-zero only under $(-,-,-)$ and its h.c.
We say these two new operators belong to
\emph{helicity sector} $\alpha$ and $\beta$ respectively:
\begin{equation}
\begin{aligned}
\label{eq:h-sector}
\alpha\text{-sector}:\quad
{\cal F}_{\cal O}^{(0),\text{min}} \neq 0\  \text{only under\ } (-,-,+),(+,+,-),
\\
\beta\text{-sector}:\quad
{\cal F}_{\cal O}^{(0),\text{min}} \neq 0 \ \text{only under\ } (-,-,-),(+,+,+) .
\end{aligned}
\end{equation}
In following context,
we require each length-3 basis operator belongs to either $\alpha$-sector or
$\beta$-sector.

The nature of ``helicity sector" for an operator is its holomorphic structure.
Taking decomposition (\ref{eq:ff-decompose}) of a length-3 operator might create
four possible components $fff$, $\bar{f}\bar{f}\bar{f}$, $ff\bar{f}$, $\bar{f}\bar{f}f$.
For operator with even $P$-parity, conjugate components always
appear in pairs so there are only two inequivalent structures:
\begin{equation}
ff\bar{f} + \bar{f}\bar{f}f \,, \qquad
fff + \bar{f}\bar{f}\bar{f} \,.
\end{equation}
We can see the former can only be detected by helicity
$(-,-,+)$ and $(+,+,-)$, while the later only by helicity $(-,-,-)$ and $(+,+,+)$.

Let us redo the classification for length-3.
First we need to enumerate all the possible spinor structures for 3-gluon form factors,
which is an analogy to enumerating primitive operators in field theory method.
There are two types of spinor structures
\begin{equation}
A_1=\la12\ra^3[13][23] \,,  \qquad A_2=\la12\ra\la13\ra\la23\ra \,,
\end{equation}
with their cyclic partners, consistent with primitive operator counting
given in (\ref{length3primitive}).
To establish the relation between $A_1,A_2$ and
${\cal O}_{\rm P1},{\cal O}_{\rm P2}$ in (\ref{length3primitive}), we
need to take linear recombination of ${\cal O}_{\rm P1},{\cal O}_{\rm P2}$ to get the new primitive basis that
can be classified into helicity sectors:
\begin{align}
 \alpha\text{-sector}: & \quad {\cal O}_{\rm P1}-{\cal O}_{\rm P2}
\rightarrow
A_1=\la 12\ra^3[13][23],
\\
 \beta\text{-sector}: & \quad \qquad \quad  {\cal O}_{\rm P2}
\rightarrow
A_2=\la12\ra\la13\ra\la23\ra .
\end{align}

The second step is to add Mandelstam scalar factors to $A_1$ and $A_2$
until certain dimension is reached,
and this is an analogy to  inserting $DD$ pairs to primitive operators in field theory method.
Below we take dimension 10 case as an example to explain the details.

\begin{enumerate}
\item Let us enumerate scalar factors for spinor structures
$A_1$ and $A_2$.
For $A_1$, dimension of scalar factor is
$\Delta-8$, so at $\Delta=10$ there are three choices:
$s_{12},s_{13},s_{23}$.
Notice that helicity setting $-,-,+$ has broken the total bosonic symmetry,
while the exchange invariance between gluon 1 and 2 is still maintained.
So 1,2-flipping property of kinematic part should be compatible with
that of color factor.
As a result, $s_{12}$ and $s_{13}+s_{23}$ correspond to $f$-sector while
 $s_{13}-s_{23}$ corresponds to $d$-sector.

For $A_2$,  scalar factor has dimension
$\Delta-6$ and must be cyclic symmetric as $A_2$.
At $\Delta=10$ there are two choices:
$s_{12}^2+s_{23}^2+s_{13}^2$ and
$s_{12} s_{23}+s_{12} s_{13}+s_{23} s_{13}$, which both belong to
$f$-sector.

In total, there are five possible expressions
for the kinematic part of a 3-gluon form factors:
\begin{align}
f\text{-sector}: \quad
&s_{12} A_1, \
(s_{13}+s_{23}) A_1,
\
(s_{12}^2+s_{23}^2+s_{13}^2)A_2, \
(s_{12} s_{23}+s_{12} s_{13}+s_{23} s_{13})A_2,
\nonumber\\
d\text{-sector}:\quad
&(s_{13}-s_{23}) A_1 \,.
\end{align}

\item
After obtaining above spinor-helicity forms,
one can apply the dictionary in Table \ref{tab:dictionary} to read out the operators.
Taking $s_{12}A_1$ as an example,
we write the bracket form back to spinors.
For an operator of primitive class $\lfloor {\cal O}_{\rm P1}\rfloor $ under
helicity setting $(1^-2^-3^+)$, one can lock the
external gluon label $i$ to block position $i$,
so we have
\begin{align}
\label{eg-O101}
&s_{12}A_1
=
(\lambda_{1\alpha} \lambda_{2}^{\ \alpha})
(\tilde{\lambda}_{1\dot{\alpha}}\tilde{\lambda}_2^{\ \dot{\alpha}})
(\lambda_{1}^{\ \beta} \lambda_{2\beta})
(\lambda_{1\gamma} \lambda_{2}^{\ \gamma})
(\lambda_{1\delta} \lambda_{2}^{\ \delta})
(\tilde{\lambda}_{1\dot{\beta}}\tilde{\lambda}_3^{\ \dot{\beta}})
(\tilde{\lambda}_{2\dot{\sigma}}\tilde{\lambda}_3^{\ \dot{\sigma}}).
\end{align}
One type of rearrangement for  (\ref{eg-O101}) is
\begin{align}
\label{eg-O101-2}
&(\lambda_{1\alpha}\tilde{\lambda}_{1\dot{\alpha}})
(\lambda_1^{\ \beta}\tilde{\lambda}_{1\dot{\beta}})
(\lambda_{1\gamma}\lambda_{1\delta})
(\lambda_2^{\ \alpha}\tilde{\lambda}_2^{\ \dot{\alpha}})
(\lambda_{2\beta}\tilde{\lambda}_{2\dot{\sigma}})
(\lambda_2^{\ \gamma}\lambda_2^{\ \delta})
(\tilde{\lambda}_3^{\ \dot{\beta}}\tilde{\lambda}_3^{\ \dot{\sigma}})&
\nonumber\\
&=D_{\alpha\dot{\alpha}}D^\beta_{\ \dot{\beta}}f_{\gamma\delta}
D^{\alpha\dot{\alpha}}D_{\beta\dot{\sigma}}f^{\gamma\delta}
\tilde{f}^{\dot{\beta}\dot{\sigma}}
=D_{\alpha\dot{\alpha}}D^\beta_{\ \dot{\beta}}(f_{\gamma\delta}\epsilon_{\dot{\gamma}\dot{\delta}})
D^{\alpha\dot{\alpha}}D^{\sigma}_{\ \dot{\sigma}}(f^{\gamma\delta}\epsilon^{\dot{\gamma}\dot{\delta}})
(\tilde{f}^{\dot{\beta}\dot{\sigma}}\epsilon_{\beta\sigma})\,.&
\end{align}
This is the $ff\tilde{f}$-component of operator
$\mathcal{O}'_{10;1}=\mathrm{Tr}(D_{12}F_{34}D_{15}F_{34}F_{25})$
derived from decomposition (\ref{eq:ff-decompose}).
However, (\ref{eg-O101-2}) is not the only way to group spinors
into counterparts of $f$ and $D$. For example, (\ref{eg-O101}) can also
be rearranged to
\begin{align}
&
(\lambda_{1\alpha}\tilde{\lambda}_{1\dot{\alpha}})
(\lambda_{1}^{\ \beta}\tilde{\lambda}_{1\dot{\beta}})
(\lambda_{1 \gamma}\lambda_{1\delta})
(\lambda_{2}^{\ \alpha}\tilde{\lambda}_{2}^{\ \dot{\alpha}})
(\lambda_{2}^{\ \gamma}\tilde{\lambda}_{2\dot{\sigma}})
(\lambda_{2 \beta}\lambda_{2}^{\ \delta})
(\tilde{\lambda}_3^{\ \dot{\beta}}\tilde{\lambda}_{3}^{\ \dot{\sigma}})
&\nonumber\\
&=D_{\alpha\dot{\alpha}}D^\beta_{\ \dot{\beta}}f_{\gamma\delta}
D^{\alpha\dot{\alpha}}D^\gamma_{\ \dot{\sigma}}f_\beta^{\ \delta}
\tilde{f}^{\dot{\beta}\dot{\sigma}}
=D_{\alpha\dot{\alpha}}D^\beta_{\ \dot{\beta}}
(
\epsilon^{\dot{\gamma}\dot{\delta}}
f_{\gamma\delta}
)
D^{\alpha\dot{\alpha}}D^{\gamma}_{\ \dot{\gamma}}
(
\epsilon_{\dot{\delta}\dot{\sigma}}
f^{\delta\sigma}
)
(
\epsilon_{\beta\sigma}
\tilde{f}^{\dot{\beta}\dot{\sigma}})\,,&\nonumber
\end{align}
which is the $ff\tilde{f}$-component of operator
$\mathrm{Tr}(D_{12}F_{34}D_{13}F_{4 5}F_{25})$ equivalent to $-\frac{1}{2}\mathcal{O}'_{10;1}$:
\begin{align}
&\mathrm{Tr}(D_{12}F_{34}D_{13}F_{45}F_{25})
=-\mathrm{Tr}(D_{12}F_{34}D_{15}F_{34}F_{25})
-\mathrm{Tr}(D_{12}F_{34}D_{14}F_{53}F_{25})
&\nonumber\\
&
=-\frac{1}{2}\mathrm{Tr}(D_{12}F_{34}D_{15}F_{34}F_{25})
=-\frac{1}{2}\mathcal{O}'_{10;1}
\,.&\nonumber
\end{align}
From above example we can see that the reconstruction of operators from
on shell minimal form factor is not an injective mapping, but the different
operators obtained from the same form factor are equivalent (up to possible overall factor) 
at the level of
length-3, so one can choose an arbitrary operator as the representative one and here
we choose (\ref{eg-O101-2}) for $s_{12}A_1$.

For an $A_2$-involved form factor we do the similar thing.
A little bit difference comes from the fact that under helicity $(-,-,-)$
the form factor has full bosonic symmetry so gluon $i$ might come from
any one of the three blocks. So first we write the cyclic symmetric
scalar back to an  asymmetric one (but maintaining $(1,2)$-flipping symmetry),
\emph{e.g.}~write $s_{12}s_{23}+s_{13}s_{23}+s_{12}s_{13}$
back to $s_{13}s_{23}$, $(s_{12}+s_{13})s_{23}$, $(s_{12}-s_{13})s_{23}$, and then
transform the spinor into operator according to dictionary, with gluon label $i$ locked
to block position $i$. So the fourth form factor in Table \ref{tab:dim10a} is mapped to
three inequivalent operators.

\begin{table}[!t]
\centering
\caption{Operator-form factor dictionary of basis operators at  dimension 10.}
\label{tab:dim10a}
\vskip 0.4 cm
\begin{tabular}{|c|c|c|}
\hline
  minimal form factor & projected operator & full operator
\\
\hline
 $-\frac{1}{2}s_{12} A_1$ &
$D_{\alpha\dot{\alpha}}D^{\beta}_{\ \dot{\beta}}f_{\gamma\delta}
D^{\alpha\dot{\alpha}}D_{\beta\dot{\epsilon}}f^{\gamma\delta}
\tilde{f}^{\dot{\beta}\dot{\epsilon}}$
& $\mathcal{O}'_{10;1}$
\\
\hline
 $-\frac{1}{2}(s_{13}+s_{23})A_1$ &
$D_{\alpha\dot{\alpha}}D^\beta_{\ \dot{\beta}}f_{\gamma\delta}
D_{\beta\dot{\epsilon}}f^{\gamma\delta}
D^{\alpha\dot{\alpha}}\tilde{f}^{\dot{\beta}\dot{\epsilon}}
+
D^\beta_{\ \dot{\beta}}f_{\gamma\delta}
D_{\alpha\dot{\alpha}}D_{\beta\dot{\epsilon}}f^{\gamma\delta}
D^{\alpha\dot{\alpha}}\tilde{f}^{\dot{\beta}\dot{\epsilon}}
$
& $\mathcal{O}'_{10;2}$
\\
\hline
 $-\frac{1}{2}(s_{13}-s_{23})A_1$ &
$
D_{\alpha\dot{\alpha}}D^\beta_{\ \dot{\beta}}f_{\gamma\delta}
D_{\beta\dot{\epsilon}}f^{\gamma\delta}
D^{\alpha\dot{\alpha}}\tilde{f}^{\dot{\beta}\dot{\epsilon}}
-
D^\beta_{\ \dot{\beta}}f_{\gamma\delta}
D_{\alpha\dot{\alpha}}D_{\beta\dot{\epsilon}}f^{\gamma\delta}
D^{\alpha\dot{\alpha}}\tilde{f}^{\dot{\beta}\dot{\epsilon}}$
& $\mathcal{O}'_{10;3}$
\\
\hline

\multirow{3}{*}{ $\frac{1}{4}(s_{12}s_{23}+\mathrm{cyclic.})A_2$.}
 &
$D^{\alpha\dot{\alpha}}f^\gamma_{\ \epsilon}
D^{\beta\dot{\beta}}f^{\delta\epsilon}
D_{\alpha\dot{\alpha}}D_{\beta\dot{\beta}}f_{\gamma\delta}$
& $\mathcal{O}'_{10;4}$
\\
\cline{2-3}
 & $D_{\alpha\dot{\alpha}}D_\beta^{\ \dot{\beta}}f_{\gamma\delta}
D_{\epsilon\dot{\beta}}f^{\gamma\delta}
D^{\alpha\dot{\alpha}}f^{\beta\epsilon}
+
D_\beta^{\ \dot{\beta}}f_{\gamma\delta}
D_{\alpha\dot{\alpha}}D_{\epsilon\dot{\beta}}f^{\gamma\delta}
D^{\alpha\dot{\alpha}}f^{\beta\epsilon}$
& $\mathcal{O}'_{10;2}$
\\
\cline{2-3}
 & $
D_{\alpha\dot{\alpha}}D_\beta^{\ \dot{\beta}}f_{\gamma\delta}
D_{\epsilon\dot{\beta}}f^{\gamma\delta}
D^{\alpha\dot{\alpha}}f^{\beta\epsilon}
-
 D_\beta^{\ \dot{\beta}}f_{\gamma\delta}
D_{\alpha\dot{\alpha}}D_{\epsilon\dot{\beta}}f^{\gamma\delta}
D^{\alpha\dot{\alpha}}f^{\beta\epsilon}$
 & $\mathcal{O}'_{10;3}$
\\
\hline

\multirow{2}{*}{$\frac{1}{4}(s_{12}^2+\mathrm{cyclic.})A_2$ }
  &
$D_{\alpha\dot{\alpha}}D_{\beta\dot{\beta}}f_{\gamma\delta}
D^{\alpha\dot{\alpha}}D^{\beta\dot{\beta}}f^\gamma_{\ \epsilon}
f^{\delta\epsilon}$
& $\mathcal{O}'_{10;5}$
\\
\cline{2-3}
  & $D_{\alpha\dot{\alpha}}D_{\beta}^{\ \dot{\beta}}f_{\gamma\delta}
D^{\alpha\dot{\alpha}}D_{\epsilon \dot{\beta}}f^{\gamma\delta}
f^{\beta\epsilon}$ & $\mathcal{O}'_{10;1}$
\\
\hline
\end{tabular}
\end{table}

In total, there are five different operators in the third column, they are equivalent to
the basis operators obtained from field theory method introduced in
Section \ref{sec:conventional-method}, see (\ref{eq:original-dim10b}).

\item
Now we need to solve the descendant relations as constraints.
One can easily find
operator relations (\ref{linearcombine1}) between $\{\mathcal{O}'_{10;i}\}$ and descendants
of ${\cal O}_{\rm P1}$ and ${\cal O}_{\rm P2}$
from scalar factor relations $s_{123}=s_{12}+s_{13}+s_{23}$.
We choose to replace $\mathcal{O}'_{10;2}$ and $\mathcal{O}'_{10;4}$ with these two
descendants and the basis set becomes (\ref{eq:original-dim10c}).

\item
Finally we need to recombine  five operators in (\ref{eq:original-dim10c}) so that
each basis operator only belongs to one helicity sector. The final operator basis is summarized
in Table \ref{tab:dim10d}.

\begin{table}[!t]
\centering
\caption{Final basis operators at  dimension 10.}
\label{tab:dim10d}
\vskip 0.4 cm
\begin{tabular}{|l|c|c|c|}
\hline
Basis operator &  ${\cal F}^{(0)}(-,-,+)$ & ${\cal F}^{(0)}(-,-,-)$  & color factor
\\
\hline
$\mathcal{O}_{10;\alpha;f;1}=\frac{1}{2}\partial^2 {\cal O}_{\rm P1}
-\frac{1}{12}\partial^4 {\cal O}_{\rm P2}$
&
 $\frac{1}{2}s_{123} A_1$ & 0 & $f^{abc}$
\\
$\mathcal{O}_{10;\alpha;f;2}=\mathcal{O}'_{10;1}-\mathcal{O}'_{10;5}$
&  $\frac{1}{2}s_{12} A_1 $ & 0 & $f^{abc}$
\\
\hline
$\mathcal{O}_{10;\alpha;d;1}=\mathcal{O}'_{10;3}$
&  $\frac{1}{2}(s_{13}-s_{23}) A_1$ & 0 & $d^{abc}$
\\
\hline
$\mathcal{O}_{10;\beta;f;1}=\frac{1}{12}\partial^4 {\cal O}_{\rm P2}$
& 0 & $\frac{1}{4}s_{123}^2 A_2$ & $f^{abc}$
\\
$\mathcal{O}_{10;\beta;f;2}=\mathcal{O}'_{10;5}$
& 0 & $\frac{1}{4}(s_{12}^2+s_{23}^2+s_{13}^2) A_2$  & $f^{abc}$
\\
\hline
\end{tabular}
\end{table}

\end{enumerate}

Let us explain the operator notation in Table \ref{tab:dim10d}.
In following context, we will label length-3 basis operators as
$\mathcal{O}_{\Delta;\alpha/\beta;f/d;i}$, where
$\Delta$ is the operator dimension, $\alpha/\beta$ labels the helicity sector,
$f/d$ labels the color factor, and $i$ is the numbering.
We also give the final basis operators for dimension 6 and 8 in
Table \ref{tab:dim6a8a}, and final basis for
dim 12-16 are given in Appendix \ref{app:newbasis}.
These operators will be the starting point for the computation of subsequent sections.

\begin{table}[!t]
\centering
\caption{Final basis operators at  dimension 6 and 8 cases.}
\label{tab:dim6a8a}
\vskip 0.4 cm
\begin{tabular}{|l|c|c|c|}
\hline
Basis operator  &  ${\cal F}^{(0)}(-,-,+)$ & ${\cal F}^{(0)}(-,-,-)$  & color factor
\\
\hline
$\mathcal{O}_{6;\beta;f;1}= \frac{1}{3}{\cal O}_{\rm P2}$
&  0& $ A_2 $  & $f^{abc}$
\\
\hline
$\mathcal{O}_{8;\alpha;f;1}={\cal O}_{\rm P1}-\frac{1}{6}\partial^2{\cal O}_{\rm P2}$
&  $A_1$ & 0 & $f^{abc}$
\\
\hline
$\mathcal{O}_{8;\beta;f;1}=\frac{1}{6}\partial^2{\cal O}_{\rm P2}$
&  0& $\frac{1}{2}s_{123}\ A_2 $  & $f^{abc}$
\\
\hline
\end{tabular}
\end{table}

\subsubsection*{length 4}

For length-4 case, we take dimension 8 as an example.
There are eight independent operators in total as shown in
(\ref{eq:len4dim8-0}).
When considering color ordered form factors,
we take coefficients of $\mathrm{tr}(1234)$ for single trace operators
and take coefficients of $\mathrm{tr}(12)\mathrm{tr}(34)$ for double trace operators.
In order to make each operator belong to only one
helicity sector, \emph{i.e.}~having color-ordered
form factors that are non-zero under only one type of helicity configuration, we recombine
above eight operators as follows:
\begin{align}
\label{eq:len4dim8-1}
&
\left(
\begin{array}{c}
\Xi_1\\  \Xi_2  \\  \Xi_3  \\  \Xi_4
\end{array}
\right)
=
\left(
\begin{array}{cccc}
0 & 1 & \frac{1}{2} & \frac{1}{4}
\\
1 & 0 & \frac{1}{2} & \frac{1}{4}
\\
0 & 1 & 0 & \frac{1}{4}
\\
1 & 0 & \frac{1}{2} & -\frac{1}{4}
\end{array}
\right)
\left(
\begin{array}{c}
\Xi'_1\\  \Xi'_2  \\  \Xi'_3  \\  \Xi'_4
\end{array}
\right),\qquad
\left(
\begin{array}{c}
\Xi_5\\  \Xi_6  \\  \Xi_7  \\  \Xi_8
\end{array}
\right)
=
\left(
\begin{array}{cccc}
 1 & 0 & \frac{1}{4} & \frac{1}{2}
\\
 0 & 1 & \frac{1}{4} & \frac{1}{2}
\\
 0 & 1 & -\frac{1}{4} & \frac{1}{2}
\\
 1 & 0 & \frac{1}{4} & 0
\end{array}
\right)
\left(
\begin{array}{c}
\Xi'_5\\  \Xi'_6  \\  \Xi'_7  \\  \Xi'_8
\end{array}
\right).
&
\end{align}

The correspondence between these newly obtained
operators and their color-ordered form factors
are summarized in  Table \ref{tab:len4spin2}.

\begin{table}[t]
\centering
\caption{Color-ordered tree-level minimal form factors of length-4 operators at dimension 8. The color factors are $\mathrm{tr}(1234)$ and $\mathrm{tr}(12)\mathrm{tr}(34)$ for single and double trace operators, respectively.}
\label{tab:len4spin2}
\vskip 0.4 cm
\begin{tabular}{|c|c|c|c|}
\hline
operator & $(-,-,+,+)$
& $(-,+,-,+)$ & $(-,-,-,-)$
\\
\hline
$\Xi_1$
& 0 &  0
&  $-\frac{1}{2}\la14\ra^2\la23\ra^2-\frac{1}{2}\la12\ra^2\la34\ra^2$

\\
\hline
$\Xi_2$
& 0   &  0  &  $-\la13\ra^2\la24\ra^2$

\\
\hline
$\Xi_3$
& $\frac{1}{2}\la12\ra^2[34]^2$ &  0  &  0

\\
\hline
$\Xi_4$
& 0   &  $\la13\ra^2[24]^2$  &  0

\\
\hline
\hline
$\Xi_5$
 & 0  & 0
&  $-\la14\ra^2\la23\ra^2-\la13\ra^2\la24\ra^2$

\\
\hline
$\Xi_6$
&  0  &  0
&  $-2\la12\ra^2\la34\ra^2$

\\
\hline
$\Xi_7$
& $2\la12\ra^2[34]^2$  &  0  &  0

\\
\hline
$\Xi_8$
& 0  &  $\la13\ra^2[24]^2$ &  0

\\
\hline
\end{tabular}
\end{table}

Notice the last column of Table \ref{tab:len4spin2} corresponds to uniform helicity,
so a total form factor has full bosonic symmetry, which means
its kinematic part and color factor have the same symmetric property
under label permutation.
For example, kinematic part of single-trace operators are invariant
under $Z_4$ cyclic permutation, and those of double trace operators are invariant
under $1\leftrightarrow2,3\leftrightarrow 4$ permutation.

Form factors under  one plus three minus or three plus one minus are not written
in the table. The results are all zero, because these eight operators
do not have $fff\tilde{f}$ or $\tilde{f}\tilde{f}\tilde{f}f$ components
under decomposition (\ref{eq:ff-decompose}).

For lengh-2 and length-3 operators, the countings of operators in these two different approaches agree with each other. However, this is not the case for higher length operators, where the evanescent operators appear.
Such evanescent operators do not show up in the 4-dim spinor approach,
on the other hand the field theory approach is valid for generic dimensions and can captures these  operators.
We leave the discussion to the future work.

\section{Two-loop form factor computation via unitarity}
\label{sec:computation}

In this section, we compute the one and two-loop form factors of the high dimensional operators discussed in the last section.
Our computation is based on the on-shell unitarity methods \cite{Bern:1994zx, Bern:1994cg, Britto:2004nc}, where the cut integrands are constructed by sewing tree-level components.  Furthermore, we combine the unitary method together with the integration by parts (IBP) reduction  \cite{Chetyrkin:1981qh, Tkachov:1981wb}. This ``unitarity-IBP" strategy not only makes the computation very efficient, but also provides important internal consistency checks for the results. Below we first outline the main strategy of the computation and then apply it to the concrete form factor computations.

The work flow of our calculation can be illustrated as follows:
\begin{align}
\mathcal{F}^{(l)}\Big|_{\mbox{cut}} &
= \prod (\textrm{tree blocks})  = \textrm{cut integrand}  \nonumber\\
& \xlongrightarrow{\mbox{IBP with cuts}}
 \sum_{\mbox{cut\ permitted}}  c_i I_i
\xlongrightarrow{\mbox{collect\ all\ cut channels}}
 \sum_{\mbox{complete}}c_i I_i = \mathcal{F}^{(l)} \,, \nonumber
\end{align}
where $I_i$ are IBP master integrals.
In the beginning, a particular cut channel (or cut  configuration) is chosen and one can calculate the cut integrand through tree-level data.
In order to avoid the issue of rational terms, here it is essential to use $D$-dimensional cut instead of four-dimensional cut.
The resulting cut integrand contains all the integrals whose topologies are permitted by the chosen cut.
As for the integral reduction, we use IBP method combined with on-shell conditions for the cut propagators.
Because terms proportional to cut inverse propagators vanish under cut condition, the expressions of IBP relations can be sharply shortened and therefore  computing efficiency is improved.
After cut-constrained IBP reduction, one obtains coefficients $c_i$ of all the cut-permitted master integrals.
Finally, by repeating the process for different cut channels, coefficients of all the master integrals are probed.
See also \cite{Jin:2019ile} for discussion.

\begin{figure}[tb]
  \centering
  \includegraphics[scale=0.4]{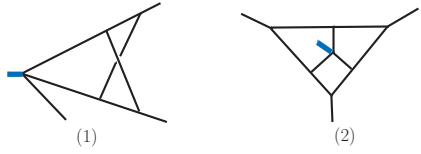}
  \caption{(1) The 2-loop non-planar topology has vanishing color factor. (2) Nonplanar topology contributing
  to leading color begins to appear at 3-loop.}
  \label{fig:no-subtriangle}
\end{figure}

In this paper we will mostly focus on the three-point form factors of length-three operators up to two-loop level. In these cases only planar integrals appear (and therefore the planar cuts are sufficient). This can be understood from a simple color analysis of Feynman diagrams that up to two-loop level
the minimal form factors of length-3 operators do not contain subleading color factors.
The only two-loop non-planar topology is shown in Fig.\,\ref{fig:no-subtriangle}(a), but its color factor is zero, for length-3 operators in both $f^{abc}$-sector and $d^{abc}$-sector.
As a comparison, at three loops
nonplanar topology (even at leading $N_c$ color) will contribute,
as shown in Fig.\,\ref{fig:no-subtriangle}(b), and therefore nonplanar cut is necessary.
Since the one-loop case is quite simple, below we will focus on the two-loop computation.

\begin{figure}[tb]
  \centering
  \includegraphics[scale=0.4]{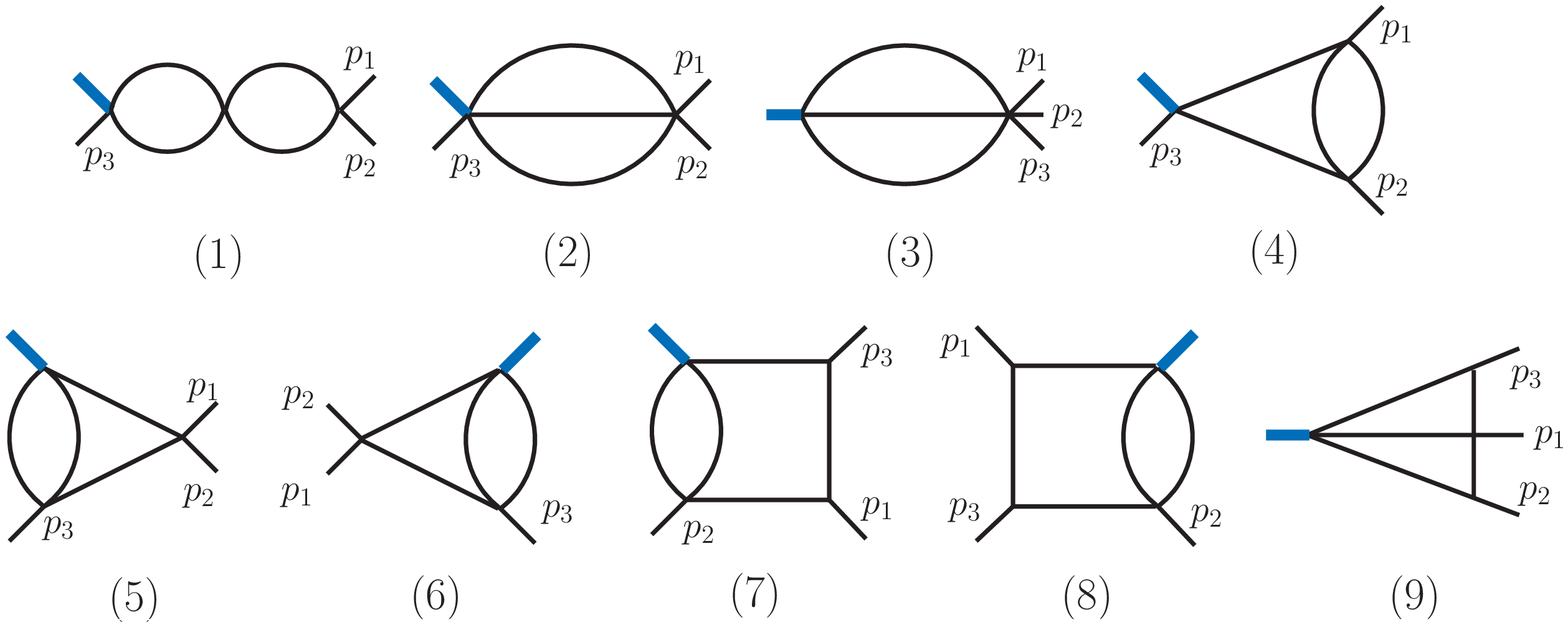}\\
  \caption{Master integrals (plus their cyclic permutations) that contribute to planar two-loop minimal form factors of length-3 operators.}
  \label{fig:all-MI}
\end{figure}

The complete set of two-loop master integrals for minimal length-3 form factors are given in Fig.\,\ref{fig:all-MI}. With color decomposition, the two-loop color-ordered form factors, associated with color factor $\mathrm{tr}(T^{a_1}T^{a_2}T^{a_3})$, can be written as a sum of master integrals $I_i$ as
\begin{align}
F^{(2)}_{\mathcal{O}}= &\Big(c_1 I_1+c_2 I_2 +c_3 I_3 +c_4 I_4 +\big[c_{5} I_{5}+c_6I_6\big] + \big[c_{7} I_{7}+c_8 I_8\big] +c_9I_9\Big) +\mathrm{cyc. perm.}(1,2,3) \,,
\label{eq:f-dabcFF}
\end{align}
where
master integrals $\{I_i\}$ strictly correspond to the topology and labeling given in Fig.\,\ref{fig:all-MI}.
The master coefficients $c_i$ are what we want to obtain using unitarity-IBP method.
Before considering that, let us discuss one important feature of the master integrals.

One can see that (5), (6) and (7), (8) in  Fig.\,\ref{fig:all-MI} are pairs of `mirror' topologies.
In color-ordered form factors, they should be considered to be independent because they are \emph{inequivalent} planar diagrams and therefore probed by different planar cuts.
On the other hand, they are closely related to each other:
graphically, (5) and (6) are related by label flipping $1\leftrightarrow 2$, while
(7) and (8) are related by flipping $3\leftrightarrow 1$.
From the planar color point of view, they are related by reversing color orientation,
 which is equivalent to a ``$C$-parity transformation" (see \emph{e.g.}~\cite{weinberg1995quantum}),
 so the kinematic parts of
a fixed color order $\mathrm{tr}(123)$ and the reversed color order $\mathrm{tr}(321)$
only differ by an overall $C$-parity factor decided
by external particles and inserted operator.
The external particles are three gluons which have $C$-parity $(-1)^3$, while  operator
from $f^{abc}/d^{abc}$ sector has $C$-parity $C_{\mathcal{O}} = +/-$, so the total $C$-parity of the form factor is $-/+$ \footnote{Considering  $f^{abc}F^a F^b F^c$, under C-parity it becomes $f^{abc}F^c F^b F^a(-1)^3$ which remains the same. }.
As a result, coefficients of integrals (5) and (6) as well as (7) and (8) are related with each other as:
\begin{equation}
c_6I_6=
\left
\{
\begin{array}{rcl}
-c_{5} I_{5}\big|_{1\leftrightarrow 2},\quad f\text{-sector}  \\
c_{5} I_{5}\big|_{1\leftrightarrow 2} ,\quad d\text{-sector}
\end{array}
\right.
,
\quad
c_8 I_8=
\left
\{
\begin{array}{rcl}
-c_{7} I_{7}\big|_{1\leftrightarrow 3} ,\quad f\text{-sector}  \\
c_{7} I_{7}\big|_{1\leftrightarrow 3} ,\quad d\text{-sector}
\end{array}
\right.
\,.
\end{equation}
Notice also that $I_3$ and its two cyclic partners share a degenerate expression,
but here we treat them as distinct ones and sum cyclic permutations together.

\begin{figure}[tb]
  \centering
  \includegraphics[scale=0.4]{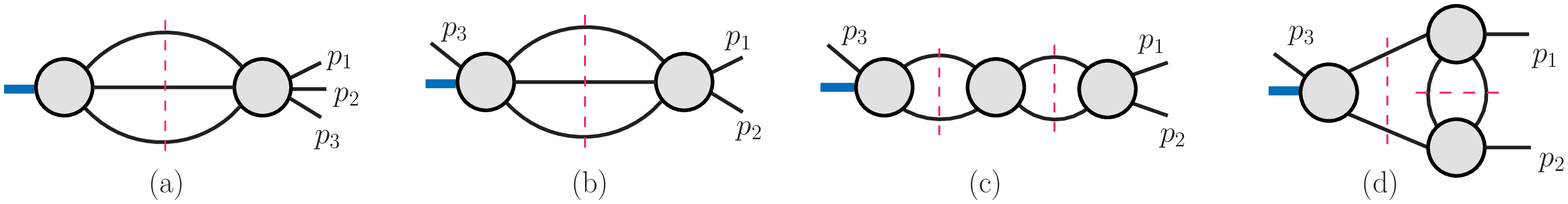}\\
  \caption{Complete set of cuts fully probing contributions
  from all the master integrals}
  \label{fig:all-cuts}
\end{figure}

A spanning set of planar cuts fully probing these master integrals are shown in Fig.\,\ref{fig:all-cuts}.
As already mentioned, a particular cut can probe only a subset of master integrals.
Among master integral coefficients,
$c_1$, $c_2$, $c_3$, $c_4$ are probed respectively by
cuts (c), (b), (a), (d) in Fig.\,\ref{fig:all-cuts}.
To probe $c_{5}$ one should apply $s_{123}$-triple-cut (a),
which also probes the coefficient of
integral $I_{6}|_{(p_3\rightarrow p_1\rightarrow p_2\rightarrow p_3)}$.
To probe $c_{7}$ and $c_9$ one can apply $s_{12}$-triple-cut (b), or $s_{312}$-triple-cut.
Notice the coefficients of  $I_{8}|_{(1\rightarrow3\rightarrow2\rightarrow1)}$
and  $I_9|_{(1\rightarrow2\rightarrow3\rightarrow1)}$
can also be probed by cut (b).
Since different cut channels can probe same or symmetry-related master integrals,
this provides strong consistency checks for the results.

\begin{figure}[tb]
  \centering
  \includegraphics[scale=0.4]{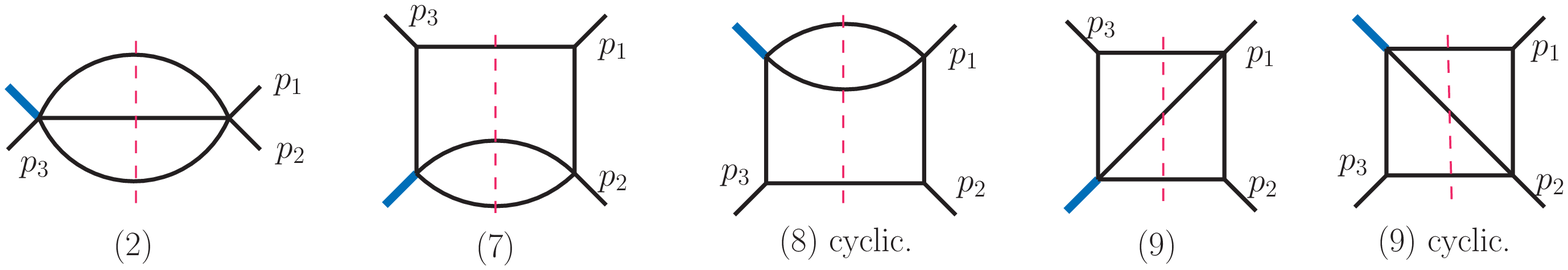}\\
  \caption{Master integrals probed by $s_{12}$-triple cut}
  \label{fig:s12triple}
\end{figure}

Below we provide some more details of the calculation by considering a particular cut channel.  Taking cut (b) in Fig.\,\ref{fig:all-cuts} as an example,
this cut allows us to determine the coefficients of master integrals as shown in Fig.\,\ref{fig:s12triple}.

The cut integrand is obtained by sewing a planar four-gluon tree form factor
together with a planar five-gluon tree amplitude.
Since we consider D-dimensional cuts, the tree results are computed via Feynman rules.
The sewing process involves the helicity sum of cut states:
\begin{align}
\label{eq:cut-tree}
\int {\rm d PS} \sum_{\text{helicities of }\epsilon_{l_1,l_2,l_3} }
F^{(0)}(p_3,-l_1, -l_2, -l_3)
A^{(0)}(p_1,p_2,l_3, l_2, l_1) \,,
\end{align}
which can be performed using
\begin{align}
\label{polarsum}
\epsilon^{\mu}(l_i)\circ\epsilon^{\nu}(l_i)
=\sum_{\mbox{helicities}} \epsilon^{\mu}(l_i)\epsilon^{\nu}(l_i)
=\eta^{\mu\nu}-\frac{q_i^\mu l_i^\nu+q_i^\nu l_i^\mu}{q_i\cdot l_i} \,,
\quad i=1,2,3,
\end{align}
where $q_i^\mu$ is an arbitrary reference momentum.
From \eqref{eq:cut-tree}, one obtains a cut integrand as a function of $\{\epsilon_i, p_i, l_a\}$.

Before performing integral reduction using IBP, it is convenient to further isolate external polarizations.
This can be achieved by choosing a set of gauge invariant basis $\{B_\alpha\} = \{ B_{\alpha}(\{\epsilon_i, p_i\})\}$ and expanding loop integrand  as (see \emph{e.g.}~\cite{Gehrmann:2011aa, Boels:2017gyc, Boels:2018nrr}):
\begin{align}
\label{Bdecompose}
\mathcal{F}_n(\{\epsilon_i\},\{p_i\},\{l_a\})\big|_{\rm cut}
=\sum_{\alpha} f_n^{\alpha}(\{p_i\},\{l_a\})\big|_{\rm cut} \cdot B_{\alpha} \,,
\end{align}
so that all the polarization information are absorbed into basis $\{B_\alpha\}$, leaving external
and loop momenta as the only variables of coefficients $\{f^\alpha_n\}$.
For the three-gluon setting, basis $\{B_\alpha\}$ has rank-4 and can be chosen as
\begin{align}
B_1=A_1 C_{23},\quad B_2=A_2 C_{31},\quad B_3=A_3 C_{21},\quad
B_4=A_1A_2A_3\,,
\end{align}
where $A_i$ and $C_{ij}$ are defined as
\begin{align}
A_i=\frac{\epsilon_i\cdot p_j}{p_i\cdot p_j}-\frac{\epsilon_i\cdot p_k}{p_i\cdot p_k},
\quad
C_{ij}=\epsilon_i\cdot\epsilon_j-\frac{(p_i\cdot \epsilon_j)(p_j\cdot \epsilon_i)}{p_i\cdot p_j} \,.
\end{align}
To get the coefficients $f^\alpha_n({p_i},{l_a})$, we only need to project the left of
(\ref{Bdecompose}) by dual basis $\{B^\alpha\}$:
\begin{align}
f_n^\alpha(\{p_i\},\{l_a\})=B^\alpha\circ \mathcal{F}_n(\{\epsilon_i\},\{p_i\},\{l_a\})\,.
\end{align}
Mutually dual basis $\{B_\alpha\}$ and $\{B^\alpha\}$ satisfy:
\begin{align}
B^\alpha\circ B_\beta=\delta^\alpha_\beta,\quad
B_\alpha=G_{\alpha\beta}B^\beta,\quad
G_{\alpha\beta}=B_\alpha\circ B_\beta\,.
\end{align}
See also \cite{Jin:2019opr} for further discussion of basis for the cases with external fermions.

It is then straightforward to perform IBP reduction for the scalar function  $\{f^\alpha_n\}$ under cut conditions, using public packages, (see \emph{e.g.}~\cite{Smirnov:2008iw, Lee:2013mka, Maierhoefer:2017hyi}).
After doing so we obtain the coefficients for five masters that are probed by  cut channel (b), such as shown in  Fig.\,\ref{fig:s12triple}.
Finally, by considering other cuts, all the master coefficients can be obtained and the full form factors are given as \eqref{eq:f-dabcFF}.

The master integrals listed  in Fig.\,\ref{fig:all-MI} are known in terms of 2d harmonic polylogarithms \cite{Gehrmann:2000zt, Gehrmann:2001jv}.
Substituting the expressions, bare form factors are obtained in explicit functional form.

\section{Anomalous dimensions of high dimensional operators}
\label{sec:anomalous-dim}

The form factors obtained in the last section contain the information of anomalous dimensions of high dimensional operators and also provide the Higgs-gluons amplitudes in the EFT.
In this section, we will focus on the anomalous dimensions, which can be extracted from the UV divergences of bare form factors.

As a brief outline, in Section \ref{sec:divergence} we first give a review of IR and UV subtraction, and
the effect of mixing between operators with different lengths is also analyzed.
Dilatation operators for length-two and three operators up to dimension 16 and their eigenvalues up to ${\cal O}(\alpha_s^2)$
are given in Section \ref{sec:dilatation}.
In Section \ref{sec:length4} we take dimension 8 case as an example to
cure the incompleteness of length truncation by considering the length-crossing mixing
between length-4 and length-3 operators, and  make correction to
two-loop anomalous dimension derived in Section \ref{sec:dilatation}.

\subsection{Subtraction of IR and UV Divergence}
\label{sec:divergence}

Bare form factors contain both IR and UV divergences. The general strategy is that by subtracting the universal IR divergences, one can extract the UV information unambiguously.
The renormalization with multi-operators in the basis generally exhibit non-trivial operator mixing effect.
We will first consider the simple case with an eigenstate operator that does not mix with other operators. Then we discuss the more general cases that involve the operator mixing effects.

\subsubsection*{Eigen operator case}

For an eigen-operator, the renormalization constant is a (non-matrix) single number,  $\mathcal{O}_R=Z_{\mathcal{O}} \mathcal{O}_B$.
The simplest example is the dimension-4 operator ${\rm tr}(F^2)$.
The analysis of UV divergence structure is relatively simple in this case, see \emph{e.g.}~\cite{Gehrmann:2011aa}.
Below we give a brief review which will also help to set up the notation.

Perturbative expansion of renormalization factors
of gauge coupling constant and operator $\mathcal{O}$ are:
\begin{align}
\label{alphaRG}
\alpha_0&=\alpha_s S^{-1}_\epsilon \frac{\mu^{2\epsilon}}{\mu_0^{2\epsilon}}
\Big[ 1-\frac{\beta_0}{\epsilon}\frac{\alpha_s}{4\pi}
+(\frac{\beta_0^2}{\epsilon^2}-\frac{\beta_1}{2\epsilon})
(\frac{\alpha_s}{4\pi})^2+{\cal O}(\alpha_s^3) \Big] \,,
\\
\label{ORG-1}
\mathcal{O}_R&= Z_{\mathcal{O}} \mathcal{O}_B
= \Big[ 1+Z_{\mathcal{O}}^{(1)}\frac{\alpha_s}{4\pi}+
Z_{\mathcal{O}}^{(2)}(\frac{\alpha_s}{4\pi})^2
+{\cal O}(\alpha_s^3) \Big]\mathcal{O}_B \,,
\end{align}
where we take renormalization under $\overline{\mathrm{MS}}$ scheme and  $S_\epsilon= (4\pi e^{- \gamma_{\text{E}}})^\epsilon$.
The normalization of beta functions are taken to be
\begin{align}
\beta_0=\frac{11 N_c}{3},\qquad \beta_1=\frac{34 N_c^2}{3}\,.
\end{align}
The perturbative expansion of anomalous dimension $\gamma$ is given as:
\begin{align}
\label{eq:eigen-AD}
\gamma=&-\frac{d\log Z_{\mathcal{O}}}{d\log\mu}
=-\frac{\partial\log Z_{\mathcal{O}}}{\partial \alpha_s}\frac{d\alpha_s}{d\log\mu}
=\sum_{i=1} \gamma^{(i)}(\frac{\alpha_s}{4\pi})^i\,,
\end{align}
and the 1-loop and 2-loop order anomalous dimension can be read out from
(\ref{alphaRG}) and (\ref{ORG-1}):
\begin{align}
\label{anomalous-dim}
\gamma^{(1)}=2\epsilon Z_{\mathcal{O}}^{(1)},\qquad
\gamma^{(2)}=4\epsilon \bigg(Z_{\mathcal{O}}^{(2)}
-\frac{(Z_{\mathcal{O}}^{(1)})^2}{2}
+\frac{\beta_0 Z_{\mathcal{O}}^{(1)}}{2\epsilon} \bigg)
=4\epsilon \Big(Z_{\mathcal{O}}^{(2)}\Big|_{\frac{1}{\epsilon}-\mathrm{part}}\Big) .
\end{align}
Anomalous dimension $\gamma^{(i)}$ should be regular as $\epsilon\rightarrow 0$, so
it is expected that $Z^{(1)}_{\mathcal{O}}\sim {\cal O}(\epsilon^{-1})$, and
the $\epsilon^{-2}$ part of $Z^{(2)}_{\mathcal{O}}$ is  $\frac{(Z_{\mathcal{O}}^{(1)})^2}{2}-\frac{Z_{\mathcal{O}}^{(1)}\beta_0}{2\epsilon}$,
which can be totally determined by 1-loop data, and resultantly the intrinsic contribution
for two-loop level is of the $\epsilon^{-1}$ part.

The coupling taken by tree-level $E$-point form factor of
length $L$ operator $\mathcal{O}$
is $g^{E-L}$.
Perturbative expansion of form factors can be written either in bare quantities
or in renormalized quantities:
\begin{align}
\label{F-in-bare-and-re}
\mathcal{F}_{\mathcal{O},R}=
\left
\{
\begin{array}{rcl}
Z_{\mathcal{O}}\mathcal{F}_{\mathcal{O},B}=
Z_{\mathcal{O}} \Big(\frac{\alpha_0}{4\pi} \Big)^{\frac{E-L}{2}}
\Big[ \mathcal{F}_{\mathcal{O}}^{(0)}+\frac{\alpha_0 }{4\pi}\mathcal{F}_{\mathcal{O},B}^{(1)}
+(\frac{\alpha_0 }{4\pi})^2 \mathcal{F}_{\mathcal{O},B}^{(2)}+ {\cal O}(\alpha^3_0) \Big]
&\ \textrm{(in bare)}\,,
&
\\
\Big(\frac{\alpha_s}{4\pi} \Big)^{\frac{E-L}{2}}
\Big[ \mathcal{F}_{\mathcal{O}}^{(0)}+\frac{\alpha_s }{4\pi}\mathcal{F}_{\mathcal{O},R}^{(1)}
+(\frac{\alpha_s }{4\pi})^2 \mathcal{F}_{\mathcal{O},R}^{(2)}
+{\cal O}(\alpha_s^3) \Big]
&\ \textrm{(in renorm.)} \,.
&
\end{array}
\right.
\end{align}

Relations between $\{F_{\mathcal{O},R}^{(l)}\}$ and $\{F_{\mathcal{O},B}^{(l)}\}$ can be obtained
by comparing above two expressions  order by order.
Plugging in gauge coupling renormalization (\ref{alphaRG}),
one can get
renormalization of form factors up to two-loop level as:
\begin{align}
\label{RG1loop}
\mathcal{F}_{\mathcal{O},R}^{(1)}=
&\mathcal{F}_{\mathcal{O},B}^{(1)}
+ \Big(Z_{\mathcal{O}}^{(1)}-\frac{\delta}{2}\frac{\beta_0}{\epsilon}
\Big) \mathcal{F}_{\mathcal{O}}^{(0)}\,,
\\
\label{RG2loop}
\mathcal{F}_{\mathcal{O},R}^{(2)}
=&\mathcal{F}_{\mathcal{O},B}^{(2)}
+\Big(
Z_{\mathcal{O}}^{(1)}
- \Big( 1+\frac{\delta}{2} \Big)\frac{\beta_0}{\epsilon}
\Big)
\mathcal{F}_{\mathcal{O},B}^{(1)}
+\Big(
Z_{\mathcal{O}}^{(2)}-\frac{\delta}{2}\frac{\beta_0}{\epsilon}Z_{\mathcal{O}}^{(1)}
-\frac{\delta}{2}\frac{\beta_1}{2\epsilon}
+\frac{\delta}{2}(\frac{\delta}{2}+1)\frac{\beta_0^2}{2\epsilon^2}
\Big)\mathcal{F}_{\mathcal{O}}^{(0)} \,.
\end{align}
Here $\delta=E-L$ accounts for the difference between number of external gluons and length of the operator.

To determine the UV divergences, one needs to subtract the IR divergences.
This can be achieved thanks to the universality of the IR divergences,
in the sense that they are independent of the type of operators but only depend on the data of external particles.
The IR subtraction formula of renormalized $E$-gluon amplitudes up to 2-loop order is known
 \cite{Catani:1998bh, Sterman:2002qn} (see also \cite{Gehrmann:2011aa}):
\begin{align}
\label{IR1loop}
\mathcal{F}_{\mathcal{O},R}^{(1)}
&=I^{(1)}(\epsilon) \mathcal{F}^{(0)}_{\mathcal{O}}+\mathcal{F}_{\mathcal{O},\mathrm{fin}}^{(1)}
+{\cal O}(\epsilon) \,,
\\
\label{IR2loop}
\mathcal{F}_{\mathcal{O},R}^{(2)}&=I^{(2)}(\epsilon) \mathcal{F}^{(0)}_{\mathcal{O}}
+ I^{(1)}(\epsilon)\mathcal{F}_{\mathcal{O},R}^{(1)}
+\mathcal{F}_{\mathcal{O},\mathrm{fin}}^{(2)}+{\cal O}(\epsilon) \,,
\end{align}
where
\begin{align}
I^{(1)}(\epsilon)&=-\frac{e^{\gamma_E\epsilon}}{\Gamma(1-\epsilon)}
\Big(\frac{N_c}{\epsilon^2}+\frac{\beta_0}{2\epsilon}
\Big)
\sum_{i=1}^E (-s_{i,i+1})^{-\epsilon} \,,
\\
I^{(2)}(\epsilon)&=
-\frac{1}{2} \big(I^{(1)}(\epsilon)\big)^2
-\frac{\beta_0}{\epsilon} I^{(1)}(\epsilon)
+\frac{e^{-\gamma_E\epsilon}\Gamma(1-2\epsilon)}{\Gamma(1-\epsilon)}
\Big(
\frac{\beta_0}{\epsilon}+\frac{67}{9}-\frac{\pi^2}{3}
\Big)
I^{(1)}(2\epsilon)
\nonumber\\
& \quad \  +E\frac{e^{\gamma_E\epsilon}}{\epsilon\Gamma(1-\epsilon)}
\Big(
\frac{\zeta_3}{2}+\frac{5}{12}+\frac{11\pi^2}{144}
\Big) \,.
\end{align}
For bare form factors, ${\cal O}(\epsilon^{-2})$ poles of 1-loop results and
 ${\cal O}(\epsilon^{-4}),{\cal O}(\epsilon^{-3})$ poles of 2-loop results
come exactly from the infrared divergence and therefore they should be
canceled according to the subtraction given in (\ref{IR1loop})and (\ref{IR2loop}),
which provides a consistency check of computation.

\subsection*{Operator mixing structure}

Next we consider the more general cases where different operators can mix with each other through loop corrections.
In such case,
a multi-operator generalization of renormalization multiplier $Z_{\mathcal{O}}$ is needed.
The main picture of the above discussion still applies, except that the renormalization constant should be taken as a \emph{matrix}:
\begin{equation}
{\cal O}_{R,i} = Z_i ^{~j} {\cal O}_{B,j} \,.
\end{equation}

Operator mixing can (and only can) take place among operators with the same canonical dimension.
Also, operators with different lengths but same dimension can mix with each other.
To put the discussion in a general context,
we will use $\mathcal{O}^{L}_{i}$ to denotes a length-$L$  operator labeled by $i$.
We denote the mixing of length-$L$ operator into length-$L'$ operator
at $\ell$-loop level as $Z^{(\ell)}_{L\rightarrow L'}$,
and the perturbative expansion of operator renormalization can be given as:%
\footnote{Here we suppose the definition of (classical) operators (as in \eqref{eq:def-general-opers}) contains no gauge coupling. One may absorb certain powers of coupling $g^m$ in operators in such way  that $O^{L}_{R,i}=\sum_{\ell=1}^\infty \Big(\frac{\alpha_s}{4\pi} \Big)^\ell \sum_j \big( Z^{(\ell)}_{L\rightarrow L'} \big)_i^{\ j} \mathcal{O}^{L'}_{B,j}$ holds for different lengths $L'$. This may be understood as a change of the definition of renormalization constants. More discussion on this will be given in Section~\ref{sec:length4} and Appendix~\ref{app:absorb-g}. \label{footnote:lengthchanging}}
\begin{align}
\label{ORG-2}
O^{L}_{R,i}&=\mathcal{O}^{L}_{B,j}
+\sum_{\ell=1}^\infty \bigg[
\Big(\frac{\alpha_s}{4\pi} \Big)^\ell \sum_j \big( Z^{(\ell)}_{L\rightarrow L} \big)_i^{\ j}
\mathcal{O}^{L}_{B,j}
+\sum_{k=1}^{\min(L-2,\ell-1)} \Big(\frac{\alpha_s}{4\pi} \Big)^{\ell-\frac{k}{2}}
\sum_{j} \big( Z^{(\ell)}_{L\rightarrow (L-k)} \big)_i^{\ j}
\mathcal{O}_{B,j}^{L-k}
\nonumber\\
& \qquad \qquad +\sum_{k=1}^\infty \Big(\frac{\alpha_s}{4\pi} \Big)^{\ell+\frac{k}{2}}
\sum_{j} \big( Z^{(\ell)}_{L\rightarrow (L+k)} \big)_i^{\ j}
\mathcal{O}_{B,j}^{L+k} \bigg] .
&
\end{align}
The upper bound of k for $L \rightarrow (L-k)$ case is based on the fact that the
length of terminal operator $L-k$ should be no shorter than 2, and the loop order
of a $L \rightarrow (L-k)$ process is no less than $k+1$.

Let us give a few general remarks regarding \eqref{ORG-2}.

\begin{enumerate}
\item
For length $L=2$, there is only one independent operator at dimension $2\Delta$, \emph{i.e.}~the descendant $\partial^{\Delta-2}\mathrm{tr}F^2$.
Since the length-2 operator is an eigenstate of anomalous dimension matrix, the length-2 operator will not mix into higher length operators along RG flow, therefore,
\begin{equation}
\label{eq:Z2toLasZero}
Z_{2\rightarrow L'}=0 \,, \qquad L'\geq3 \,.
\end{equation}

\item

On the other hand, high length operator can mix into length-2 ones, which starts from two-loop order (since
$Z^{(1)}_{L\rightarrow 2}=0$ for $L>2$).  The nonzero 2-loop mixing matrix elements $Z^{(2)}_{3\rightarrow 2}$ can
be probed by two-gluon form factors of length-3 operators.

\item
Similarly, operators of high length $L>3$ begin to mix into length-3 ones at $L-2$ loop order, since $Z^{(\ell<L-2)}_{L\rightarrow 3}=0$ (which can be seen from a Feynman diagram analysis).
So up to 2-loop order, the only higher length operators contributing to $Z^{(2)}_{L\rightarrow 3}$
are length-4 ones.
Length changing elements $Z^{(2)}_{4\rightarrow 3}$ can be probed by 3-gluon form factors of length-4 operators.
\end{enumerate}

For form factors with operator mixing,
the previous renormalization formulae \eqref{RG1loop}-\eqref{RG2loop} are
generalized to
\begin{align}
\label{formfactor-RG-1}
\mathcal{F}_{\mathcal{O}_i;R}^{(1)}&=
\mathcal{F}_{\mathcal{O}_i;B}^{(1)}
+ \sum_j (Z^{(1)})_i^{\ j}\mathcal{F}_{\mathcal{O}_j;B}^{(0)}
-\frac{\delta_i}{2}\frac{\beta_0}{\epsilon}
\mathcal{F}_{\mathcal{O}_i;B}^{(0)} \,,
\\
\label{formfactor-RG-2}
\mathcal{F}_{\mathcal{O}_i;R}^{(2)}&=
\mathcal{F}_{\mathcal{O}_i;B}^{(2)}
+ \sum_j (Z^{(1)})_i^{\ j}\mathcal{F}_{\mathcal{O}_j;B}^{(1)}
-\big(1+\frac{\delta_i}{2}\big)\frac{\beta_0}{\epsilon}\mathcal{F}_{\mathcal{O}_i;B}^{(1)}
&
\\
& \quad \ +\sum_j \Big[ (Z^{(2)})_i^{\ j}
-\frac{\delta_j}{2}\frac{\beta_0}{\epsilon}(Z^{(1)})_i^{\ j} \Big]
\mathcal{F}_{\mathcal{O}_j;B}^{(0)}
+ \Big[ -\frac{\delta_i}{2}\frac{\beta_1}{2\epsilon}
+(1+\frac{\delta_i}{2})\frac{\delta_i}{2}
\frac{\beta_0^2}{2\epsilon^2} \Big] \mathcal{F}_{\mathcal{O}_i;B}^{(0)} \,,
\nonumber
\end{align}
where $\delta_k = E - L_k$ with $L_k$ the length of ${\mathcal{O}_k}$.
In the case of minimal form factors of length-3 operators (with $E=L_i=3$), they reduce to
\begin{align}
\label{RG1loop-multi}
\mathcal{F}_{\mathcal{O}_i,R}^{(1)}&
=
\mathcal{F}_{\mathcal{O}_i,B}^{(1)}
+\sum_j (Z_{3\rightarrow 3}^{(1)})_{i}^{\ j}\mathcal{F}_{\mathcal{O}_j}^{(0)} \,,
\\
\label{RG2loop-multi}
\mathcal{F}_{\mathcal{O}_i,R}^{(2)}&
= \mathcal{F}_{\mathcal{O}_i,B}^{(2)}
+
\sum_j
(Z_{3\rightarrow 3}^{(1)})_{i}^{\ j}
\mathcal{F}_{\mathcal{O}_j,B}^{(1)}
-\frac{\beta_0}{\epsilon}
\mathcal{F}_{\mathcal{O}_i,B}^{(1)}
+
\sum_j
(Z^{(2)}_{3\rightarrow 3})_{i}^{\ j}
\mathcal{F}_{\mathcal{O}_j}^{(0)}+
(Z^{(2)}_{3\rightarrow 2})_{i}^{\ 0}
\mathcal{F}_{\mathcal{O}_0}^{(0)} \,.
\end{align}
Here we label the only length-2 operator as $\mathcal{O}_0$.

In the operator mixing case, we generalize the anomalous dimension \eqref{eq:eigen-AD} by introducing the dilation operator as
\begin{equation}
\mathds{D}=-\frac{d\log Z}{d\log\mu} \,,
\end{equation}
and the eigenvalues of the dilatation operator are anomalous dimensions. The expansion of dilatation operator may contain terms with odd power of coupling $g \sim \alpha_s^{1/2}$:
\begin{align}
\mathds{D}=\sum_{n=1}^\infty \bigg[ \Big(\frac{\alpha_s}{4\pi} \Big)^n
\mathds{D}^{(n)}
+ \Big(\frac{\alpha_s}{4\pi} \Big)^{n+\frac{1}{2}}
 \mathds{D}^{(n+\frac{1}{2})} \bigg] \,.
\end{align}
Up to ${\cal O}(\alpha_s^2)$ order, $\mathds{D}^{(1)},\mathds{D}^{(\frac{3}{2})},\mathds{D}^{(2)}$ can be read out from
(\ref{alphaRG}) and (\ref{ORG-2}):
\begin{align}
\label{Doperator1}
\mathds{D}^{(1)}&=2\epsilon Z^{(1)}_{L\rightarrow L},\qquad
\mathds{D}^{(\frac{3}{2})}=3\epsilon\Big(
Z^{(1)}_{L\rightarrow (L+1)}+Z^{(2)}_{L\rightarrow (L-1)}
\Big),&
\\
\label{Doperator2}
\mathds{D}^{(2)}&
=4\epsilon\Big(
Z^{(2)}_{L\rightarrow L}-\frac{1}{2} (Z^{(1)}_{L\rightarrow L})^2
+\frac{1}{2\epsilon} {\beta_0 Z^{(1)}_{L\rightarrow L}}
+Z^{(1)}_{L\rightarrow (L+2)}+Z^{(3)}_{L\rightarrow(L-2)}
\Big)&
\nonumber\\
&=4\epsilon\Big(
Z^{(2)}_{L\rightarrow L}\big|_{\frac{1}{\epsilon}-\mathrm{part}}
+Z^{(1)}_{L\rightarrow (L+2)}+Z^{(3)}_{L\rightarrow(L-2)}
\Big)\,.
\end{align}

The requirement that $\mathbb{D}^{(\ell)}$ is regular as $\epsilon\rightarrow 0$ predicts
several analytical structures of renormalization matrices:
\begin{enumerate}
\item
Two-loop order length-changed mixing $Z^{(2)}_{L\rightarrow(L-1)}$ has no $\epsilon^{-2}$
pole.
\item
The  $\epsilon^{-2}$ pole of $Z^{(2)}_{L\rightarrow L}$ is totally determined
by the 1-loop data:
\begin{equation}\label{Z22-Z1}
Z^{(2)}_{L\rightarrow L}\big|_{\frac{1}{\epsilon^2}-\mathrm{part}}
-\frac{1}{2}(Z^{(1)}_{L\rightarrow L})^2+\frac{\beta_0}{2\epsilon}
Z^{(1)}_{L\rightarrow L}=0 \,.
\end{equation}
\end{enumerate}
These provide important consistency checks of the calculation.

\subsection{Anomalous dimension matrices and eigenvalues}
\label{sec:dilatation}

Now we apply the above strategy to compute the results of dilation operators of length-3 operators up to canonical
 dimension 16 and up to 2 loop order.
Their eigenvalues, \emph{i.e.}~anomalous dimensions, are also computed up to ${\cal O}(\alpha_s^2)$.
The renormalization of the dimension-four operator is well known, see \emph{e.g.}~\cite{Gehrmann:2011aa}, and one-loop renormalization for dimension 6 and 8 operators were considered in \cite{Morozov:1985ef, Gracey:2002he, Neill:2009tn, Harlander:2013oja, Dawson:2014ora}.\footnote{Operator renormalization has also been considered in higher dimensional (6D and 8D) gauge theories in \cite{Gracey:2015xmw, Gracey:2017nly}. Techniques of using six dimensional spinor helicity formalism have been developed to compute form factors in pure YM theory in \cite{Huber:2019fea}.}
The two-loop renormalization for dimension-6 operators (also with quark operators) were obtained recently in \cite{Jin:2018fak, Jin:2019opr}. Other results for operators with dimension 8-16 are given for the first time to our knowledge.
These results have passed various non-trivial consistency checks, which are listed in the end of this subsection.

The basis operators we choose are given in Appendix \ref{app:newbasis}, labeled as
$\mathcal{O}_{\Delta,\alpha/\beta,f/d,i}$
which means it has dimension $\Delta$, color factor $f^{abc}/d^{abc}$, numbering $i$
and belongs to helicity sector $\alpha/\beta$ as
introduced in (\ref{eq:h-sector}).
As we consider operators only up to length-$3$,  the dilatation operator \eqref{Doperator2} in this section is defined as a truncated version as:
\begin{align}
\mathds{D}^{(1)}&=2\epsilon \Big(Z^{(1)}_{2\rightarrow 2}+Z^{(1)}_{3\rightarrow 3} \Big),
\nonumber\\
\label{Dforlen3-2}
\mathds{D}^{(\frac{3}{2})}&=3\epsilon\Big(
Z^{(1)}_{2\rightarrow 3}+Z^{(2)}_{3\rightarrow 2}\Big)
=3\epsilon Z^{(2)}_{3\rightarrow 2},
\\
\mathds{D}^{(2)}
&=4\epsilon\Big(
Z^{(2)}_{2\rightarrow 2}\big|_{\frac{1}{\epsilon}-\mathrm{part}}
+Z^{(2)}_{3\rightarrow 3}\big|_{\frac{1}{\epsilon}-\mathrm{part}}
\Big).
\nonumber
\end{align}
Note that $Z^{(1)}_{2\rightarrow 3}$ in $\mathds{D}^{(\frac{3}{2})}$ vanishes as mentioned in \eqref{eq:Z2toLasZero}.
Correction with high length operators will be discussed in  Section~\ref{sec:length4} and Appendix~\ref{app:absorb-g}.

\subsubsection*{Dimension 4}

At dimension 4 there is only one independent operator ${\cal O}_{4} = {\rm Tr}(F^2)$, which has been thoroughly studied.
Here we briefly review the result and also help to clarify the notations.
The renormalization constants up to two loops are:\footnote{In fact, $\mathcal{O}_4$ is just the Lagrangian of Yang-Mills theory, so an insertion of
$\mathcal{O}_4$ is equivalent to adding a vertex in Feynman diagrams.
The scaling behavior of renormalizable Yang-Mills Lagrangian is completely
determined by the running of gauge coupling, and
therefore one cannot not read any additional information
from anomalous dimension of $\mathcal{O}_4$
except for beta function order by order.}
\begin{equation}
Z_{{\cal O}_{4}}^{(1)} = - {11 N_c \over 3 \epsilon}
= -\frac{\beta_0}{\epsilon} \,, \qquad
Z_{{\cal O}_{4}}^{(2)} =  {121 N_c^2\over 9 \epsilon^2} - {34 N_c^2\over 3 \epsilon}
= {\beta_0^2 \over \epsilon^2} - {\beta_1 \over \epsilon} \,.
\end{equation}

The double pole term of $Z^{(2)}$ is determined by the one-loop result as expected.
According to (\ref{anomalous-dim}) the anomalous dimension at 1-loop and
2-loop level can be read out:
\begin{align}
\gamma^{(1)}_{\mathcal{O}_4}=-2\beta_0 = -\frac{22}{3} N_c \,, \qquad
\gamma^{(2)}_{\mathcal{O}_4}=-4\beta_1 = -\frac{136}{3} N_c \,.
\end{align}

\subsubsection*{Dimension 6}

At dimension 6 there is one length-2 descendent operator $\partial^2\mathrm{tr}(F^2)$
and one length-3 operator $\mathrm{tr}(F^3)$ belonging to $\beta$-helicity sector
according to (\ref{eq:h-sector}):
\begin{equation}
{\cal O}_{6;0} =\partial^2\mathcal{O}_4= \partial^2 {\rm Tr}(F^2) \,,
\qquad {\cal O}_{6;\beta;f;1} = \frac{1}{3}{\rm Tr}(F^3)\,.
\end{equation}
Two-loop minimal form factor of $\mathrm{tr}(F^3)$ was calculated in \cite{Jin:2018fak}.
The renormalization matrix at one and two-loop level are:
\begin{align}
Z_{{\cal O}_6}^{(1)} = {N_c\over \epsilon} \begin{pmatrix} - {11 \over 3} & 0
\\
0 &  {1 \over 2} \end{pmatrix} \,, \qquad
 Z_{{\cal O}_6}^{(2)} \big|_{{1\over\epsilon}\textrm{-part.}}
={N_c^2\over \epsilon} \begin{pmatrix}  - {34\over 3 }   & 0 \\ -1 &  {25\over 12} \end{pmatrix} \,.
\label{eq:dim6-Zmatrix}
\end{align}
At one-loop level there is no mixing, and as expected $Z^{(2)}_{2\rightarrow 3}=0$.
As defined in the perturbative expansion (\ref{Doperator1})-(\ref{Doperator2}),
the off-diagonal elements of the first column belong to
$Z^{(2)}_{3\rightarrow 2}$ so they are associated with $\alpha_s^{3/2}$ and contribute to $\mathds{D}^{(3/2)}$,
while the diagonal elements are associated with $\alpha_s^{2}$ and contribute to $\mathds{D}^{(2)}$.

The dilation matrix is straightforward to obtain using \eqref{Dforlen3-2},
and it reads:
\begin{align}
\mathds{D}_{\mathcal{O}_6} & =
\Big(\frac{\alpha_s}{4\pi} \Big)
\mathds{D}^{(1)} + \Big(\frac{\alpha_s}{4\pi} \Big)^{3/2}
\mathds{D}^{(3/2)} + \Big(\frac{\alpha_s}{4\pi} \Big)^2
\mathds{D}^{(2)} + {\cal O}(\alpha_s^{5/2}) \nonumber\\
& =
\left(
\begin{array}{cc}
 -\frac{22 }{3}\hat{\lambda} -\frac{136}{3} \hat{\lambda}^2 & 0 \\
 -3 \frac{\hat{\lambda}^{2}}{\hat{g}} & 1 \hat{\lambda}+\frac{25}{3} \hat{\lambda}^2 \\
\end{array}
\right) + {\cal O}(\frac{\hat{\lambda}^{3}}{\hat{g}}) \,,
\label{dilatation-dim6}
\end{align}
where  for the convenience of notation, we introduce newly normalized `t Hooft coupling $\hat{\lambda}$
and gauge coupling $\hat{g}$:
\begin{equation}
\hat{\lambda}:= N_c \frac{\alpha_s}{4\pi}\,,\qquad
\hat{g}:=\frac{g}{4\pi}\,.
\end{equation}
By diagonalizing the matrix, one obtains the anomalous dimension (eigenvalues) as:
\begin{align}
\label{eigenvalue-dim8}
\hat\gamma^{(1)}_{\mathcal{O}_6} = \left\{-\frac{22}{3}; 1\right\} \,, \qquad
\hat\gamma^{(2)}_{\mathcal{O}_6} = \left\{-\frac{136}{3}; \frac{25}{3}\right\} \,.
\end{align}

\subsubsection*{Dimension 8}

There are two length-3 basis operators at dimension 8, which are given in Table \ref{tab:dim6a8a}.
Together with $\mathcal{O}_{8;0}=\frac{1}{2} \partial^4\mathcal{O}_4$, they can be classified
into two helicity sectors according to (\ref{eq:h-sector}):
\begin{equation}
\begin{aligned}
\label{eq:dim8len23}
(\mathrm{f}^{123};-,-,+)&:\ \mathcal{O}_{8;\alpha;f;1}, \ \mathcal{O}_{8;0} \,, &
\\
(\mathrm{f}^{123};-,-,-)&:\ \mathcal{O}_{8;\beta;f;1}, \ \mathcal{O}_{8;0} \,. &
\end{aligned}
\end{equation}

An observation from (\ref{RG1loop-multi}) and (\ref{RG2loop-multi}) is that $(Z^{(1)})_i^{\ j}$
and $(Z^{(2)})_i^{\ j}$ representing mixing from $\mathcal{O}_i$ to $\mathcal{O}_j$ can
only be probed when $\mathcal{F}_{\mathcal{O}_j}^{(0)}$ does not vanish, i.e., the external helicity
setting matches the helicity sector of $\mathcal{O}_j$.
Since $\mathcal{O}_{8;0}$ belongs to both helicity sectors,
mixing from other operators to it can be probed under both $1^-2^-3^+$
and $1^-2^-3^-$.
This will provide another consistency check of final results:
different helicity settings must produce the same $3\rightarrow 2$ elements of
renormalization matrix, namely
$(Z^{(2)})_{\mathcal{O}_{8;i}}^{\ \ \mathcal{O}_{8;0}}$.

Let us show the mixing aspects of dimension-8 operators in detail
by analyzing UV divergence structure of their form factors.
For external gluon helicity case $(-,-,+)$, the 2-loop UV divergences
of $\mathcal{O}_{8;\alpha;f;1}$ and $\mathcal{O}_{8;\beta;f;1}$
at order $\mathcal{O}(\epsilon^{-1})$ are
\begin{align}
\label{eq:sameO0-1}
{\mathcal{F}^{(2)}_{\mathcal{O}_{8;\alpha;f;1}}(1^-, 2^-, 3^+)}
\Big|_{\frac{1}{\epsilon}\,\textrm{UV-div.}}
= &
{\mathcal{F}^{(0)}_{\mathcal{O}_{8;\alpha;f;1}}(1^-, 2^-, 3^+)}
\times \frac{N_c^2}{\epsilon}
\Big( -\frac{1}{3 v w}+\frac{269}{72} \Big)
\,, \\
{\mathcal{F}^{(2),\alpha}_{\mathcal{O}_{8;\beta;f;1}}(1^-, 2^-, 3^+)}
\Big|_{\frac{1}{\epsilon}\,\textrm{UV-div.}}
= &
{\mathcal{F}^{(0)}_{\mathcal{O}_{8;\alpha;f;1}}(1^-, 2^-, 3^+)}
\times \frac{N_c^2}{\epsilon}
\Big( -\frac{1}{v w} \Big)\,.
\end{align}
Here ${1 \over v w}= {s_{123}^2 \over s_{23}s_{13}} $ is the ratio between tree form factors of $\mathcal{O}_{8;0}$ and
$\mathcal{O}_{8;\alpha;f;1}$, and from these one reads the renormalization matrix elements:
\begin{align}
(Z^{(2)})_{\mathcal{O}_{8;\alpha;f;1}}^{~~~~\mathcal{O}_{8;0}}= - \frac{N_c^2}{3\epsilon} \,,  \quad
(Z^{(2)})_{\mathcal{O}_{8;\alpha;f;1}}^{~\mathcal{O}_{8;\alpha;f;1}}= \frac{ 269 N_c^2}{72\epsilon} \,,  \quad
(Z^{(2)})_{\mathcal{O}_{8;\beta;f;1}}^{~~~~\mathcal{O}_{8;0}}=  -\frac{N_c^2}{\epsilon} \,.
\label{eq:dim8-example-1}
\end{align}
For external gluon helicity case $(-,-,-)$, the 2-loop UV divergences at order
$\mathcal{O}(\epsilon^{-1})$ are
\begin{align}
\label{eq:sameO0-2}
{\mathcal{F}^{(2)}_{\mathcal{O}_{8;\alpha;f;1}}(1^-, 2^-, 3^-)}
\Big|_{\frac{1}{\epsilon}\,\textrm{UV-div.}}
= &
{\mathcal{F}^{(0)}_{\mathcal{O}_{8;\beta;f;1}}(1^-, 2^-, 3^-)}
\times\frac{N_c^2}{\epsilon} \Big( -\frac{1}{3 u v w}+\frac{5}{2} \Big)\,,
\\
{\mathcal{F}^{(2)}_{\mathcal{O}_{8;\beta;f;1}}(1^-, 2^-, 3^-)}
\Big|_{\frac{1}{\epsilon}\,\textrm{UV-div.}}
= &
{\mathcal{F}^{(0)}_{\mathcal{O}_{8;\beta;f;1}}(1^-, 2^-, 3^-)}
\times \frac{N_c^2}{\epsilon} \Big( -\frac{1}{u v w}+\frac{25}{12} \Big)\,.
\end{align}
Here ${1 \over u v w}= {s_{123}^3 \over s_{12} s_{23}s_{13}} $ is the ratio between tree form factors of $\mathcal{O}_{8;0}$ and
$\mathcal{O}_{8;\beta;f;1}$, and  one reads the renormalization matrix elements:
\begin{align}
(Z^{(2)})_{\mathcal{O}_{8;\alpha;f;1}}^{~~~~\mathcal{O}_{8;0}}= - \frac{N_c^2}{3\epsilon} \,,  & \qquad
(Z^{(2)})_{\mathcal{O}_{8;\alpha;f;1}}^{~\mathcal{O}_{8;\beta;f;1}}= \frac{ 5 N_c^2}{2\epsilon} \,,  \nonumber\\
(Z^{(2)})_{\mathcal{O}_{8;\beta;f;1}}^{~~~~\mathcal{O}_{8;0}}=  -\frac{N_c^2}{\epsilon} \,,  & \qquad
(Z^{(2)})_{\mathcal{O}_{8;\beta;f;1}}^{\mathcal{O}_{8;\beta;f;1}}=  \frac{25 N_c^2}{12 \epsilon} \,.
\label{eq:dim8-example-2}
\end{align}
As expected,
(\ref{eq:dim8-example-1}) and  (\ref{eq:dim8-example-2}) give the same  $3\rightarrow 2$ elements
$(Z^{(2)})_{\mathcal{O}_{8;\alpha;f;1}}^{~~~~\mathcal{O}_{8;0}}$ and
$(Z^{(2)})_{\mathcal{O}_{8;\beta;f;1}}^{~~~~\mathcal{O}_{8;0}}$, which is a non-trivial check of the result.

We arrange the operators into a vector $\{\mathcal{O}_{8;0},\mathcal{O}_{8;\alpha;f;1},
\mathcal{O}_{8;\beta;f;1}\}$.
According to this order, the one-loop renormalization matrix is:
\begin{equation}
Z_{{\cal O}_8}^{(1)}
= \frac{N_c}{\epsilon}
\begin{pmatrix}
-\frac{11}{3} & 0 & 0\\
0 &\frac{7}{6} & 0 \\
0 & 0 & \frac{1}{2}
\end{pmatrix}.
\end{equation}
At two-loop level,  the $Z^{(2)}$ matrix is:
\begin{equation}
\label{Z21of8dim}
Z_{{\cal O}_8}^{(2)}\Big|_{\frac{1}{\epsilon}\mbox{-part.}}
=\frac{N_c^2}{\epsilon}
\begin{pmatrix}
 -\frac{34}{3} & 0 & 0\\
 -\frac{1}{3} & \frac{269}{72} & \frac{5}{2} \\
 -1 & 0 & \frac{25}{12}
 \end{pmatrix} \,.
\end{equation}

Using \eqref{Dforlen3-2}, the dilation operator is given as
\begin{equation}
\label{dilatation-dim8}
\mathds{D}_{\mathcal{O}_8}=
\left(
\begin{array}{ccc}
 -\frac{22 }{3}\hat{\lambda} -\frac{136}{3} \hat{\lambda}^2 & 0 & 0 \\
 - \frac{\hat{\lambda}^{2}}{\hat{g}} & \frac{7}{3} \hat{\lambda}+\frac{269}{18} \hat{\lambda}^2
 & 10 \hat{\lambda}^2 \\
 -3 \frac{\hat{\lambda}^{2}}{\hat{g}} & 0 & \hat{\lambda}+\frac{25}{3} \hat{\lambda}^2 \\
\end{array}
\right) \,.
\end{equation}

Note that the off-diagonal elements of the first column belong to $Z^{(2)}_{3\rightarrow 2}$ and thus have a different coupling, as discussed below \eqref{eq:dim6-Zmatrix}.
Computing the  eigenvalues of \eqref{dilatation-dim8}, one obtains the anomalous dimensions up to ${\cal O}(\hat{\lambda}^2)$:
\begin{align}
\label{eigenvalue-dim8}
\hat\gamma^{(1)}_{\mathcal{O}_8} = \left\{-\frac{22}{3}; 1; \frac{7}{3}\right\} \,, \qquad
\hat\gamma^{(2)}_{\mathcal{O}_8} = \left\{-\frac{136}{3}; \frac{25}{3}; \frac{269}{18}\right\} \,.
\end{align}
From now on we sort eigenvalues according to the lowest dimensions they emerge. For example,
$\mathcal{O}(\hat{\lambda})$ anomalous dimension $-\frac{22}{3}$ appears at dimension four,
$1$ appears at dimension six, and $\frac{7}{3}$ appears at dimension eight, so they are listed
in the order of $\{-\frac{22}{3};1;\frac{7}{3}\}$.

\subsubsection*{Dimension 10}

There are five length-3 basis operators at dimension 10, as shown in Table \ref{tab:dim10}.
Together with $\mathcal{O}_{10;0}=\frac{1}{4}{\partial^6}\mathcal{O}_4$, they can be classified
into three sectors:
\begin{equation}
\begin{aligned}
(\mathrm{f}^{123};-,-,+)&:\ \mathcal{O}_{10;0},\ \mathcal{O}_{10;\alpha;f;1}, \ \mathcal{O}_{10;\alpha;f;2} \,, &
\\
(\mathrm{f}^{123};-,-,-)&:\ \mathcal{O}_{10;0}, \ \mathcal{O}_{10;\beta;f;1}, \ \mathcal{O}_{10;\beta;f;2} \,. &
\\
(\mathrm{d}^{123};-,-,+)&:\ \mathcal{O}_{10;\alpha;d;1} \,. &
\end{aligned}
\end{equation}
Operators with different color factors will never mix with each other because of
their opposite $C$-parities, so renormalization matrices of $f^{abc}$ and $d^{abc}$ sectors
can be written separately.

The computation of renormalization constant is the same as explained in the dimension-8 case
and therefore not repeated here,
see the discussion around (\ref{eq:dim8-example-1}) and  (\ref{eq:dim8-example-2}).
For the $f^{abc}$-sector, we arrange the operators as
$\{\mathcal{O}_{10;0},\mathcal{O}_{10;\alpha;f;1},\mathcal{O}_{10;\alpha;f;2},
\mathcal{O}_{10;\beta;f;1},\mathcal{O}_{10;\beta;f;2}\}$.
Renormalization matrices of $f^{abc}$ and $d^{abc}$ sector at one-loop level are
\begin{equation}
Z_{{\cal O}_{10,f}}^{(1)} = \frac{N_c}{\epsilon}
\left(
\begin{array}{c|cc|cc}
 -\frac{11}{3} & 0 & 0 & 0 & 0\\
 \hline
 0  &\frac{7}{6}& 0 & 0 & 0  \\
0 &-\frac{3}{5} & \frac{71}{30} &  0 & 0  \\
\hline
0& 0 & 0 &  \frac{1}{2} & 0 \\
0& 0 & 0 &  -1 & \frac{17}{6} \\
\end{array}
\right), \qquad
Z^{(1)}_{\mathcal{O}_{10;d}}=\frac{N_c}{\epsilon}\frac{13}{6} \,.
\end{equation}
We add the lines in the matrix to separate length-2 operator and also different helicity sectors.
We can see that the operator mixing starts to appear even at one-loop order, but only within the same helicity sector.

\begin{figure}[tb]
  \centering
  \includegraphics[scale=0.65]{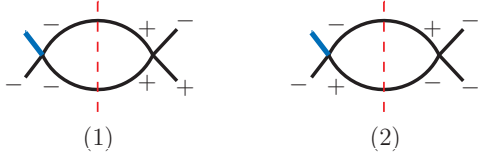}\\
  \caption{The helicity changing amplitude vanishes for 4-dimensional spacetime.}
  \label{fig:helicitychanging}
\end{figure}

Mixing between different helicity sectors does not happen at one-loop level, which can be understood
from unitarity cut.
A double cut divides one-loop minimal form factor $\mathcal{F}^{(1)}_3$ into a
tree level 3-gluon form factor $\mathcal{F}^{(0)}_3$ and
a tree level 4-gluon amplitude $\mathcal{A}^{(0)}_4$.
As shown in Figure~\ref{fig:helicitychanging}, a 4-dimensional $\mathcal{A}^{(0)}_4$ component of
helicity changing process has either $(+,-,-,-)$ or $(-,+,+,+)$ configuration which vanishes,
while at dimension $d$ it is of order $\mathcal{O}(\epsilon)$
and finally contributes to rational term without UV divergence.
Whereas helicity changing process does contain non-vanishing UV divergence at two-loop,
and therefore mixing between two helicity sectors are expected.

The intrinsic two-loop renormalization matrices of
 $f^{abc}$ and $d^{abc}$ sector are
\begin{equation}
\label{eq:Z2dim10}
Z_{{\cal O}_{10,f}}^{(2)}\Big|_{\frac{1}{\epsilon}\mbox{-part.}}
= \frac{N_c^2}{\epsilon}
\left(
\begin{array}{c|cc|cc}
-\frac{34}{3} &0 & 0 &  0 & 0
\\
\hline
-\frac{1}{3} & \frac{269}{72}& 0 & \frac{5}{2} & 0
\\
-\frac{209}{900} &-\frac{5579}{18000} &\frac{712}{125} & \frac{1493}{1200} & \frac{5}{36}
\\
 \hline
-1 & 0 & 0 &  \frac{25}{12} & 0
 \\
-\frac{19}{36}& \frac{139}{2400} & \frac{499}{800} &  -\frac{143}{288} & \frac{2195}{288}
\end{array}
\right), \quad
Z^{(2)}_{\mathcal{O}_{10;d}}\Big|_{\frac{1}{\epsilon}\mbox{-part.}}
=\frac{575}{144}\frac{N_c^2}{\epsilon} \,.
\end{equation}
Still, the off-diagonal elements of the first column of (\ref{eq:Z2dim10})
 belong to $Z^{(2)}_{3\rightarrow 2}$,
so they are associated with $\hat{\lambda}^{2}/\hat{g}$, while rest of the elements are
associated with $\hat{\lambda}^2$.
And we can see that operators of different helicity sectors can indeed mix with each other at two loops.

The dilation operator matrix can be obtained from \eqref{Dforlen3-2},
and for $f^{abc}$ sector it is
\begin{footnotesize}
\begin{align}
\label{dilatation-dim10}
&\hspace{-0.5cm}
\mathds{D}_{\mathcal{O}_{10,f}}=
\left(
\begin{array}{ccccc}
 -\frac{22 \hat{\lambda}}{3}-\frac{136}{3} \hat{\lambda}^2 & 0 & 0 & 0 & 0 \\
 - \frac{\hat{\lambda}^{2}}{\hat{g}} & \frac{7 \hat{\lambda}}{3}+\frac{269}{18} \hat{\lambda}^2 & 0
 & 10 \hat{\lambda}^2 & 0
   \\
 -\frac{209}{300} \frac{\hat{\lambda}^{2}}{\hat{g}} & -\frac{6 \hat{\lambda}}{5}-\frac{5579 \hat{\lambda}^2}{4500} &
   \frac{71 \hat{\lambda}}{15}+\frac{2848}{125} \hat{\lambda}^2 & \frac{1493}{300} \hat{\lambda}^2 &
   \frac{5}{9} \hat{\lambda}^2 \\
 -3 \frac{\hat{\lambda}^{2}}{\hat{g}} & 0 & 0 & \hat{\lambda}+\frac{25}{3} \hat{\lambda}^2 & 0 \\
 -\frac{19}{12} \frac{\hat{\lambda}^{2}}{\hat{g}} & \frac{139}{600} \hat{\lambda}^2
 & \frac{499}{200} \hat{\lambda}^2 &
  -2 \hat{\lambda} -\frac{143}{72} \hat{\lambda}^2
  & \frac{17 \hat{\lambda}}{3}+\frac{2195}{72} \hat{\lambda}^2
   \\
\end{array}
\right)\,.
\end{align}
\end{footnotesize}
Its eigenvalues give the anomalous dimensions:
\begin{align}
\label{eigenvalue-dim10}
\hat\gamma^{(1)}_{\mathcal{O}_{10,f}} =  \left\{-\frac{22}{3}; 1; \frac{7}{3};   \frac{71 }{15},
   \frac{17}{3} \right\} \,,
\qquad
\hat\gamma^{(2)}_{\mathcal{O}_{10,f}} =  \left\{-\frac{136}{3}; \frac{25}{3}; \frac{269}{18};
  \frac{2848}{125} , \frac{2195}{72} \right\} \,.
\end{align}
Here eigenvalues emerging at different dimensions are divided by semicolons and
those emerging at the same dimension are divided by commas.

For the singlet operator in $d^{abc}$ sector, one has:
\begin{equation}
\hat{\gamma}^{(1)}_{\mathcal{O}_{10,d}}=\frac{13 }{3} \,, \qquad
\hat{\gamma}^{(2)}_{\mathcal{O}_{10,d}}=\frac{575}{36} \,.
\end{equation}

\subsubsection*{Dimension 12}

There are 10 length-3 basis operators at dimension 12, as shown in Table \ref{tab:dim12}.
Together with $\mathcal{O}_{12;0}=\frac{1}{8}\partial^8\mathcal{O}_4$, they can be classified
into four sectors: $(f^{123};-,-,+),\ (f^{123};-,-,-)$,
$(d^{123};-,-,+)$, $(d^{123};-,-,-)$.

We arrange the operators as
$\{\mathcal{O}_{12;0},\mathcal{O}_{12;\alpha;f;1},...,\mathcal{O}_{12;\alpha;f;4},
\mathcal{O}_{12;\beta;f;1},...,\mathcal{O}_{12;\beta;f;3}\}$ for $f^{abc}$-sector,
and $\{\mathcal{O}_{12;\alpha;d;1},\mathcal{O}_{12;\alpha;d;2},
\mathcal{O}_{12;\beta;d;1}\}$ for $d^{abc}$-sector.
Renormalization matrices of $f^{abc}$ and $d^{abc}$ sectors at one-loop level are
\begin{equation}
Z_{{\cal O}_{12,f}}^{(1)}
 = \frac{N_c}{\epsilon}
\left(
\begin{array}{c|cccc|ccc}
-\frac{11}{3} &0 & 0 & 0 & 0 & 0 & 0 & 0 \\
\hline
0& \frac{7}{6} & 0 & 0 & 0 & 0 & 0 & 0 \\
0& -\frac{3}{5} &\frac{71}{30} & 0 & 0 & 0 & 0 & 0\\
0 &0 & -\frac{5}{4} & \frac{221}{60} &  -\frac{1}{6} & 0 & 0 & 0 \\
0& -1 & \frac{1}{10} & -\frac{19}{30} & \frac{37}{10} & 0 & 0 & 0\\
\hline
0 & 0 & 0 & 0 & 0 & \frac{1}{2} & 0 & 0\\
0 & 0 & 0 & 0 & 0 & -1 & \frac{17}{6} & 0\\
0 & 0 & 0 & 0 & 0 & 0 & -2 & \frac{9}{2}
\end{array}
\right) , \qquad
Z^{(1)}_{\mathcal{O}_{12,d}}
= \frac{N_c}{\epsilon}
\left(
\begin{array}{cc|c}
\frac{13}{6} & 0 & 0\\
 -\frac{1}{2} & \frac{41}{12} & 0\\
 \hline
0 & 0 & \frac{9}{2}
\end{array}
\right).
\end{equation}
The intrinsic two-loop renormalization matrices of
 $f^{abc}$ and $d^{abc}$ sector are
\begin{equation}
Z^{(2)}_{\mathcal{O}_{12,f}}\Big|_{\frac{1}{\epsilon}\mbox{-part.}}
= \frac{N_c^2}{\epsilon}
\left(
\begin{array}{c|cccc|ccc}
-\frac{34}{3} & 0 & 0 & 0 & 0 & 0 & 0 & 0
\\
\hline
-\frac{1}{3} &  \frac{269}{72}  & 0 & 0& 0 & \frac{5}{2} & 0 & 0
\\
-\frac{209}{900} & -\frac{5579}{18000} &\frac{712}{125} & 0 & 0
& \frac{1493}{1200} & \frac{5}{36} &0
\\
-\frac{31}{180} & \frac{53}{3600} & -\frac{36227}{28800} & \frac{3575983}{432000} & \frac{9793}{21600}
& \frac{13}{16} & \frac{16877}{14400} & -\frac{7319}{14400}
\\
-\frac{181}{900} & -\frac{60979}{36000} & \frac{78487}{72000} & -\frac{2177}{2000} & \frac{704167}{72000}
& \frac{1299}{1200} & \frac{115501}{43200} & -\frac{9803}{43200}
\\
\hline
-1 & 0 & 0 & 0 & 0 & \frac{25}{12} & 0 & 0
\\
-\frac{19}{36} & \frac{139}{2400} & \frac{499}{800} & 0 & 0
& -\frac{143}{288} &\frac{2195}{288} & 0
\\
-\frac{1}{3} & \frac{4}{15} & \frac{121}{400} & \frac{637}{800} & -\frac{211}{800} &
\frac{119}{120} & -\frac{15643}{7200} & \frac{79313}{7200}
\end{array}
\right),
\end{equation}
\begin{equation}
Z^{(2)}_{\mathcal{O}_{12,d}}\Big|_{\frac{1}{\epsilon}\mbox{-part.}}
= \frac{N_c^2}{\epsilon}
\left(
\begin{array}{cc|c}
\frac{575}{144} & 0 & 0\\
-\frac{23347}{14400} & \frac{46517}{5760} & \frac{487}{1800}\\
\hline
\frac{3349}{7200} & -\frac{2591}{2400} & \frac{150391}{14400}
\end{array}
\right).
\end{equation}

The dilation operator matrix is straightforward to obtain using \eqref{Dforlen3-2}. The anomalous dimensions are given by the eigenvalues, which read
\begin{align}
\label{eigenvalue-dim12f}
\hat\gamma^{(1)}_{\mathcal{O}_{12,f}} & =
\{-\frac{22}{3}; 1; \frac{7}{3};  \frac{71 }{15},  \frac{17}{3};  \frac{241}{30}, \frac{101 }{15},   9 \} \,,
\\
\hat\gamma^{(2)}_{\mathcal{O}_{12,f}} & =
\{-\frac{136}{3}; \frac{25}{3}; \frac{269}{18};  \frac{2848}{125} ,  \frac{2195}{72};  \frac{49901119 }{1404000}, \frac{8585281 }{234000}, \frac{79313 }{1800}\} \,,
\\
\label{eigenvalue-dim12d}
\hat\gamma^{(1)}_{\mathcal{O}_{12,d}} & =
\left\{\frac{13 }{3}; \frac{41 }{6},  9  \right\} \,, \qquad
\hat\gamma^{(2)}_{\mathcal{O}_{12,d}}  =
\left\{ \frac{575}{36}; \frac{46517 }{1440}, \frac{150391}{3600}\right\}  \,.
\end{align}

\subsubsection*{Dimension 14}

There are 15 length-3 basis operators at dimension 14, as shown in Table \ref{tab:dim14}.
Together with $\mathcal{O}_{14;0}=\frac{1}{16}\partial^{10}\mathcal{O}_4$, they can be classified
into four sectors: $(f^{123};-,-,+),\ (f^{123};-,-,-)$,
$(d^{123};-,-,+)$, $(d^{123};-,-,-)$.

We arrange the operators as
$\{\mathcal{O}_{14;0},\mathcal{O}_{14;\alpha;f;1},...,\mathcal{O}_{14;\alpha;f;6},
\mathcal{O}_{14;\beta;f;1},...,\mathcal{O}_{14;\beta;f;4}\}$ for $f^{abc}$-sector,
and $\{\mathcal{O}_{14;\alpha;d;1},...,\mathcal{O}_{14;\alpha;d;4},
\mathcal{O}_{14;\beta;d;1}\}$ for $d^{abc}$-sector.
Renormalization matrices of $f^{abc}$ and $d^{abc}$ sectors at one-loop level are
\begin{equation}
Z^{(1)}_{\mathcal{O}_{14,f}}=\frac{N_c}{\epsilon}
\left(
\begin{array}{c|cccccc|cccc}
-\frac{11}{3} & 0 & 0 & 0 & 0 & 0 & 0 & 0 & 0 & 0 & 0
\\
\hline
0 & \frac{7}{6} & 0 & 0 & 0 & 0 & 0 & 0 & 0 & 0 & 0
\\
0 & -\frac{3}{5} & \frac{71}{30} & 0 & 0 & 0 & 0 & 0 & 0 & 0 & 0
\\
0 & 0 & -\frac{5}{4} & \frac{221}{60} & -\frac{1}{6} & 0 & 0 & 0 & 0 & 0 & 0
\\
0 & -1 & \frac{1}{10} & -\frac{19}{30} & \frac{37}{10} & 0 & 0 & 0 & 0 & 0 & 0
\\
0 & \frac{17}{84} & -\frac{17}{28} & -\frac{47}{70} & -\frac{17}{28} & \frac{337}{84} & \frac{5}{14} & 0 & 0 & 0 & 0
\\
0 & -\frac{3}{20} & \frac{9}{20} & -1 & -\frac{31}{20} & -\frac{1}{4} & \frac{31}{6} & 0 & 0 & 0 & 0
\\
\hline
0 & 0 & 0 & 0 & 0 & 0 & 0 & \frac{1}{2} & 0 & 0 & 0
\\
0 & 0 & 0 & 0 & 0 & 0 & 0 & -1 & \frac{17}{6} & 0 & 0
\\
0 & 0 & 0 & 0 & 0 & 0 & 0 & 0 & -2 & \frac{9}{2} & 0
\\
0 & 0 & 0 & 0 & 0 & 0 & 0 & \frac{1}{3} & -2 & \frac{1}{3} & \frac{43}{10}
\end{array}
\right),
\end{equation}
\begin{equation}
Z^{(1)}_{\mathcal{O}_{14,d}}=\frac{N_c}{\epsilon}
\left(
\begin{array}{cccc|c}
 \frac{13}{6} & 0 & 0 & 0 & 0
\\
 -\frac{1}{2} & \frac{41}{12} & 0 & 0 & 0
\\
 \frac{1}{2} & -2 & \frac{301}{60} & -\frac{2}{3} & 0
\\
 -1 & 1 & -\frac{3}{10} & \frac{25}{6} & 0
\\
\hline
 0 & 0 & 0 & 0 & \frac{9}{2}
\end{array}
\right).
\end{equation}
The intrinsic two-loop renormalization matrices of
 $f^{abc}$ and $d^{abc}$ sector are:
 {\small
\begin{align}
& Z^{(2)}_{\mathcal{O}_{14,f}}\Big|_{\frac{1}{\epsilon}-\mathrm{part.}}
=\frac{N_c^2}{\epsilon} \times \\
& \left(
\begin{array}{c|cccccc|cccc}
-\frac{34}{3} & 0 & 0 & 0 & 0 & 0 & 0 & 0 & 0 & 0 & 0
\\
\hline
-\frac{1}{3} & \frac{269}{72} & 0 & 0 & 0 & 0 & 0 & \frac{5}{2} & 0 & 0 & 0
\\
-\frac{209}{900} & -\frac{5579}{18000} & \frac{712}{125} & 0 & 0 & 0 & 0 &
\frac{1493}{1200} & \frac{5}{36} & 0 & 0
\\
-\frac{31}{180} & \frac{53}{3600} & -\frac{36227}{28800} &
\frac{3575983}{432000} & \frac{9793}{21600} & 0 & 0 &
\frac{13}{16} & \frac{16877}{14400} & -\frac{7319}{14400} & 0
\\
-\frac{181}{900} & -\frac{60979}{36000} & \frac{78487}{72000} &
-\frac{2177}{2000} & \frac{704167}{7200} & 0 & 0 &
\frac{1299}{1200} & \frac{115501}{43200} & -\frac{9803}{43200} & 0
\\
-\frac{523}{3920} & -\frac{2201287}{29635200} & \frac{605939}{1975680} &
-\frac{64128769}{24696000} & \frac{3303367}{9878400} &
\frac{332422343}{29635200} & \frac{6699071}{14817600} &
\frac{37547}{78400} & \frac{75071}{39200} & -\frac{497}{576} & \frac{103}{1440}
\\
-\frac{809}{5600} & -\frac{12166789}{21168000} & \frac{11202299}{7056000} &
-\frac{73487}{36750} & -\frac{9182209}{7056000} & \frac{37249}{156800} & \frac{26302879}{2116800} &
\frac{1613}{3360} & \frac{17401}{6720} & \frac{19}{225} & \frac{1187}{2880}
\\
\hline
-1 & 0 & 0 & 0 & 0 & 0 & 0 &
\frac{25}{12} & 0 & 0 & 0
\\
-\frac{19}{36} & \frac{139}{2400} & \frac{499}{800} & 0 & 0 & 0 & 0 &
-\frac{143}{288} & \frac{2195}{288} & 0 & 0
\\
-\frac{1}{3} & \frac{4}{15} & \frac{121}{400} & \frac{637}{800} & -\frac{211}{800} & 0 & 0 &
\frac{119}{120} & -\frac{15643}{7200} & \frac{79313}{7200} & 0
\\
-\frac{209}{900} & \frac{6299}{21168} & \frac{6767}{35280} & \frac{71063}{88200} &
-\frac{34723}{176400} & \frac{77523}{176400} & -\frac{36091}{264600} &
\frac{22723}{21600} & -\frac{35}{48} & -\frac{2861}{5400} & \frac{443801}{36000}
\end{array}
\right).
\nonumber
\end{align}
}
\begin{equation}
Z^{(2)}_{\mathcal{O}_{14,d}}\Big|_{\frac{1}{\epsilon}-\mathrm{part.}}
=\frac{N_c^2}{\epsilon}
\left(
\begin{array}{cccc|c}
 \frac{575}{144} & 0 & 0 & 0 & 0
\\
 -\frac{23347}{14400} & \frac{46517}{5760} & 0 & 0 & \frac{487}{1800}
\\
 \frac{3883}{4032} & -\frac{171823}{37800} & \frac{36597791}{3024000} &
 -\frac{29581}{16800} & -\frac{1789}{4800}
\\
 -\frac{9271}{11200} & -\frac{35239}{50400} & \frac{74209}{168000} & \frac{188599}{18900} &
 \frac{2101}{4800}
\\
\hline
 \frac{3349}{7200} & -\frac{2591}{2400} & 0 & 0 & \frac{150391}{14400}
\end{array}
\right).
\end{equation}

The dilation operator matrix is straightforward to obtain using \eqref{Dforlen3-2} and we will not provide here. The anomalous dimensions are given by the eigenvalues, which read
{\small
\begin{align}
\label{eigenvalue-dim14f}
\hat\gamma^{(1)}_{\mathcal{O}_{14,f}} & =
\Big\{-\frac{22}{3}; 1; \frac{7}{3};  \frac{71 }{15},  \frac{17}{3};  \frac{241}{30}, \frac{101 }{15},   9;
\frac{61}{6}, \frac{172 }{21}, \frac{43 }{5} \Big\} \,,
\\
\hat\gamma^{(2)}_{\mathcal{O}_{14,f}} & =
\Big\{-\frac{136}{3}; \frac{25}{3}; \frac{269}{18};  \frac{2848}{125} ,  \frac{2195}{72};  \frac{49901119 }{1404000}, \frac{8585281 }{234000}, \frac{79313 }{1800};
\frac{4392073141}{87847200}, \frac{685262197 }{15373260}, \frac{443801}{9000}
\Big\},
\\
\label{eigenvalue-dim14d}
\hat\gamma^{(1)}_{\mathcal{O}_{14,d}} & =
\Big\{\frac{13 }{3}; \frac{41 }{6},  9 ;
\frac{1}{60}(551+3 \sqrt{609}), \frac{1}{60} (551-3 \sqrt{609})
 \Big\} \,,
\\
\hat\gamma^{(2)}_{\mathcal{O}_{14,d}} & =
\Big\{ \frac{575}{36}; \frac{46517 }{1440}, \frac{150391}{3600};
\frac{(5809305897+19635401 \sqrt{609}) }{131544000},
 \frac{(5809305897-19635401 \sqrt{609}) }{131544000}
\Big\}  \,.
\end{align}
}
At operator dimension 14,
the two-loop anomalous dimensions in $d^{abc}$ sector start to involve irrational numbers.

\subsubsection*{Dimension 16}

There are 22 length-3 basis operators at dimension 16, as shown in Table \ref{tab:dim16}.
Together with $\mathcal{O}_{16;0}=\frac{1}{32}\partial^{12}\mathcal{O}_4$, they can be classified
into four sectors: $(f^{123};-,-,+),\ (f^{123};-,-,-)$,
$(d^{123};-,-,+)$, $(d^{123};-,-,-)$.

We arrange the operators as
$\{\mathcal{O}_{16;0},\mathcal{O}_{16;\alpha;f;1},...,\mathcal{O}_{16;\alpha;f;9},
\mathcal{O}_{16;\beta;f;1},...,\mathcal{O}_{16;\beta;f;5}\}$ for $f^{abc}$-sector
and $\{\mathcal{O}_{16;\alpha;d;1},...,\mathcal{O}_{16;\alpha;d;6},
\mathcal{O}_{16;\beta;d;1},\mathcal{O}_{16;\beta;d;2}\}$ for $d^{abc}$-sector.
Renormalization matrices of $f^{abc}$ and $d^{abc}$ sector at one-loop level are
\begin{align}
\label{eq:Z1dim16f}
Z^{(1)}_{\mathcal{O}_{16,f}}=\frac{N_c}{\epsilon}
\left(
\begin{array}{c|ccccccccc|ccccc}
-\frac{11}{3}& 0 & 0 & 0 & 0 & 0 & 0 & 0 & 0 & 0 & 0 & 0 & 0 & 0 & 0
\\
\hline
0& \frac{7}{6} & 0 & 0 & 0 & 0 & 0 & 0 & 0 & 0 & 0 & 0 & 0 & 0 & 0
\\
0&  -\frac{3}{5} & \frac{71}{30} & 0 & 0 & 0 & 0 & 0 & 0 & 0 & 0 & 0 & 0 & 0 & 0
\\
0& 0 & -\frac{5}{4} & \frac{221}{60} & -\frac{1}{6} & 0 & 0 & 0 & 0 & 0 & 0 & 0 & 0 & 0 & 0
   \\
0& -1 & \frac{1}{10} & -\frac{19}{30} & \frac{37}{10} & 0 & 0 & 0 & 0 & 0 & 0 & 0 & 0 & 0 &
   0 \\
0& \frac{17}{84} & -\frac{17}{28} & -\frac{47}{70} & -\frac{17}{28} & \frac{337}{84} &
   \frac{5}{14} & 0 & 0 & 0 & 0 & 0 & 0 & 0 & 0
   \\
0& -\frac{3}{20} & \frac{9}{20} & -1 & -\frac{31}{20} & -\frac{1}{4} & \frac{31}{6} & 0 & 0
   & 0 & 0 & 0 & 0 & 0 & 0
   \\
0& \frac{13}{30} & -\frac{13}{15} & \frac{13}{10} & -\frac{13}{10} & -\frac{5}{2} &
   \frac{13}{15} & \frac{961}{210} & \frac{8}{15} & 0 & 0 & 0 & 0 & 0 & 0
   \\
 0& \frac{71}{105} & -\frac{212}{105} & \frac{141}{35} & -\frac{71}{35} & -\frac{141}{35} &
   \frac{79}{105} & -\frac{38}{35} & \frac{223}{35} & \frac{5}{14} & 0 & 0 & 0 & 0 & 0
   \\
 0& \frac{17}{70} & \frac{19}{105} & -\frac{19}{70} & -\frac{121}{70} & -\frac{11}{42} &
   \frac{16}{105} & -\frac{6}{5} & \frac{127}{210} & \frac{559}{105} & 0 & 0 & 0 & 0 & 0
   \\
   \hline
 0& 0 & 0 & 0 & 0 & 0 & 0 & 0 & 0 & 0 & \frac{1}{2} & 0 & 0 & 0 & 0 \\
 0& 0 & 0 & 0 & 0 & 0 & 0 & 0 & 0 & 0 & -1 & \frac{17}{6} & 0 & 0 & 0 \\
 0& 0 & 0 & 0 & 0 & 0 & 0 & 0 & 0 & 0 & 0 & -2 & \frac{9}{2} & 0 & 0 \\
 0& 0 & 0 & 0 & 0 & 0 & 0 & 0 & 0 & 0 & \frac{1}{3} & -2 & \frac{1}{3} & \frac{43}{10} & 0
   \\
 0& 0 & 0 & 0 & 0 & 0 & 0 & 0 & 0 & 0 & \frac{1}{2} & -\frac{5}{2} & \frac{5}{2} &
   -\frac{11}{4} & \frac{67}{12} \\
\end{array}
\right) ,
\end{align}
\begin{align}
\label{eq:Z1dim16d}
Z^{(1)}_{\mathcal{O}_{16,d}}=\frac{N_c}{\epsilon}
\left(
\begin{array}{cccccc|cc}
 \frac{13}{6} & 0 & 0 & 0 & 0 & 0 & 0 & 0 \\
 -\frac{1}{2} & \frac{41}{12} & 0 & 0 & 0 & 0 & 0 & 0 \\
 \frac{1}{2} & -2 & \frac{301}{60} & -\frac{2}{3} & 0 & 0 & 0 & 0 \\
 -1 & 1 & -\frac{3}{10} & \frac{25}{6} & 0 & 0 & 0 & 0 \\
 -\frac{2}{5} & \frac{1}{5} & 0 & -\frac{1}{5} & \frac{307}{60} & \frac{7}{20} & 0 & 0 \\
 \frac{1}{3} & -1 & \frac{1}{2} & -\frac{7}{3} & \frac{13}{12} & \frac{67}{12} & 0 & 0 \\
 \hline
 0 & 0 & 0 & 0 & 0 & 0 & \frac{9}{2} & 0 \\
 0 & 0 & 0 & 0 & 0 & 0 & \frac{7}{12} & \frac{67}{12} \\
\end{array}
\right) .
\end{align}

The intrinsic two-loop renormalization matrices of
 $f^{abc}$ and $d^{abc}$ sector are
\begin{align}
Z^{(2)}_{\mathcal{O}_{16,f}}\Big|_{\frac{1}{\epsilon}-\mathrm{part}.}=
\frac{N_c^2}{\epsilon}
\left(
\begin{array}{c|c}
M & N
\end{array}
\right) \,,
\end{align}
where two block matrices $M$, $N$ are
\begin{footnotesize}
\begin{align}
&M=
\left(
\begin{array}{c|ccccccccc}
 -\frac{34}{3} & 0 & 0 & 0 & 0 & 0 & 0 & 0 & 0 & 0 \\
 \hline
 -\frac{1}{3} & \frac{269}{72} & 0 & 0 & 0 & 0 & 0 & 0 & 0 & 0 \\
 -\frac{209}{900} & -\frac{5579}{18000} & \frac{712}{125} & 0 & 0 & 0 & 0 & 0 & 0 & 0 \\
 -\frac{31}{180} & \frac{53}{3600} & -\frac{36227}{28800} & \frac{3575983}{432000} &
   \frac{9793}{21600} & 0 & 0 & 0 & 0 & 0 \\
 -\frac{181}{900} & -\frac{60979}{36000} & \frac{78487}{72000} & -\frac{2177}{2000} &
   \frac{704167}{72000} & 0 & 0 & 0 & 0 & 0 \\
 -\frac{523}{3920} & -\frac{2201287}{29635200} & \frac{605939}{1975680} &
   -\frac{64128769}{24696000} & \frac{3303367}{9878400} & \frac{332422343}{29635200} &
   \frac{6699071}{14817600} & 0 & 0 & 0 \\
 -\frac{809}{5600} & -\frac{12166789}{21168000} & \frac{11202299}{7056000} &
   -\frac{73487}{36750} & -\frac{9182209}{7056000} & \frac{37249}{156800} &
   \frac{26302879}{2116800} & 0 & 0 & 0 \\
 -\frac{269}{2520} & \frac{125599}{10584000} & \frac{50369}{1323000} &
   -\frac{98317}{1176000} & \frac{73489}{392000} & -\frac{8625329}{3528000} &
   -\frac{97913}{756000} & \frac{90760559}{7408800} & \frac{25354501}{21168000} &
   \frac{40519}{56448} \\
 -\frac{19717}{176400} & \frac{3374557}{7408800} & -\frac{102465523}{74088000} &
   \frac{5260289}{1764000} & -\frac{6201763}{4939200} & -\frac{115070197}{24696000} &
   \frac{10687837}{9261000} & \frac{6498287}{9261000} & \frac{1025255701}{74088000} &
   -\frac{25511}{493920} \\
 -\frac{19717}{176400} & -\frac{2733089}{9261000} & \frac{88146899}{74088000} &
   -\frac{5678651}{3528000} & -\frac{1966229}{12348000} & \frac{17842339}{18522000} &
   -\frac{6878309}{4630500} & -\frac{58976629}{37044000} & \frac{8569667}{9261000} &
   \frac{179275483}{12348000} \\
   \hline
 -1 & 0 & 0 & 0 & 0 & 0 & 0 & 0 & 0 & 0 \\
 -\frac{19}{36} & \frac{139}{2400} & \frac{499}{800} & 0 & 0 & 0 & 0 & 0 & 0 & 0 \\
 -\frac{1}{3} & \frac{4}{15} & \frac{121}{400} & \frac{637}{800} & -\frac{211}{800} & 0 &
   0 & 0 & 0 & 0 \\
 -\frac{209}{900} & \frac{6299}{21168} & \frac{6767}{35280} & \frac{71063}{88200} &
   -\frac{34723}{176400} & \frac{25841}{58800} & -\frac{36091}{264600} & 0 & 0 & 0 \\
 -\frac{31}{180} & \frac{13843}{105840} & \frac{8317}{15120} & -\frac{797}{35280} &
   \frac{5477}{35280} & \frac{2417}{3528} & \frac{611}{105840} & \frac{13975}{14112} &
   -\frac{5377}{10584} & -\frac{3581}{10080} \\
\end{array}
\right)
\nonumber\\
&
N=\left(
\begin{array}{ccccc}
 0 & 0 & 0 & 0 & 0 \\
 \hline
 \frac{5}{2} & 0 & 0 & 0 & 0 \\
 \frac{1493}{1200} & \frac{5}{36} & 0 & 0 & 0 \\
 \frac{13}{16} & \frac{16877}{14400} & -\frac{7319}{14400} & 0 & 0 \\
 \frac{1229}{1200} & \frac{115501}{43200} & -\frac{9803}{43200} & 0 & 0 \\
 \frac{37547}{78400} & \frac{75071}{39200} & -\frac{497}{576} & \frac{103}{1440} & 0 \\
 \frac{1613}{3360} & \frac{17401}{6720} & \frac{19}{225} & \frac{1187}{2880} & 0 \\
 \frac{184259}{1058400} & \frac{65297}{23520} & -\frac{420373}{211680} &
   \frac{248791}{235200} & -\frac{2747}{9408} \\
 \frac{347437}{1764000} & \frac{863371}{302400} & -\frac{230747}{105840} &
   \frac{938797}{705600} & -\frac{78243}{196000} \\
 \frac{28489}{661500} & \frac{54403}{14700} & -\frac{228689}{88200} &
   \frac{687461}{264600} & -\frac{485507}{5292000} \\
   \hline
 \frac{25}{12} & 0 & 0 & 0 & 0 \\
 -\frac{143}{288} & \frac{2195}{288} & 0 & 0 & 0 \\
 \frac{119}{120} & -\frac{15643}{7200} & \frac{79313}{7200} & 0 & 0 \\
 \frac{22723}{21600} & -\frac{35}{48} & -\frac{2861}{5400} & \frac{443801}{36000} & 0 \\
 \frac{114221}{151200} & \frac{6017}{15120} & \frac{121}{216} & -\frac{3661627}{1411200}
   & \frac{63879443}{4233600} \\
\end{array}
\right),
\end{align}
\end{footnotesize}
and
{\small
\begin{align}
Z^{(2)}_{\mathcal{O}_{16,d}}\Big|_{\frac{1}{\epsilon}-\mathrm{part}.}=
\frac{N_c^2}{\epsilon}
\left(
\begin{array}{cccccccc}
 \frac{575}{144} & 0 & 0 & 0 & 0 & 0 & 0 & 0 \\
 -\frac{23347}{14400} & \frac{46517}{5760} & 0 & 0 & 0 & 0 & \frac{487}{1800} & 0 \\
 \frac{3883}{4032} & -\frac{171823}{37800} & \frac{36597791}{3024000} &
   -\frac{29581}{16800} & 0 & 0 & -\frac{1789}{4800} & 0 \\
 -\frac{9271}{11200} & -\frac{35239}{50400} & \frac{74209}{168000} & \frac{188599}{18900}
   & 0 & 0 & \frac{2101}{4800} & 0 \\
 \frac{3287}{84000} & -\frac{2048479}{1176000} & \frac{422283}{392000} &
   -\frac{2501309}{1764000} & \frac{49211483}{3528000} & \frac{293221}{392000} &
   \frac{2764807}{2116800} & -\frac{61}{20160} \\
 \frac{947587}{1058400} & -\frac{1555357}{705600} & \frac{16831}{29400} &
   -\frac{239641}{75600} & -\frac{381527}{2116800} & \frac{5839021}{423360} &
   -\frac{5807}{201600} & \frac{118933}{1411200} \\
 \frac{3349}{7200} & -\frac{2591}{2400} & 0 & 0 & 0 & 0 & \frac{150391}{14400} & 0 \\
 -\frac{45083}{44100} & \frac{16564}{11025} & \frac{5447}{117600} & \frac{380791}{176400}
   & \frac{1063}{29400} & -\frac{545189}{352800} & \frac{1176541}{1058400} &
   \frac{174229}{12600} \\
\end{array}
\right) .
\end{align}
}
The dilation operator matrix is straightforward to obtain using \eqref{Dforlen3-2} and we will not provide here. The anomalous dimensions are given by the eigenvalues.
We summarize the anomalous dimensions in Table
\ref{tab:AD}.
Irrational numbers also appear in the two-loop anomalous dimensions in $f^{abc}$ sector at dimension 16.

\begin{table}[t]
\centering
\caption{Summary of anomalous dimensions for length-2 and length-3 operators. The lower dimension operators will appear as descendants in the high dimension operators.  }
\label{tab:AD}
\vskip 0.3 cm
{\footnotesize
\begin{tabular}{c|c|c|c|c|c|c|c}
\hline
$\text{dim}$ & $4$ & $6$ & $8$ & $10$ & $12$  & $14$ & $16$
 \\
\hline
$\gamma^{(1)}_{f, \alpha}$ &
$-\frac{22}{3}$ & $/$ & $\frac{7}{3}$ & $\frac{71 }{15}$ &
$\frac{241}{30}, \frac{101 }{15}$ &
$\frac{61}{6}, \frac{172 }{21}$ &
$\frac{331 }{35},\frac{1212\pm\sqrt{3865}}{105}$
 \\
\hline
\multirow{2}{*}{$\gamma^{(2)}_{f, \alpha}$} &
\multirow{2}{*}{$-\frac{136}{3}$} & \multirow{2}{*}{$/$} &
\multirow{2}{*}{$\frac{269}{18}$} &  \multirow{2}{*}{$\frac{2848}{125} $} &
\multirow{2}{*}{$\frac{49901119 }{1404000}, \frac{8585281 }{234000}$} &
\multirow{2}{*}{$\frac{4392073141}{87847200}, \frac{685262197 }{15373260}$} &
$\frac{231568398949 }{4253886000}$,
\\
 & & & & & & & $\frac{355106171452034\pm 95588158951 \sqrt{3865}}{6576507756000}$
 \\
\hline
\hline
$\gamma^{(1)}_{f, \beta}$ &
$-\frac{22}{3}$ & $1$ & $/$ & $\frac{17}{3}$ & $9$ &
$\frac{43}{5}$ & $\frac{67}{6}$
\\
\hline
$\gamma^{(2)}_{f, \beta}$ &
$-\frac{136}{3}$ & $\frac{25}{3}$ & $/$ & $\frac{2195}{72}$ &
$\frac{79313 }{1800}$ & $\frac{443801 }{9000}$ &
$\frac{63879443 }{1058400}$
\\
\hline
\hline
$\gamma^{(1)}_{d,\alpha}$ & $/$ & $/$ & $/$ &  $\frac{13 }{3}$ &
$\frac{41 }{6}$ & $\frac{551 \pm 3 \sqrt{609}}{60}$ &
$\frac{321\pm\sqrt{1561}}{30} $
\\
\hline
$\gamma^{(2)}_{d, \alpha}$ & $/$ & $/$ & $/$ &  $\frac{575}{36}$ &
$\frac{46517 }{1440}$ & $\frac{5809305897 \pm 19635401 \sqrt{609}}{131544000}$ &
$\frac{229162584707\pm 225658792 \sqrt{1561}}{4130406000}$
\\
\hline
\hline
$\gamma^{(1)}_{d,\beta}$ & $/$ & $/$ & $/$ &  $/$ & $ 9$ & $/$ & $\frac{67}{6}$
\\
\hline
$\gamma^{(2)}_{d, \beta}$ & $/$ & $/$ & $/$ &  $/$ & $\frac{150391}{3600}$ & $/$ &
$\frac{174229 }{3150}$
\\
\hline
\end{tabular}
}
\end{table}

\subsubsection*{Checks and analysis}

Some consistency checks for our calculation have been mentioned above, and here we make a summary:
\begin{enumerate}
\item
The ${\cal O}(\epsilon^{-2})$ poles of one-loop bare form factors and
the ${\cal O}(\epsilon^{-3}),{\cal O}(\epsilon^{-4})$ poles of two-loop bare form factors
have infrared origin and therefore should be totally canceled after IR subtraction
procedure shown in (\ref{IR1loop}), (\ref{IR2loop}).
\item
The ${\cal O}(\epsilon^{-2})$ poles of two-loop UV divergences are totally determined by
one-loop UV divergences and $\beta_0$, as shown in (\ref{Z22-Z1}).
\item
At a given dimension, mixing from descendent operators to
non-descendent operators never takes place, such as length-2 to higher length operators in \eqref{eq:Z2toLasZero}.

\item
As explained in the dimension eight case, mixing from general length-3 operators to the unique length-2 operator
can be probed by form factors with both $(-,-,+)$ and $(-,-,-)$.
So form factors under these two helicity settings should give the same length-changing matrix elements
$Z^{(2)}_{3\rightarrow 2}$.

\end{enumerate}
Our results satisfy all these requirements. Some further consistency checks will be also mentioned for the computation of finite remainder function in next section.

Let us make a few comments on the anomalous dimensions and dilatation matrix.
\begin{itemize}
\item

In Table \ref{tab:AD}, the irrational number appears in the dimension 14 and 16 cases.
As eigenvalues of dilatation operators,
anomalous dimensions can be obtained straightforwardly by solving characteristic equation.
Alternatively, one can get their series expansions in $\hat{\lambda}$ up to arbitrary
finite order through perturbation method introduced in quantum mechanics,
which is equivalent to treat dilatation operator as a Hamiltonian of a finite system, see \emph{e.g.} \cite{griffiths2018introduction}.
From perturbative calculation, one can find that whether irrational numbers
appear in perturbative expansions  is determined by
characteristic equation of the first non-degenerate order, which is the one-loop order here.
As a result, if dilatation operator $\mathbb{D}^{(1)}$  is of lower-triangular, then eigenvalues and eigenvectors
at $\mathcal{O}(\hat{\lambda})$ must be rational, which guarantees the series expansions
up to arbitrary order to be rational.
On the contrary, if $Z^{(1)}$ contains sufficiently many non-vanishing upper-right elements,
then the $\mathcal{O}(\hat{\lambda})$
characteristic equation might be complicated and irrational solution may appear.
We find more and more non-vanishing upper-right elements of $Z^{(1)}$ emerge
as operator dimension increases,
so irrational numbers  are expected to appear when operator dimension is high enough,
which is just the case we find.

\item

There are large integer numbers appearing in the denominator of the dilatation matrix elements.
They can be decomposed in terms of prime integer factors. We find that these prime numbers are relatively small: at one-loop no larger than $(\Delta /2)$ and at two-loop no larger than $(\Delta /2)^3$, as shown in Table \ref{tab:primedenominator}.
The pattern of denominators indicates that the elements are related to Harmonic numbers:
 $S_\Delta^{(1)}$ in one-loop element $Z_\Delta^{(1)}$, and  $S_\Delta^{(3)}$ in the two-loop elements $Z_\Delta^{(2)}$, where
\begin{equation}
S_\Delta^{(i)}=\sum_{k=1}^\Delta \frac{1}{k^i} \,.
\end{equation}

It would be interesting to see if the  matrix elements of dilatation operator could be organizable in
certain generic analytic form involving harmonic numbers.

\begin{table}[!t]
\centering
\caption{Largest prime number in the denominator of dilatation matrix elements.}
\label{tab:primedenominator}
\vskip 0.4 cm
\begin{tabular}{c|c|c|c|c|c|c|c}
\hline
$\Delta$ & 4 & 6 & 8 & 10 & 12 & 14 & 16\\
\hline
1-loop & 3 & 3 & $3$ & $5$ & $5$ & $7$ & $7$ \\
\hline
2-loop & 3 & 3 & $3^2$ & $5^3$ & $5^3$ & $7^3$ & $7^3$ \\
\hline
\end{tabular}
\end{table}

\item

It is important to keep in mind that the operators we consider contain only length-2 and length-3 cases. A full basis should contain length-4 and higher length operators. By including these operators, the eigenvalues will be changed.
We will  investigate the effect by adding higher length operators in next subsection.
To study the analytic structure of anomalous dimensions, it would be meaningful only after including the full basis operators, which we hope to discuss in the future work.

\end{itemize}

\subsection{Correction from higher length operators}
\label{sec:length4}

So far we consider only operators up to length-3.
As discussed in Section \ref{sec:operatorbasis},
to simplify operator classification, we introduce an extra equivalence relation
apart from E.o.M and Bianchi identity, by identifying two length-3 operators if
their difference turns out to be higher length operators:
\begin{equation}
\mathcal{O}'_{L=3}-\mathcal{O}_{L=3}={\mathcal{O}}_{L\geq 4} \ \Longrightarrow \
\mathcal{O}'_{L=3}\sim\mathcal{O}_{L=3}
\end{equation}
At two-loop order, the length-3 operators can also mix with length-4 operators, which can modify the dilatation operator and also the anomalous dimensions computed in previous subsection.
In this subsection we consider the basis operators
of dimension 8 by adding length-4 ones.
This allows us to compare the correct anomalous dimensions
with the former result. We will find that the correction from length-change mixing between
length-3 and length-4 is small at order $\mathcal{O}(10^{-1})$.

Let us first consider the basis of dimension-8 operators. For simplicity, we will also consider the large $N_c$ limit, thus we can ignore
the double-trace length-4 operators.
Besides the previously considered  three lower length operators $\{\mathcal{O}_{8;0},\mathcal{O}_{8;\alpha;f;1}\mathcal{O}_{8;\beta;f;1}\}$,
one needs to add four single-trace length-4 operators to form a complete basis.

The four single-trace length-4 operators of dimension 8 have been given in
(\ref{eq:len4dim8-0}), and here we use the linear recombination (\ref{eq:len4dim8-1})
which have been classified according to helicity sectors as
new length-4 basis.
After adding these four operators, leading color basis at dimension 8
are enlarged from (\ref{eq:dim8len23}) to:
\begin{equation}
\begin{aligned}
L=2:&\quad \mathcal{O}_{8;0}=\frac{1}{2}\partial^4\mathrm{tr}F^2,
\\
\label{eq:dim8len234}
L=3:&\quad \mathcal{O}_{8;\alpha;f;1}=
\mathrm{tr}(D_1F_{23}D_4F_{23}F_{14})
-\mathcal{O}_{8;\beta;f;1},\
\mathcal{O}_{8;\beta;f;1}=
\frac{1}{6}\partial^2\mathrm{tr}(F_{12}F_{13}F_{23}),
\\
L=4:&\quad
g \Xi_1,\ g \Xi_2,\
g \Xi_3,\ g \Xi_4\,.
\end{aligned}
\end{equation}
Note that we have included a coupling $g$ in the definition of length-4 operators.
As we mentioned in Footnote \ref{footnote:lengthchanging}, this will change the formula for renormalization constant and the corresponding form factors.
We leave the detailed discussion to Appendix \ref{app:absorb-g}.

From the form factor results, one can extract the dilatation operators up to two loops as
\begin{small}
\begin{align}
\label{eq:dilatation-dim8-I}
\mathds{D}_{\mathcal{O}_8;I}=
\left(
\begin{array}{ccccccc}
 -\frac{22 \hat{\lambda}}{3}-\frac{136 \hat{\lambda}^2}{3} & 0 & 0 & 0 & 0 & 0 & 0
 \\
 - \frac{\hat{\lambda}^2}{\hat{g}} & \frac{7 \hat{\lambda}}{3} +\frac{269 \hat{\lambda}^2}{18} & 10 \hat{\lambda}^2
 & -\frac{10 \hat{\lambda}}{3}+z_{24} \hat{\lambda}^2 & -\frac{14 \hat{\lambda}}{3} +z_{25} \hat{\lambda}^2
 & 3 \hat{\lambda} +z_{26} \hat{\lambda}^2 & \frac{13 \hat{\lambda}}{3} +z_{27} \hat{\lambda}^2
   \\
 -3 \frac{\hat{\lambda}^2}{\hat{g}}  & 0 & \hat{\lambda}+\frac{25 \hat{\lambda}^2}{3}
 & z_{34} \hat{\lambda}^2 & z_{35} \hat{\lambda}^2 & z_{36} \hat{\lambda}^2 & z_{37} \hat{\lambda}^2
 \\
 0 & 0 & -4 \hat{\lambda}^2
 & -5 \hat{\lambda}+z_{44} \hat{\lambda}^2  & 2 \hat{\lambda} +z_{45} \hat{\lambda}^2
 &  z_{46} \hat{\lambda}^2  &  z_{47} \hat{\lambda}^2
 \\
 0 & 0 & -8 \hat{\lambda}^2
 & -4 \hat{\lambda}+ z_{54} \hat{\lambda}^2 & 9 \hat{\lambda} + z_{55} \hat{\lambda}^2
 & z_{56} \hat{\lambda}^2 & z_{57} \hat{\lambda}^2
 \\
 0 & \hat{\lambda}^2 & 0
 &  z_{64} \hat{\lambda}^2 &  z_{65} \hat{\lambda}^2
 & -\frac{8 \hat{\lambda}}{3} + z_{66} \hat{\lambda}^2 & -\hat{\lambda} + z_{67} \hat{\lambda}^2
 \\
 0 & \frac{26 \hat{\lambda}^2}{9} & 0
 &  z_{74} \hat{\lambda}^2 &  z_{75} \hat{\lambda}^2
 & -2 \hat{\lambda} + z_{76} \hat{\lambda}^2 & \frac{17 \hat{\lambda}}{3}+ z_{77} \hat{\lambda}^2 \\
\end{array}
\right)
\end{align}
\end{small}
where the order of operators corresponds to (\ref{eq:dim8len234}).

The last four columns contain undetermined $z_{ij}$ terms
contributed from uncalculated $Z^{(2)}_{4\rightarrow 4}$ elements.
Compute the eigenvalues up to $\mathcal{O}(\hat{\lambda}^2)$,
one can find three eigenvalues (out of seven) are independent of $z_{ij}$ which are:
\begin{align}
\label{eigenvalue-dim8-define1}
\hat\gamma^{(1)}_{\mathcal{O}_8} = \left\{-\frac{22}{3}; 1; \frac{7}{3}\right\}  ,\qquad
\hat{\gamma}^{(2)}_{\mathcal{O}_8}=\left\{
-\frac{136}{3};\frac{25}{3};\frac{235}{18}
\right\} .
\end{align}

Consider another length-3 operator (belonging to $\alpha$-sector) defined as
\begin{align}
\mathcal{O}_{8;\alpha;f;2}={\cal O}_{\rm P1}-\mathrm{Tr}(D_1F_{23}D_1F_{24}F_{34}) \,,
\end{align}
which differs from $\mathcal{O}_{8;\alpha;f;1}$ by a length-4 operator:
\begin{align}
\label{eq:O8-rel34}
\mathcal{O}_{8;\alpha;f;2}-\mathcal{O}_{8;\alpha;f;1}
=g(\Xi_2-\Xi_1)\,.
\end{align}
One can replace $\mathcal{O}_{8;\alpha;f;1}$ by $\mathcal{O}_{8;\alpha;f;2}$ in the basis (\ref{eq:dim8len234}), and find that this does not change the eigenvalues in \eqref{eigenvalue-dim8-define1}.
Comparing \eqref{eigenvalue-dim8-define1} with previous result (\ref{eigenvalue-dim8}),
the first two eigenvalues stay unchanged as expected, since they have already appeared
at lower dimensions and should not be affected by operators intrinsically emerging
at dimension 8.
Furthermore, the third eigenvalue is corrected only at $\mathcal{O}(\hat{\lambda}^2)$ order
from $\frac{269}{18}$ to $\frac{235}{18}$ with amount of
$\mathcal{O}(10^{-1})$. We expect this to be a general feature, as the mixing between different length operators are suppressed physically. As for our consideration, the results in Section \ref{sec:dilatation} by truncating higher length operators are expected to provide a good approximation for the anomalous dimension of length-3 operators.

\section{Finite remainder functions}
\label{sec:finite}

As mentioned before, the form factors can be understood as the Higgs to three-gluon amplitudes, in which the high dimension operators correspond to the interaction vertices in the Higgs EFT.
In this section, we compute the finite remainder functions of form factors, which contain the essential information of the Higgs amplitudes.

\subsubsection*{Conventions}

Form factors have two independent helicity configurations: $(-,-,+)$ and $(-,-,-)$, and each can be
written as a Lorentz scalar function times $\la12\ra^3[13][23]$ and $\la12\ra\la13\ra\la23\ra$ respectively.
For convenience, we will use upper subscript $\mmp$ and $\mmm$ to denote external helicity configuration $(-,-,+)$ and $(-,-,-)$.

In section \ref{sec:onshell-method},
we require each basis operator belongs to a certain helicity sector.
An operator in $\alpha$-sector has nonzero tree-level form factor of helicty
$(-,-,+)$ and vanishing tree form factor of helicty $(-,-,-)$, namely
\begin{equation}
{\cal F}^{(0), \mmp}_{{\cal O}_{\alpha\textrm{-sector}}} \neq 0 \,, \qquad {\cal F}^{(0), \mmm}_{{\cal O}_{\alpha\textrm{-sector}}} = 0 \,.
\end{equation}
Thus we call $(-,-,+)$  the matched helicity and $(-,-,-)$  the unmatched
helicity for $\alpha$-sector.
Similarly, we call $(-,-,-)$ the matched helicity and $(-,-,+)$  the unmatched
helicity for $\beta$-sector.
In the following context we discuss factors under matched and unmatched helicities
separately.

\begin{table}[t]
\caption{Notation of form factors with three gluons, where $\pm$ indicates positive or negative helicity gluons.
 $f^{(0), \pm}_{{\cal O}}$ are scalar factors that depend on the dimension of the operators.}
\label{table:FF-notation}
\begin{center}
\begin{tabular}{c|c|c}
\hline
external particles & $(1^-, 2^-, 3^+)$ & $(1^-, 2^-, 3^-)$  \\ \hline
form factors & ${\cal F}^{(l), \mmp}_{{\cal O}}$ & ${\cal F}^{(l), \mmm}_{{\cal O}}$  \\ \hline
tree form factors & $\la12\ra^3[13][23] f^{(0), \mmp}_{{\cal O}}$ & $\la12\ra\la13\ra\la23\ra f^{(0), \mmm}_{{\cal O}}$  \\
\hline
\end{tabular}
\end{center}
\end{table}

By subtracting the IR and UV divergences,
the finite part of the form factor $\mathcal{F}_{\mathcal{O},\mathrm{fin}}^{(\ell)}$ is defined as in \eqref{IR1loop} - \eqref{IR2loop}.
We introduce the finite remainder function $\mathcal{R}_{\mathcal{O}}^{(2),\pm}$ as follows:
\begin{align}
\label{eq:normalization1}
\mathcal{F}_{\mathcal{O},\mathrm{fin}}^{(2),\mmp}=
\la12\ra^3[13][23]\times
\mathcal{R}_{\mathcal{O}}^{(2),\mmp}
\times
\left
\{
\begin{array}{rcl}
f^{(0),+}_{\mathcal{O}}     &      & \alpha\text{-sector} \\
f^{(0),+}_{\mathcal{O}_{\textrm{L=2}}}     &      & \beta\text{-sector}
\end{array}
\right.,
\\
\label{eq:normalization2}
\mathcal{F}_{\mathcal{O},\mathrm{fin}}^{(2),\mmm}=
\la12\ra\la13\ra\la23\ra\times
\mathcal{R}_{\mathcal{O}}^{(2),\mmm}
\times
\left
\{
\begin{array}{rcl}
f^{(0),-}_{\mathcal{O}_{\textrm{L=2}}}     &      & \alpha\text{-sector}  \\
f^{(0),-}_{\mathcal{O}}     &      & \beta\text{-sector}
\end{array}
\right..
\end{align}
Note that for the unmatched helicity cases, \emph{i.e.}~$\alpha$-sector under $(-,-,-)$ and $\beta$-sector under $(-,-,+)$,
the tree form factors are zero, so we use the scalar factors of the length-2 operator ${\mathcal{O}_{\textrm{L=2}}}$ to normalize the remainder function.

One can further decompose the two-loop remainder according to their trancendentality degree as:
\begin{equation}
\label{eq:remainderdecomp}
\mathcal{R}_{\mathcal{O}}^{(2),\pm} =
\sum_{n=0}^4\mathcal{R}_{\mathcal{O}}^{(2),\pm}\Big|_{\textrm{deg-}n}
+\mathcal{R}^{(2),\pm}_{\mathcal{O}}\Big|_{\log^2(-q^2)}
+\mathcal{R}^{(2),\pm}_{\mathcal{O}}\Big|_{\log(-q^2)} \,.
\end{equation}
Here $q^2 = s_{123}=s_{12}+s_{23}+s_{13}$, and we separate the $q^2$-dependent terms into
 $\mathcal{R}^{(2)}_{\mathcal{O}}\big|_{\log^2(-q^2)}$ and
$\mathcal{R}^{(2)}\big|_{\log(-q^2)}$, so the rest terms
$\{\mathcal{R}^{(2)}_{\mathcal{O}}\big|_{\textrm{deg-}n}\}$ only depend on
ratio variables:
\begin{equation}
u=\frac{s_{12}}{s_{123}}\,, \quad v =\frac{s_{23}}{s_{123}} \,, \quad w=\frac{s_{13}}{s_{123}} \,.
\end{equation}


\subsection{Transcendentality structure of remainder}

In this subsection, we discuss the two-loop remainders according their transcendentality degrees.
Explicit results of two-loop finite remainders are given in the ancillary file submitted together with this paper.
As an example, the result of $\mathcal{O}_{8;\alpha;f;1}$ is explicitly given in
Appendix \ref{app:O81remainder}.

\subsubsection*{Universal building blocks}

For two-loop remainders under matched helicities,
we find the transcendentality degree-4 part of two-loop minimal form factors (under match helicity) always
share a universal expression:
\begin{align}
\mathcal{R}_{\mathcal{O}}^{(2),\pm}\Big|_{\textrm{deg-}4}
= &  -{3\over2} {\rm Li}_4(u) + {3\over4} {\rm Li}_4\left(-{u v \over w} \right) - {3\over4} \log(w) \left[ {\rm Li}_3 \left(-{u\over v} \right) + {\rm Li}_3 \left(-{v\over u} \right)  \right] \nonumber\\
& + {\log^2(u) \over 32} \left[ \log^2(u) + \log^2(v) + \log^2(w) - 4\log(v)\log(w) \right] \nonumber\\
& + {\zeta_2 \over 8} \left[ 5\log^2(u) - 2 \log(v)\log(w) \right]- {1\over4} \zeta_4 + \textrm{perms}(u,v,w) \,,
\label{eq:R2L3-def}
\end{align}
which is expected and also appears in previous computations of lower dimension operators \cite{Brandhuber:2017bkg, Jin:2018fak, Brandhuber:2018xzk, Brandhuber:2018kqb, Jin:2019ile, Jin:2019opr}.\footnote{The computation here in QCD using Catani IR subtraction scheme, and the expression is slightly different (as purely a scheme change) from the ${\cal N}=4$ results which are based on the BDS subtraction scheme \cite{Bern:2005iz}.}
This implies the two-loop minimal form factor of a length-3 operator with
arbitrary dimension in pure Yang-Mills theory always
 obeys the maximal transcendentality principle.

For two-loop remainders under unmatched helicities, there are no degree 4 or 3
parts, in accord with the vanishing  of
$\epsilon^{-4}$, $\epsilon^{-3}$ poles in
bare form factors.
Finite remainders and poles of the same degree originate from the same
term in the master integral coefficients, so they usually vanish simultaneously.
The absence of degree 4 and 3 poles at two-loop level can be traced back
to the absence of one-loop divergence.
As mentioned in section \ref{sec:dilatation}, under unmatched helicity
the tree-level form factor is zero and the
one-loop form factor only has rational term, so
divergence subtraction formula from (\ref{RG2loop}) and (\ref{IR2loop})
becomes
\begin{equation}
\begin{aligned}
\mathcal{F}_{\mathcal{O},\mathrm{fin}}^{(2)}
=&\mathcal{F}_{\mathcal{O},B}^{(2)}
+\Big(
Z_{\mathcal{O}}^{(1)}
- \Big( 1+\frac{\delta}{2} \Big)\frac{\beta_0}{\epsilon}
-I^{(1)}(\epsilon)
\Big)
\mathcal{F}_{\mathcal{O},B}^{(1)}
 \,,
\end{aligned}
\end{equation}
which explicitly shows the leading singularity is of $\mathcal{O}(\epsilon^{-2})$
from $I^{(1)}(\epsilon)\mathcal{F}^{(1)}_B$, and no term can contribute to
$\epsilon^{-3}$, $\epsilon^{-4}$.

Apart from maximal transcendental universality,
degree-3  and degree-2 parts also signify some universal structure,
in the sense that complicated transcendental functions  can always be
absorbed into a set of universal building blocks,
and no other polylogarithm functions like $\mathrm{Li}_2,\mathrm{Li}_3$ are left outside these
basis functions.\footnote{When quark is added, $T_3,T_2$ no longer compose complete basis for polylogarithms, see \cite{Jin:2019opr}.}

Building blocks of degree-3 part are six functions
$\{T_3[\sigma(x),\sigma(y),\sigma(z)]|\sigma\in S_3\}$ together
with $\pi^2\log$ and $\zeta_3$, where
$T_3(u, v, w)$ is given as
\begin{align}
T_3(u,v,w) := & \Big[ -{\rm Li}_3 \left(-{u\over w} \right) + \log(u) {\rm Li}_2\left({v \over 1-u} \right) - {1\over2} \log(u) \log(1-u) \log\left({w^2\over 1-u}\right) \nonumber\\
& + {1\over2} {\rm Li}_3\left(-{uv \over w}\right) + {1\over2} \log(u)\log(v)\log(w) + {1\over12}\log^3(w) + (u\leftrightarrow v) \Big] \nonumber\\
& +  {\rm Li}_3(1-v) - {\rm Li}_3(u) + {1\over2} \log^2(v) \log\left({1-v\over u}\right) - \zeta_2 \log\left( {u v \over w} \right) \,.
\label{eq:T3-def}
\end{align}
Similar function has appeared in the ${\cal N}=4$ form factors \cite{Loebbert:2015ova, Loebbert:2016xkw, Brandhuber:2017bkg}.
Building blocks of degree-2 part are three functions
$\{T_2[\sigma(x),\sigma(y)]|\sigma\in Z_3\}$ together with $\log^2$ and $\pi^2$,
where $T_2(u, v)$ is given as (see also \cite{Jin:2019opr})
\begin{align}
T_2(u,v) := & \text{Li}_2(1-u)+\text{Li}_2(1-v)+\log (u) \log (v)- \zeta_2
\label{eq:T2-def}  \,.
\end{align}
When expanding the form factor remainders in these building blocks,
the coefficients in front of them are just
rational functions of $u, v, w$, see examples in Appendix \ref{app:O81remainder}.

\subsubsection*{Comment on $\log(q^2)$ terms}

As mentioned in \eqref{eq:remainderdecomp}, remainder functions also contain two type of terms that are linear and quadratic in $\log(q^2)$ respectively.
All these terms have simple universal structures which we explain below.

Coefficient of $\log^2(-q^2)$ term of $\mathcal{O}_i$
equals to the  $\epsilon^{-2}$ residue of $\sum_j(Z^{(2)})_i^{\ j} \mathcal{F}^{(0)}_{\mathcal{O}_j}$.
 This is because every master integral contains an overall factor $(-q^2)^{-2\epsilon}$,
whose $\epsilon$-expansion produces $1$ and $\log^2(-q^2)\epsilon^{2}$ at the same time.
So leading singularity $1/\epsilon^2$ can be recovered back to
$1/\epsilon^2(1+\log^2(-q^2)\epsilon^{2})$ and this is the only origin of
$\log^2(-q^2)$ term.
Plugging in (\ref{Z22-Z1}), we can write $\log^2(-q^2)$ term  as
\begin{equation}
\label{eq:logs-square}
\mathcal{R}_{\mathcal{O}_i}^{(2),\pm}\Big|_{\log^2(-q^2)}
= \log^2(-q^2)  \sum_{j} \epsilon^2 \Big[
\frac{1}{2}( Z^{(1)})_i^{\ k} ( Z^{(1)})_k^{\ j}
-\frac{\beta_0}{2\epsilon} ( Z^{(1)})_i^{\ j}
\Big] f^{(0), \pm}_{{\cal O}_j}
\,,
\end{equation}
where $f^{(0), \pm}_{{\cal O}_j}$ is the tree-level scalar
factor of $\mathcal{O}_j$, as introduced in Table \ref{table:FF-notation}.
Since matrix elements of $Z^{(1)}$ is nonzero only for operators within same $\alpha$ or $\beta$ sector,
the $\log^2(-q^2)$ terms vanish under unmatched helicity.

Terms linear in $\log(-q^2)$ can be determined by data of
one-loop finite remainders and renormalization matrices $Z^{(1)},Z^{(2)}$.
Coefficients of linear $\log(-q^2)$ terms contain degree 3,2,1,0 parts.
The generic expression of this part is:
\begin{align}
\label{eq:logs-linear}
\mathcal{R}^{(2),\pm}_{\mathcal{O}_i}\Big|_{\log(-q^2)}
= \log(-q^2) \sum_{j} \Big\{
& \epsilon
\Big[
(Z^{(1)})_i^{\ j}-\frac{\beta_0}{\epsilon}\delta_i^j
\Big]
\mathcal{R}^{(1),\pm}_{\mathcal{O}_j}\Big|_{\log(-q^2)=0}
\nonumber\\
&
+
\Big[
2\epsilon (Z^{(2)})_i^{\ j}\Big|_{\frac{1}{\epsilon}\text{-part}}
- \Big(
\frac{11\pi^2}{24}+3\zeta_3+\frac{5}{2}
\Big)\delta_i^j
\Big]
f_{\mathcal{O}_j}^{(0),\pm}
\Big\} \,,
\end{align}
where $\mathcal{R}^{(1),\pm}_{\mathcal{O}_j}$ is the one-loop
finite remainder, whose $\log(-q^2)$ coefficient is
$\sum_j(\epsilon Z^{(1)})_i^{\ j} f_{\mathcal{O}_j}^{(0),\pm}$.
The degree 3 and 2 parts of $\log(-q^2)$ coefficient are universal.
Both (\ref{eq:logs-square}) and (\ref{eq:logs-linear}) provide  nice consistency checks
for our calculation.

\subsection{Cancellation of spurious poles}

While the transcendentality functions in the remainder take very simple universal structures as discussed in last subsection, their rational coefficients depend on the dimension of operators and are the main complication of the remainders. In particular, they contain spurious poles which cancel only after summarizing the results.

Spurious poles exist at transcendentality degree 3,2,1,0. They contain high order poles of $u^n, v^n, w^n$ with $n>1$,
as well as linear combination poles $u+v$ and $u+w$.
In Table \ref{tab:spurious} we summarize the leading spurious poles in the form factor remainders.
\begin{table}[tb]
\centering
\caption{Leading spurious poles contained by individual degree parts. $\Delta_0$ denotes the canonical dimension of the operator.}
\label{tab:spurious}
\vskip 0.4 cm
\begin{tabular}{|c|c|c|c|c|}
\hline
operator & external & $u$ & $v,w$ & $u+v, u+w$  \\
\hline
 \multirow{2}{*}{$\mathcal{O}_{\Delta_0,\alpha,f}$}
 & $(-,-,+)$ & $\frac{\Delta_0}{2}+2$ & $\frac{\Delta_0}{2}+1$ &
$2$  \\
\cline{2-5}
 & $(-,-,-)$ & $\frac{\Delta_0}{2}+1$ & $\frac{\Delta_0}{2}+1$ &
 0  \\
\hline
 \multirow{2}{*}{$\mathcal{O}_{\Delta_0,\beta,f}$}
 & $(-,-,+)$ & $\frac{\Delta_0}{2}+1$ & $\frac{\Delta_0}{2}$ &
2  \\
\cline{2-5}
 & $(-,-,-)$ & $\frac{\Delta_0}{2}$ & $\frac{\Delta_0}{2}$ &
 0  \\
\hline
 \multirow{2}{*}{$\mathcal{O}_{\Delta_0,\alpha,d}$}  & $(-,-,+)$ &
 $\frac{\Delta_0}{2}+1$ & $\frac{\Delta_0}{2}$ &
 1
 \\
\cline{2-5}
 & $(-,-,-)$ & $\frac{\Delta_0}{2}-1$& $\frac{\Delta_0}{2}-1$ &
 0  \\
\hline
 \multirow{2}{*}{$\mathcal{O}_{\Delta_0,\beta,d}$}  & $(-,-,+)$ &
 $\frac{\Delta_0}{2}$ & $\frac{\Delta_0}{2}-1$ &
1
\\
\cline{2-5}
 & $(-,-,-)$ & $\frac{\Delta_0}{2}-2$ & $\frac{\Delta_0}{2}-2$  &
 0   \\
\hline
\end{tabular}
\end{table}

A non-trivial feature is that
the cancellation of these spurious poles takes place across different transcendentality degrees.
As a concrete non-trivial example,  we consider the remainder function of $\mathcal{O}_{8;\alpha;f;1}$ under match helicity $(-,-,+)$.
Explicit expressions of two-loop remainder of $\mathcal{O}_{8;\alpha;f;1}$ (with degree 3,2,1,0 parts)  can be found in Appendix \ref{app:O81remainder}.
We first summarize the property of pole structures:
\begin{enumerate}
\item
Leading poles of ${1\over u^m}, {1\over v^m}, {1\over w^m}$, with powers $6,5,5$ respectively, appear only in coefficients of
$T_2$ functions in degree-2 part.

\item
degree-3 part contains only one spurious pole ${1\over u^3}$.

\item
Spurious poles ${1\over (u+v)^m}$ and ${1\over (u+w)^m}$ appear in degree 1 and 0 parts,
with powers up to $2$.

\item
Due to the symmetry of remainder under $v\leftrightarrow w$, residues of poles  ${1\over v^n}$ and ${1\over (u+v)^m}$ are identical
to residues of poles ${1\over w^n}$ and ${1\over (u+w)^m}$.
\end{enumerate}
In the following we discuss these spurious poles separately and show that they explicitly cancel in the full remainder.

\subsubsection*{${1/ u^m}$-pole}

To analyze the ${1\over u^m}$-poles, one needs to consider the limit $u\rightarrow 0$. Polylogarithm functions can be expanded
in this limit, for example:
\begin{equation}
\begin{aligned}
& {\rm Li}_2(u+v)= {\rm Li}_2(v)-\frac{\log(1-v)}{v}u+\mathcal{O}(u^2) \,, \\
&T_2(v,w)=\Bigl(-\frac{\log v}{1-v}-\frac{\log(1-v)}{v}\Bigr)u+\mathcal{O}(u^2) \,. \\
\end{aligned}
\end{equation}

From the expression in \eqref{eq:O81deg2} ,
it seems that the degree-2 part of $\mathcal{R}^{(2),\mmp}_{\mathcal{O}_{8;\alpha;f;1}}$ has leading pole at order $\mathcal{O}(\frac{1}{u^6})$:
\begin{equation}
 \mathcal{R}^{(2),\mmp}_{\mathcal{O}_{8;\alpha;f;1}}\Big|_{\mathrm{deg} 2}=
   T_2(v,w) \Big(
  \frac{v^2 w^2}{2 u^4}-\frac{5v w(v^2+w^2)}{3 u^4}
   +\frac{11  v^2 w^2(v+w)}{6 u^5}  +\frac{5 v^3 w^3}{ u^6} \Big) +\mathcal{O}(\frac{1}{u^3}) \,.
\end{equation}
However, since $T(v,w)\sim \mathcal{O}(u)$, it is actually $\mathcal{O}(u^{-5})$
\begin{equation}
\begin{aligned}
& \mathcal{R}^{(2),\mmp}_{\mathcal{O}_{8;\alpha;f;1}}\Big|_{\mathrm{deg} 2}= \frac{5v^2 (1-v)^2}{u^5} \Bigl(-v\log v-(1-v)\log(1-v)\Bigr)+\mathcal{O}(u^{-4}) \,.
\end{aligned}
\end{equation}
To cancel the $\mathcal{O}(u^{-5})$ term, one needs to consider the contribution from the degree-1 part. Concretely, one can extract the residue for $1/u^5$ pole terms from various degree parts as:
\begin{equation}
\begin{aligned}
\text{deg-}0&:\ 0\,,
\\
\text{deg-}1&:\ 5 (v-1)^2 v^3 \log (v)-5 (v-1)^3 v^2 \log (1-v)\,,
\\
\text{deg-}2&:\ -5 (v-1)^2 v^3 \log (v)+5 (v-1)^3 v^2 \log (1-v)\,,
\\
\text{deg-}3&:\ 0\,.
\end{aligned}
\end{equation}

One can check that after expanding the polylogarithm functions to appropriate orders in $u$,
all $u^{-k}$ poles vanish except for the physical $1/u$ pole.
Explicitly, residues of
sub-leading poles contained
by different transcendentality degree parts are:
\begin{footnotesize}
\begin{align}
& \underline{{1/ u^4}\text{-pole}} \nonumber\\
&\text{deg-}0:\  \frac{5}{2}v^2 (v-1)^2 \,,
\nonumber\\
&\text{deg-}1:\  -5 v^2 (v-1)^2+\frac{1}{6} v^2 (75 v-11) (v-1) \log (v)-\frac{1}{6} v (75 v+4) (v-1)^2 \log (1-v)\,,
\nonumber   \\
&\text{deg-}2:\   \frac{5}{2} v^2 (v-1)^2-\frac{1}{6} v^2 (75 v-11) (v-1) \log (v)+\frac{1}{6} v (75 v+4) (v-1)^2 \log(1-v)\,,
\nonumber   \\
&\text{deg-}3:\    0\,; \nonumber \\
& \underline{{1/ u^3}\text{-pole}}  \nonumber \\
&\text{deg-}0:\ \frac{1}{12} (v-1) v (60 v-1)\,,
\nonumber\\
&\text{deg-}1:\  -\frac{1}{12} (v-1) (86 v^2+41 v-11) \log (1-v)+\frac{1}{12} v (86 v^2-9 v-20)
   \log (v)-\frac{2}{3} (v-1) v (15 v+1)\,,
\nonumber   \\
&\text{deg-}2:\   \frac{1}{12} (v-1) (86 v^2+41 v-11) \log (1-v)-\frac{1}{12} v (86 v^2-9 v-20)
   \log (v)+\frac{1}{4} (v-1) v (20 v+3)\,,
\nonumber\\
&\text{deg-}3:\    0\,; \nonumber \\
&\underline{{1/ u^2}\text{-pole}} \nonumber \\
&\text{deg-}0:\  \frac{1}{72} (93 v^2+81 v-52),
\nonumber\\
&\text{deg-}1:\  \frac{1}{12} (-31 v^2-37 v+11)
+\frac{(63 v^3-311 v^2+187 v-22) }{36 v} \log (1-v)
   -\frac{(63 v^3-302 v^2+246 v-29) }{36 (v-1)} \log (v) ,
\nonumber   \\
&\text{deg-}2:\  \frac{1}{3} (1-2 v) \text{Li}_2(v)-\frac{1}{6} v^2 \log^2(1-v)
  +\frac{1}{3} (v-1)^2 \log (1-v)\log (v)
  -\frac{1}{6} (v-1)^2 \log ^2(v)
   -\frac{1}{18} \pi ^2 (v-1)^2
\nonumber  \\ & \qquad
   -\frac{(63 v^3-347 v^2+223 v-34) }{36 v}\log (1-v)
   +\frac{63 v^3-338 v^2+282 v-41}{36 (v-1)}\log (v)
   +\frac{1}{72} (93 v^2+141 v-14),
\nonumber   \\
&\text{deg-}3:\   \frac{1}{3} (2 v-1) \text{Li}_2(v)+\frac{1}{6} v^2 \log ^2(1-v)
   -\frac{1}{3} (v-1)^2 \log (1-v)\log (v)
   +\frac{1}{6} (v-1)^2 \log ^2(v)
   +\frac{1}{18} \pi ^2 (v-1)^2
\nonumber   \\& \qquad
   -\frac{(3 v^2-3 v+1)}{3 v} \log (1-v)
    +\frac{3 v^2-3 v+1}{3 (v-1)}\log (v) .\nonumber
\end{align}
\end{footnotesize}
All these ${1/ u^m}$ poles do not cancel within single transcendentality degree, but only after the sum of different
degree parts.

\subsubsection*{${1/ v^m}$-pole}

For ${1/ v^m}$-poles the analysis is similar, one can take limit $v\rightarrow 0$ and expand polylogarithm functions
in $v$. After doing so, ${1\over v^5}$ pole in degree-2 part vanishes.
Residues of sub-leading poles contained by four parts become:
\begin{footnotesize}
\begin{align}
& \underline{ {1/ v^4}\text{-pole}} \\
& \text{deg-}0:\  0\,,\nonumber
\\
& \text{deg-}1:\  -3 (u-1)^3 u \log (1-u) + 3 (u-1)^2 u^2 \log (u)\,,\nonumber
\\
& \text{deg-}2:\  3 (u-1)^3 u \log (1-u)-3 (u-1)^2 u^2 \log (u)\,,\nonumber
\\
& \text{deg-}3:\  0\,; \nonumber\\
& \underline{ {1/ v^3}\text{-pole}} \\
& \text{deg-}0:\   \frac{3}{2} (u-1)^2 u\,,\nonumber
\\
& \text{deg-}1:\  -3 u (u-1)^2-\frac{1}{12} (17 u+7) (u-1)^2 \log (1-u)+\frac{1}{12} u (17 u-11) (u-1) \log (u)\,,\nonumber
\\
& \text{deg-}2:\  \frac{3}{2} u (u-1)^2+\frac{1}{12} (17 u+7) (u-1)^2 \log (1-u)-\frac{1}{12} u (17 u-11) (u-1) \log (u)\,,\nonumber
\\
& \text{deg-}3:\  0\,; \nonumber\\
& \underline{ {1/ v^2}\text{-pole} }\\
& \text{deg-}0:\   -\frac{1}{24} (u-1)^2\,,\nonumber
\\
& \text{deg-}1:\   \frac{\left(133 u^2-98 u+1\right) (u-1) \log (1-u)}{72 u}+\frac{1}{72} \left(-133 u^2+137 u-40\right)
   \log (u)+\frac{1}{12} (u-7) (u-1)\,,\nonumber
\\
& \text{deg-}2:\  -\frac{\left(133 u^2-98 u+1\right) (u-1) \log (1-u)}{72 u}+\frac{1}{72} \left(133 u^2-137 u+40\right)
   \log (u)-\frac{1}{24} (u-13) (u-1)\,,\nonumber
   \\
& \text{deg-}3:\  0\,.\nonumber
\end{align}
\end{footnotesize}
These ${1\over v^m}$ poles cancel after summing over degree 2,1,0 parts.  Poles of ${1\over w^m}$ are related by symmetry and are cancelled in the same way.

\subsubsection*{${1/(u+v)^m}$-pole}

Similarly, take limit $v\rightarrow -u$ and expand polylogarithm functions in
$(u+v)$. After doing so, leading poles ${1\over (u+v)^2}$ in degree 1 and 0 parts
cancel within their own degree, and residues of
sub-leading $1/(u+v)$ pole cancel when summing over degree 1 and 0 parts:
\begin{equation}
\text{deg-}0:\  -\frac{5}{18 u  }\,,\qquad
\text{deg-}1:\  \frac{5}{18 u  }\,.
\end{equation}

\section{Summary and outlook}
\label{sec:conclusion}

In this paper we study the two-loop renormalization of high dimensional QCD local operators and related Higgs amplitudes using Higgs effective action.
We discuss the classification of operators and the construction of their basis, for which we apply both field theory method and on-shell spinor helicity formalism. 
Given the operator basis, we compute the two-loop minimal form factors of length-3 operators up to dimension 16, by using the efficient unitarity-IBP method.
The on-shell methods have played important roles for both operator construction and loop computation. Our results have passed various very stringent checks: including the unitarity check of different cut channels, the consistency of IR and UV divergences, as well as the spurious pole cancellation in the finite parts.

Based on the new form factor results, we are able to extract the renormalization constant matrix up to two-loop order, most of which are obtained for the first time.
By diagonalizing the dilatation matrix, we also obtain the anomalous dimensions, \emph{i.e.}~the eigenvalues.
To study the effect of high length operators, we consider the dimension eight case and include the length-4 operators to form a complete set of basis. 
We find that the anomalous dimension only receives a small correction,
which implies that the mixing with high length operators are small. 

The form factor results also provide the Higgs to three-gluon amplitudes using the Higgs EFT. The results with high dimension operators correspond to Higgs amplitudes with high order of top mass corrections. They can be used to improved the precision for the cross section of Higgs plus a jet production, which is not yet available by other means at N$^2$LO QCD order. The major amplitude information is captured by the finite remainder functions, for which the analytic results are obtained up to dimension-16 operators. 
We study their analytic structure in detail and find that the maximal transcendentality part is always universal and independent of the operators. This provides further evidence to the maximal transcendentality principle for form factors. 
For the lower transcendental parts, two universal building blocks that capture all non-trivial polylogarithm functions are also identified. Besides, we also provide a detail analysis for the cancellation of spurious poles.

There are several straightforward generalizations based on this paper.
In this paper, we focus on the pure gluon sector of QCD. It is also important to consider the operators with quark fields, similar to the dimension-6 operators considered to two-loop order in \cite{Jin:2019opr}.
It would be also interesting to consider general parity odd operators, such examples contain the Weinberg's operator \cite{Weinberg:1989dx}, see \emph{e.g.}~\cite{deVries:2019nsu} for a recent study.
There have been many studies of constructing operator basis in standard model EFT (see \emph{e.g.}~\cite{Grzadkowski:2010es, Lehman:2014jma, Lehman:2015coa, Henning:2015alf, Liao:2016hru, Li:2020gnx, Murphy:2020rsh, Li:2020xlh, Liao:2020jmn}) and consider their renormalization using both conventional and on-shell formalism (see \emph{e.g.}~\cite{Elias-Miro:2013gya,Elias-Miro:2013mua,Jenkins:2013zja,Jenkins:2013wua,Alonso:2013hga,Liao:2019tep, EliasMiro:2020tdv, Baratella:2020lzz, Jiang:2020mhe}).
While the state-of-the-art computation is mostly at one-loop order, it would be worth extending the method developed in this paper to to general two-loop renormalization in SMEFT (see also \cite{Bern:2020ikv}).

\acknowledgments

We would like to thank Rui Yu for discussions.
This work is supported in part by the National Natural Science Foundation of China (Grants No.~11822508, 11935013, 11947302, 11947301),
and by the Key Research Program of Frontier Sciences of CAS under Grant No.~QYZDB-SSW-SYS014.
We also thank the support of the HPC Cluster of ITP-CAS.

\appendix

\section{Primitive length-3 operators}
\label{app:miniprimOper}

In this appendix, we provide some details about the derivation of primitive length-3 operators given in (\ref{length3primitive}).

Recall that an operator is called \emph{primitive} if it contains no  $DD$ contraction.
Every non-primitive operator  belongs to a certain \emph{primitive class} represented by
the primitive operator obtained from taking off all its $DD$ pairs.
We denote the number of $D$s in an primitive operator as $\hat{n}_d$, so all these $\hat{n}_d$
$D$s are contracted with $F$s.
For length-3 cases $\hat{n}_d$ might take values $0,2,4,6$.
We will show there are only two independent primitive classes at length-3:
\begin{equation}
\label{eq:primlen3}
\lfloor {\cal O}_{\rm P1} \rfloor = \lfloor \mathrm{Tr}\big(D_1F_{23}D_4 F_{23}F_{14}\big) \rfloor,
\qquad
\lfloor {\cal O}_{\rm P2} \rfloor = \lfloor \mathrm{Tr}\big(F_{12}F_{13}F_{23}\big) \rfloor\,.
\end{equation}

First, for $\hat{n}_d=6$, the three field strengths must take forms like
$F_{ij}$, $F_{kl}$, $F_{mn}$.
There is at least one block contains nonzero covariant derivatives, so
one can rewrite this block using Bianchi identity. In this way
one Lorentz index will be moved from a $D$ to the $F$,
which means $\hat{n}_d$ is reduced from 6 to 4.  For example:
\begin{equation}
D_{35}F_{12}D_{16}F_{34}D_{24}F_{56}
=D_{35}F_{12}D_{16}F_{34}D_{25}F_{46}
-D_{35}F_{12}D_{16}F_{34}D_{26}F_{45} \,.
\nonumber
\end{equation}

Second, for $\hat{n}_d=4$, the three field strengths must take forms like $F_{ij}$, $F_{jk}$, $F_{mn}$.
Among them $F_{mn}$ is the special one because it shares
no Lorentz index with the other two. We consider following two cases:
\begin{description}
\item[1]
If there is at least one covariant derivative, say $D_i$ or $D_k$, acting on $F_{mn}$, we can
make use of Bianchi identity to move the Lorentz index from this $D$ to the $F$ so
that $\hat{n}_d$ is reduced from 4 to 2.
For example:
\begin{equation}
D_4F_{12}D_{15}F_{23}D_3F_{45}
=D_4F_{12}D_{15}F_{23}D_4F_{35}-D_4F_{12}D_{15}F_{23}D_5F_{34} \,.
\nonumber
\end{equation}
\item[2]
If there is no covariant derivative acting on $F_{mn}$, then other two blocks must both contain
nonzero $D$s. We can rewrite one of those two blocks using Bianchi identity, For example:
\begin{equation}
D_{34} F_{12}D_{15}F_{23}F_{45}
=-D_{31}F_{24}D_{15}F_{23}F_{45}+D_{32}F_{14}D_{15}F_{23}F_{45} \,.
\nonumber
\end{equation}
The first term on the r.h.s is already reduced to $\hat{n}_d=2$. The second term still has $\hat{n}_d=4$, but
now $F_{23}$ becomes the special $F_{mn}$, and there are nonzero $D$s acting on it,
so this actually belongs to the previous case.
\end{description}
In summary, every primitive operator with $\hat{n}_d=4$ can be reduced to $\hat{n}_d=2$.

Third, for $\hat{n}_d=2$, the three field strengths may take forms like
$(F_{ij}$ $F_{kl}$, $F_{kl})$ or $(F_{ij}$, $F_{jk}$, $F_{kl})$. We consider various different cases as follows:
\begin{description}
\item[a]
For configuration $(F_{ij}$, $F_{kl}$, $F_{kl})$, the only permitted operator is
$ D_3F_{12}D_4F_{12}F_{34} $.
Other seemingly qualified operators like
$ D_{34}F_{12}F_{12}F_{34}$ and
$ F_{12}D_{34}F_{12}F_{34} $
actually have higher length and therefore should not
be taken into account, \emph{e.g.},
\begin{equation}
\begin{aligned}
D_{34}F_{12}F_{12}F_{34}=\frac{1}{2}
\big(D_{34}F_{12}F_{12}F_{34}
-D_{43}F_{12}F_{12}F_{34}
\big)
=[F_{34},F_{12}]F_{12}F_{34}.
\nonumber
\end{aligned}
\end{equation}

\item[b]
For configuration $(F_{ij}$, $F_{jk}$, $F_{kl})$, we can see
$F_{jk}$ is the special one because it shares two
Lorentz indices with the other two $F$s.

b.1) If there is only one covariant derivative, say $D_i$ or $D_l$, acting on $F_{jk}$,
we can rewrite this block using Bianchi identity, relabel the Einstein indices
and finally obtain an operator belonging to case a. For example:
\begin{equation}
\begin{aligned}
&F_{12}D_4F_{23}D_1F_{34}=F_{12}D_3F_{24}D_1F_{34}-F_{12}D_2F_{34}D_1F_{34}
\\
=&-F_{12}D_4F_{23}D_1F_{34}-F_{12}D_2F_{34}D_1F_{34}
=-\frac{1}{2}F_{12}D_2F_{34}D_1F_{34} \,.
\nonumber
\end{aligned}
\end{equation}

b.2) If there are two $D$s acting on $F_{jk}$, the operator turns out
 to be  a higher length one, so it should be dropped out:
\begin{equation}
\begin{aligned}
F_{12}D_{14}F_{23}F_{34}&=-F_{12}D_{12}F_{34}F_{34}+F_{12}D_{13}F_{24}F_{34}
=-F_{12}D_{12}F_{34}F_{34}-F_{12}D_{14}F_{23}F_{34}
\\
=&-\frac{1}{2}F_{12}D_{12}F_{34}F_{34}
=-\frac{1}{4} F_{12}[F_{12},F_{34}]F_{34}\sim \mathcal{O}(F^4) \,.
\nonumber
\end{aligned}
\end{equation}

b.3) If there is no $D$s acting on $F_{jk}$, it must take the form
$\mathrm{Tr}\big(D_4 F_{12}F_{23}D_1F_{34}\big)$. This  can be rewritten as follows:
\begin{equation}
\begin{aligned}
\mathrm{Tr}\big(D_4F_{12}F_{23}D_1F_{34}\big)
=-\mathrm{Tr}\big(D_1F_{24}F_{23}D_1F_{34}\big)+
\mathrm{Tr}\big(D_2F_{14}F_{23}D_1F_{34}\big) \,.
\nonumber
\end{aligned}
\end{equation}
The first term on the r.h.s is reduced to $\hat{n}_d=0$ class
$[\ \mathrm{Tr}\big(F_{24}F_{23}F_{34}\big)\ ]$.
The second term on the r.h.s belongs to the case b.1, since now $F_{34}$ becomes the
special $F_{jk}$ and there is a $D_1$ acting on it.
According to former discussion, this term eventually belongs to case a.
\end{description}

For $\hat{n}_d=2$ operators, we prefer configuration $(F_{ij}$, $F_{kl}$, $F_{kl})$
instead of $(F_{ij}$, $F_{jk}$, $F_{kl})$, because the former one has a simpler expression under
 decomposition $F_{\alpha\dot{\alpha}\beta\dot{\beta}}=\epsilon_{\alpha\beta}
 \bar{f}_{\dot{\alpha}\dot{\beta}}+\epsilon_{\dot{\alpha}\dot{\beta}}f_{\alpha\beta}$.
The components from two $F_{kl}$'s must be both self-dual or both anti-self-dual
because contraction between $f$ or $\bar{f}$ and antisymmetric tensor gives zero.
If we probe the minimal form factor under configuration $(-,-,+)$ or $(+,+,-)$,
the particle with the opposite helicity must be emitted from $F_{ij}$. That is why
we choose the only independent $\hat{n}_d=2$ primitive operator from case a.

Above analysis shows there are only two independent primitive class as shown in (\ref{eq:primlen3}).
Based on this, one can construct basis of length-3 operators for any given dimension $\Delta_0$
systematically by inserting pairs of identical $D_i$s into the primitive ones
until $\Delta_0$ is reached.

\section{Preliminary operator basis ${\cal O}''_{\Delta_0, i}$}
\label{app:oldbasis}

In this appendix, we provide length-3 basis operators constructed by inserting DD pairs to primary operators, as described  in Section \ref{sec:conventional-method}.
The underlined number $\underline{(n+m)}$ represents that
there are $n$ operators created from primitive operator ${\cal O}_{\rm P1} = \mathrm{Tr}(D_1F_{23}D_4F_{23}F_{14})$
and $m$ ones from the primitive operator ${\cal O}_{\rm P2} = \mathrm{Tr}(F_{12}F_{13}F_{23})$.

\begin{footnotesize}
\begin{align}
\label{old-dim6}
& \underline{\mbox{dim\ 6}\ (0+1)}&\nonumber\\
&
\mathcal{O}''_{6;1}=\frac{1}{3}\mathrm{Tr}(F_{12}F_{13}F_{23}) \,.
\\
\label{old-dim8}
& \underline{\mbox{dim\ 8}\ (1+1)}&\nonumber\\
&
\mathcal{O}''_{8;1}=\mathrm{Tr}(D_1F_{23}D_4F_{23}F_{14});\
\mathcal{O}''_{8;2}=\mathrm{Tr}(D_1F_{23}D_1F_{24}F_{34}) \,.
\\
\label{old-dim10}
& \underline{\mbox{dim\ 10}\ (3+2)}&\nonumber\\
&
\mathcal{O}''_{10;1}=\mathrm{Tr}(D_{12}F_{34}D_{15}F_{34}F_{25}),\
\mathcal{O}''_{10;2}=\mathrm{Tr}(D_{12}F_{34}D_5F_{34}D_1F_{25}),\
\mathcal{O}''_{10;3}=\mathrm{Tr}(D_2 F_{34}D_{15}F_{34}D_1 F_{25});&\nonumber\\
&
\mathcal{O}''_{10;4}=\mathrm{Tr}(D_{12}F_{34}D_1F_{35}D_2F_{45}),\
\mathcal{O}''_{10;5}=\mathrm{Tr}(D_{12}F_{34}D_{12}F_{35}F_{45}) \,.
\\
\label{old-dim12}
& \underline{\mbox{dim\ 12}\ (6+4)}&\nonumber\\
&
\mathcal{O}''_{12;1}=\mathrm{Tr}(D_{123}F_{45}D_{126}F_{45}F_{36}),\
\mathcal{O}''_{12;2}=\mathrm{Tr}(D_{123}F_{45}D_{16}F_{45}D_2F_{36}),\
\mathcal{O}''_{12;3}=\mathrm{Tr}(D_{13}F_{45}D_{126}F_{45}D_2F_{36}),&
\nonumber\\
&
\mathcal{O}''_{12;4}=\mathrm{Tr}(D_{123}F_{45}D_6F_{45}D_{12}F_{36}),\
\mathcal{O}''_{12;5}=\mathrm{Tr}(D_{13}F_{45}D_{26}F_{45}D_{12}F_{36}),\
\mathcal{O}''_{12;6}=\mathrm{Tr}(D_3F_{45}D_{126}F_{45}D_{12}F_{36});
&\nonumber\\
&
\mathcal{O}''_{12;7}=\mathrm{Tr}(D_{12}F_{45}D_{13}F_{46}D_{23}F_{56}),\
\mathcal{O}''_{12;8}=\mathrm{Tr}(D_{12}F_{45}D_{123}F_{46}D_3F_{56}),\
\mathcal{O}''_{12;9}=\mathrm{Tr}(D_{123}F_{45}D_{12}F_{46}D_3F_{56}),&
\nonumber\\
&
\mathcal{O}''_{12;10}=\mathrm{Tr}(D_{123}F_{45}D_{123}F_{46}F_{56}) \,.
\\
\label{old-dim14}
& \underline{\mbox{dim\ 14}\ (10+5)}&\nonumber\\
&
\mathcal{O}''_{14;1}=\mathrm{Tr}(D_{1234}F_{56}D_{1237}F_{56}F_{47}),\
\mathcal{O}''_{14;2}=\mathrm{Tr}(D_{1234}F_{56}D_{127}F_{56}D_3F_{47}),\
\mathcal{O}''_{14;3}=\mathrm{Tr}(D_{124}F_{56}D_{1237}F_{56}D_3F_{47}),&
\nonumber\\
&
\mathcal{O}''_{14;4}=\mathrm{Tr}(D_{1234}F_{56}D_{17}F_{56}D_{23}F_{47}),\
\mathcal{O}''_{14;5}=\mathrm{Tr}(D_{124}F_{56}D_{137}F_{56}D_{23}F_{47}),\
\mathcal{O}''_{14;6}=\mathrm{Tr}(D_{14}F_{56}D_{1237}F_{56}D_{23}F_{47})&
\nonumber\\
&
\mathcal{O}''_{14;7}=\mathrm{Tr}(D_{1234}F_{56}D_7F_{56}D_{123}F_{47}),\
\mathcal{O}''_{14;8}=\mathrm{Tr}(D_{124}F_{56}D_{37}F_{56}D_{123}F_{47}),\
\mathcal{O}''_{14;9}=\mathrm{Tr}(D_{14}F_{56}D_{237}F_{56}D_{123}F_{47}),&
\nonumber\\
&
\mathcal{O}''_{14;10}=\mathrm{Tr}(D_4F_{56}D_{1237}F_{56}D_{123}F_{47});&
\nonumber\\
&
\mathcal{O}''_{14;11}=\mathrm{Tr}(D_{123}F_{56}D_{124}F_{57}D_{34}F_{67}),\
\mathcal{O}''_{14;12}=\mathrm{Tr}(D_{1234}F_{56}D_{12}F_{57}D_{34}F_{67}),\
\mathcal{O}''_{14;13}=\mathrm{Tr}(D_{123}F_{56}D_{1234}F_{57}D_4F_{67}),&
\nonumber\\
&
\mathcal{O}''_{14;14}=\mathrm{Tr}(D_{1234}F_{56}D_{123}F_{57}D_4F_{67}),\
\mathcal{O}''_{14;15}=\mathrm{Tr}(D_{1234}F_{56}D_{1234}F_{57}F_{67}) \,.
\\
\label{old-dim16}
& \underline{\mbox{dim\ 16}\ (15+7)}&\nonumber\\
&
\mathcal{O}''_{16;1}=\mathrm{Tr}(D_{12345}F_{67}D_{12348}F_{67}F_{58}),\
\mathcal{O}''_{16;2}=\mathrm{Tr}(D_{12345}F_{67}D_{1238}F_{67}D_4F_{58}),\
\mathcal{O}''_{16;3}=\mathrm{Tr}(D_{1235}F_{67}D_{12348}F_{67}D_4F_{58}),&
\nonumber\\
&
\mathcal{O}''_{16;4}=\mathrm{Tr}(D_{12345}F_{67}D_{128}F_{67}D_{34}F_{58}),\
\mathcal{O}''_{16;5}=\mathrm{Tr}(D_{1235}F_{67}D_{1248}F_{67}D_{34}F_{58}),\
\mathcal{O}''_{16;6}=\mathrm{Tr}(D_{125}F_{67}D_{12348}F_{67}D_{34}F_{58})&
\nonumber\\
&
\mathcal{O}''_{16;7}=\mathrm{Tr}(D_{12345}F_{67}D_{18}F_{67}D_{234}F_{58}),\
\mathcal{O}''_{16;8}=\mathrm{Tr}(D_{1235}F_{67}D_{148}F_{67}D_{234}F_{58}),\
\mathcal{O}''_{16;9}=\mathrm{Tr}(D_{125}F_{67}D_{1348}F_{67}D_{234}F_{58}),&
\nonumber\\
&
\mathcal{O}''_{16;10}=\mathrm{Tr}(D_{15}F_{67}D_{12348}F_{67}D_{234}F_{58}),\
\mathcal{O}''_{16;11}=\mathrm{Tr}(D_{12345}F_{67}D_8F_{67}D_{1234}F_{58}),\
\mathcal{O}''_{16;12}=\mathrm{Tr}(D_{1235}F_{67}D_{48}F_{67}D_{1234}F_{58}),&
\nonumber\\
&
\mathcal{O}''_{16;13}=\mathrm{Tr}(D_{125}F_{67}D_{348}F_{67}D_{1234}F_{58}),\
\mathcal{O}''_{16;14}=\mathrm{Tr}(D_{15}F_{67}D_{2348}F_{67}D_{1234}F_{58}),\
\mathcal{O}''_{16;15}=\mathrm{Tr}(D_5F_{67}D_{12348}F_{67}D_{1234}F_{58});&
\nonumber\\
&
\mathcal{O}''_{16;16}=\mathrm{Tr}(D_{1234}F_{67}D_{125}F_{68}D_{345}F_{78}),\
\mathcal{O}''_{16;17}=\mathrm{Tr}(D_{123}F_{67}D_{12345}F_{68}D_{45}F_{78}),\
\mathcal{O}''_{16;18}=\mathrm{Tr}(D_{1234}F_{67}D_{1235}F_{68}D_{45}F_{78}),&
\nonumber\\
&
\mathcal{O}''_{16;19}=\mathrm{Tr}(D_{12345}F_{67}D_{123}F_{68}D_{45}F_{78}),\
\mathcal{O}''_{16;20}=\mathrm{Tr}(D_{1234}F_{67}D_{12345}F_{68}D_5F_{78}),\
\mathcal{O}''_{16;21}=\mathrm{Tr}(D_{12345}F_{67}D_{1234}F_{68}D_5F_{78}),&
\nonumber\\
&
\mathcal{O}''_{16;22}=\mathrm{Tr}(D_{12345}F_{67}D_{12345}F_{68}F_{78}) \,.
\end{align}
\end{footnotesize}

As discussed in Section~\ref{sec:operatorbasis}, the above operators are not the final basis choice,
we still need to solve descendant relations as constraints and symmetrize
the operators which do not have symmetric properties.
As shown in Appendix \ref{app:newbasis}, the final basis operators we choose
are linear combinations of these preliminary ones.

\section{Operator basis up to dimension 16}

\label{app:newbasis}

In Section \ref{sec:conventional-method} and \ref{sec:onshell-method} we've explained how to find
length-3 basis operators from both field theory method and on-shell method by showing an example at dimension 10,
and these two strategies apply for general operator dimensions.

As mentioned in (\ref{linearcombine1}) there is a freedom in choosing which non-descendant operators to be
replaced by the descendant ones.  We require that the scalar factors $f_{\mathcal{O}}^{(0),\pm}$
(see Table \ref{table:FF-notation}, here $+$ for $\alpha$-sector, $-$ for $\beta$-sector) of
the chosen operators  should have the following forms:
\begin{enumerate}
\item
$f^{abc},(-,-,+)$:
$u^n(u^m+v^m+w^m)\ s_{123}^{(\Delta_0-8)/2}$,
 $\quad(m\neq 1,m+n\leq\frac{\Delta_0-8}{2})$
\item
$f^{abc},(-,-,-)$:
$u^m+v^m+w^m\ s_{123}^{(\Delta_0-6)/2}$,
$\quad(m\neq 1,m\leq\frac{\Delta_0-6}{2})$
\item
$d^{abc},(-,-,+)$:
$u^n(w^m-v^m)\ s_{123}^{(\Delta_0-8)/2}$,
$\quad(m\neq 0,m+n\leq\frac{\Delta_0-8}{2})$
\item
$d^{abc},(-,-,-)$:
$(u-w)(u-v)(v-w)(u^m+v^m+w^m)\ s_{123}^{(\Delta_0-6)/2}$,
$\quad(m\neq 1,m\leq\frac{\Delta_0-12}{2})$
\end{enumerate}
Here $u=\frac{s_{12}}{s_{123}},v=\frac{s_{23}}{s_{123}},w=\frac{s_{13}}{s_{123}}$,
and $\Delta_0$ is the dimension of the operator.

Under this constraint we can reduce the number of candidate operators and do not
violate the completeness. However, there is still some choice freedom.
For example, the number of independent non-descendent operators for dimension-12
$(d^{abc},--+)$-sector is 1, but there are two non-descendent scalar factors
satisfying the stated forms: $u(w-v)s_{123}^2$ and $w^2-v^2 s_{123}^2$.
Here during computation we choose the former.
Generally speaking, the choice freedom can not be avoided, and the different operator choices
correspond to different tree-level scalar factor.

The chosen basis operators are listed in following tables, labeled as
$\mathcal{O}_{\Delta_0,\alpha/\beta,f/d,i}$, where $\Delta_0$ is
the dimension operator, $\alpha/\beta$ denotes helicity sector (introduced in (\ref{eq:h-sector}) ),
$f/d$ denotes color factor $f^{abc}/d^{abc}$,  and $i$ denotes numbering.
All the new basis operators are linear combinations of
old ones given  in (\ref{old-dim6})-(\ref{old-dim16})
labeled as $\mathcal{O}''_{\Delta_0,i}$.

In the following tables, we show the tree level minimal form
factor of every basis operator in spinor-helicity formalism.
As for length-3 operators,
spinor structure of minimal form factor is universal within each
given helicity sector. Concretely, $\alpha$- and $\beta$-sectors, namely $(-,-,+)$-
and $(-,-,-)$-sectors, correspond to spinor factors:
\begin{equation}
\alpha:\
A_1=\la 12\ra^3[13][23],\qquad
\beta:\
A_2=\la 12\ra^3\la13\ra\la23\ra\,.
\end{equation}
Apart from that, each scalar factor is written as a power of $s_{123}$
times a polynomial of ratio variables
$u=\frac{s_{12}}{s_{123}}$, $v=\frac{s_{23}}{s_{123}}$, $w=\frac{s_{13}}{s_{123}}$.

For dimension 6, the only independent length-3 operator is
\begin{align}
\mathcal{O}_{6;\beta;f;1}=\mathcal{O}''_{6;1}
\end{align}
with tree-level minimal form factor $A_2$.
Basis operators at dimension 8, 10, 12, 14, 16 are given in
Table \ref{tab:dim8}, \ref{tab:dim10}, \ref{tab:dim12}, \ref{tab:dim14},
\ref{tab:dim16}.

\begin{table}[!t]
\centering
\caption{Final  basis operators  at  dimension 8}
\label{tab:dim8}
\vskip 0.4 cm
\begin{tabular}{|c|c|c|c|}
\hline
  basis operator & ${\cal F}^{(0)}(-,-,+)$ & ${\cal F}^{(0)}(-,-,-)$  &
  \begin{tabular}{c}  color \\ factor  \\ \end{tabular}
\\
\hline
$\mathcal{O}_{8;\alpha;f;1}=\mathcal{O}''_{8;1}-\frac{1}{2}\partial^2\mathcal{O}_{6;\beta;f;1}$
& $A_1$ & 0 & $f^{abc}$

\\
\hline
$\mathcal{O}_{8;\beta;f;1}=\frac{1}{2}\partial^2 \mathcal{O}_{6;\beta;f;1}$
& 0& $\frac{1}{2}s_{123}\ A_2 $  & $f^{abc}$

\\
\hline
\end{tabular}
\end{table}


\begin{table}[!t]
\centering
\caption{Final  basis operators  at  dimension 10}
\label{tab:dim10}
\vskip 0.4 cm
\begin{tabular}{|c|c|c|c|}
\hline
basis operator &  ${\cal F}^{(0)}(-,-,+)$ & ${\cal F}^{(0)}(-,-,-)$  &
  \begin{tabular}{c}  color \\ factor  \\ \end{tabular}
\\
\hline
$\mathcal{O}_{10;\alpha;f;1}=\frac{1}{2}\partial^2 \mathcal{O}_{8;\alpha;f;1}$
& $\frac{1}{2}s_{123} A_1$ & 0 & $f^{abc}$

\\
$\mathcal{O}_{10;\alpha;f;2}=\mathcal{O}''_{10;1}-\mathcal{O}''_{10;5}$
& $\frac{1}{2}s_{123} A_1\ u$ & 0 & $f^{abc}$

\\
\hline
$\mathcal{O}_{10;\alpha;d;1}=\mathcal{O}''_{10;2}-\mathcal{O}''_{10;3}$
& $\frac{1}{2}s_{123} A_1\ (w-v)$ & 0 & $d^{abc}$

\\
\hline
$\mathcal{O}_{10;\beta;f;1}=\frac{1}{4}\partial^4 \mathcal{O}_{6;\beta;f;1}$
& 0 & $\frac{1}{4}s_{123}^2 A_2$ & $f^{abc}$

\\
$\mathcal{O}_{10;\beta;f;2}=\mathcal{O}''_{10;5}$
& 0 & $\frac{1}{4}s_{123}^2 A_2\ (u^2+v^2+w^2)$  & $f^{abc}$

\\
\hline
\end{tabular}

\end{table}


\begin{table}[!t]
\centering
\caption{Final  basis operators  at  dimension 12}
\label{tab:dim12}
\vskip 0.4 cm
\begin{tabular}{|c|c|c|c|}
\hline
 basis operator & ${\cal F}^{(0)}(-,-,+)$ & ${\cal F}^{(0)}(-,-,-)$  &
  \begin{tabular}{c}  color \\ factor  \\ \end{tabular}
\\
\hline
\footnotesize{$\mathcal{O}_{12;\alpha;f;1}=\frac{1}{4}\partial^4 \mathcal{O}_{8;\alpha;f;1}$}
& \footnotesize{$\frac{1}{4}s_{123}^2 A_1$} & 0 & $f^{abc}$

\\
\footnotesize{$\mathcal{O}_{12;\alpha;f;2}=\frac{1}{2}\partial^2 \mathcal{O}_{10;\alpha;f;2}$}
& \footnotesize{$\frac{1}{4}s_{123}^2 A_1\ u$} & 0 & $f^{abc}$

\\
\footnotesize{$\mathcal{O}_{12;\alpha;f;3}=\mathcal{O}''_{12;1}-\mathcal{O}''_{12;10}$}
& \footnotesize{$\frac{1}{4}s_{123}^2 A_1\ u^2$ }
& 0 & $f^{abc}$

\\
\footnotesize{$\mathcal{O}_{12;\alpha;f;4}=\mathcal{O}''_{12;1}+\mathcal{O}''_{12;4}+\mathcal{O}''_{12;6}$}
& \footnotesize{$\frac{1}{4}s_{123}^2A_1\ (u^2+v^2+w^2)$ }
& 0 & $f^{abc}$

\\
\footnotesize{$\hspace{1.7cm}-\frac{1}{2}\partial^2\mathcal{O}_{10;\beta;f;2}$}
& & &
\\
\hline
\footnotesize{$\mathcal{O}_{12;\alpha;d;1}=\frac{1}{2}\partial^2\mathcal{O}_{10;\alpha;d;1}$}
&\footnotesize{ $\frac{1}{4}s_{123}^2 A_1\ (w-v)$ }
& 0 & $d^{abc}$

\\
\footnotesize{$\mathcal{O}_{12;\alpha;d;2}=\mathcal{O}''_{12;2}-\mathcal{O}''_{12;3}+\mathcal{O}''_{12;8}-\mathcal{O}''_{12;9}$}
& \footnotesize{$\frac{1}{4}s_{123}^2 A_1\ u(w-v)$}
 & 0 & $d^{abc}$

 \\
\hline
\footnotesize{$\mathcal{O}_{12;\beta;f;1}=\frac{1}{8}\partial^6 \mathcal{O}_{6;\beta;f;1}$}
& 0 & \footnotesize{$\frac{1}{8}s_{123}^3 A_2$} & $f^{abc}$

\\
\footnotesize{$\mathcal{O}_{12;\beta;f;2}=\frac{1}{2}\partial^2 \mathcal{O}_{10;\beta;f;2}$}
& 0 & \footnotesize{$\frac{1}{8}s_{123}^3A_2\ (u^2+v^2+w^2)$}  & $f^{abc}$

\\
 \footnotesize{$\mathcal{O}_{12;\beta;f;3}=\mathcal{O}''_{12;10}$}
& 0 &\footnotesize{ $\frac{1}{8}s_{123}^3A_2\ (u^3+v^3+w^3)$} & $f^{abc}$

\\
\hline
\footnotesize{$\mathcal{O}_{12;\beta;d;1}=\mathcal{O}''_{12;8}-\mathcal{O}''_{12;9}$}
& 0 & \footnotesize{$\frac{1}{8}s_{123}^3 A_2 (u-v)(u-w)(v-w)$}
 & $d^{abc}$
\\
\hline
\end{tabular}
\end{table}


\begin{table}[!t]
\centering
\caption{Final  basis operators  at  dimension 14}
\label{tab:dim14}
\vskip 0.4 cm
\begin{tabular}{|c|c|c|c|}
\hline
basis operator &  ${\cal F}^{(0)}(-,-,+)$ & ${\cal F}^{(0)}(-,-,-)$  &
  \begin{tabular}{c}  color \\ factor  \\ \end{tabular}
\\
\hline
\footnotesize{$\mathcal{O}_{14;\alpha;f;1}=\frac{1}{8}\partial^6 \mathcal{O}_{8;\alpha;f;1}$}
& \footnotesize{$\frac{1}{8}s_{123}^3 A_1$} & 0 & $f^{abc}$

\\
\footnotesize{$\mathcal{O}_{14;\alpha;f;2}=\frac{1}{4}\partial^4 \mathcal{O}_{10;\alpha;f;2}$}
& \footnotesize{$\frac{1}{8}s_{123}^3 A_1\ u$} & 0 & $f^{abc}$

\\
\footnotesize{$\mathcal{O}_{14;\alpha;f;3}=\frac{1}{2}\partial^2 \mathcal{O}_{12;\alpha;f;3}$}
& \footnotesize{$\frac{1}{8}s_{123}^3 A_1\ u^2$ } & 0 & $f^{abc}$

\\
\footnotesize{$\mathcal{O}_{14;\alpha;f;4}=\frac{1}{2}\partial^2 \mathcal{O}_{12;\alpha;f;4} $}
& \footnotesize{$\frac{1}{8}s_{123}^3 A_1\ (u^2+v^2+w^2)$ } & 0 & $f^{abc}$

\\
\footnotesize{$\mathcal{O}_{14;\alpha;f;5}=\mathcal{O}''_{14;1}-\mathcal{O}''_{14;15}$}
& \footnotesize{$\frac{1}{8}s_{123}^3 A_1\ u^3$ } & 0 & $f^{abc}$

\\
\footnotesize{$\mathcal{O}_{14;\alpha;f;6}=\mathcal{O}''_{14;1}+\mathcal{O}''_{14;7}+\mathcal{O}''_{14;10}$}
& \footnotesize{$\frac{1}{8}s_{123}^3 A_1\ (u^3+v^3+w^3)$ } & 0 & $f^{abc}$

\\
\footnotesize{\hspace{1.7cm}$-\frac{1}{2}\partial^2\mathcal{O}_{12;\beta;f;3}$}
& & &

\\
\hline
\footnotesize{$\mathcal{O}_{14;\alpha;d;1}=\frac{1}{4}\partial^4\mathcal{O}_{10;\alpha;d;1}$}
&\footnotesize{ $\frac{1}{8}s_{123}^3 A_1\ (w-v)$ } & 0 & $d^{abc}$

\\
 \footnotesize{$\mathcal{O}_{14;\alpha;d;2}=\frac{1}{2}\partial^2 \mathcal{O}_{12;\alpha;d;2}$}
& \footnotesize{$\frac{1}{8}s_{123}^3 A_1\ u(w-v)$} & 0 & $d^{abc}$

\\
\footnotesize{$\mathcal{O}_{14;\alpha;d;3}=\mathcal{O}''_{14;2}-\mathcal{O}''_{14;3}+\frac{1}{2}\partial^2\mathcal{O}_{12;\beta;d;1}$}
& \footnotesize{$\frac{1}{8}s_{123}^3 A_1\ u^2(w-v)$} & 0 & $d^{abc}$

\\
\footnotesize{$\mathcal{O}_{14;\alpha;d;4}=\mathcal{O}''_{14;7}-\mathcal{O}''_{14;10}-\frac{1}{2}\partial^2\mathcal{O}_{12;\beta;d;1}$}
& \footnotesize{$\frac{1}{8}s_{123}^3 A_1\ (w^3-v^3)$} & 0 & $d^{abc}$

\\
\hline
\footnotesize{$\mathcal{O}_{14;\beta;f;1}=\frac{1}{16}\partial^8 \mathcal{O}_{6;\beta;f;1}$}
& 0 & \footnotesize{$\frac{1}{16}s_{123}^4 A_2$} & $f^{abc}$

\\
\footnotesize{$\mathcal{O}_{14;\beta;f;2}=\frac{1}{4}\partial^4 \mathcal{O}_{10;\beta;f;2}$}
& 0 & \footnotesize{$\frac{1}{16}s_{123}^4 A_2\ (u^2+v^2+w^2)$}  & $f^{abc}$

\\
\footnotesize{$\mathcal{O}_{14;\beta;f;3}=\frac{1}{2}\partial^2 \mathcal{O}_{12;\beta;f;3}$}
&0 & \footnotesize{ $\frac{1}{16}s_{123}^4 A_2\ (u^3+v^3+w^3)$} & $f^{abc}$
\\

\footnotesize{$\mathcal{O}_{14;\beta;f;4}=\mathcal{O}''_{14;15}$}
& 0 & \footnotesize{ $\frac{1}{16}s_{123}^4 A_2\ (u^4+v^4+w^4)$} & $f^{abc}$
\\
\hline
\footnotesize{$\mathcal{O}_{14;\beta;d;1}=\frac{1}{2}\partial^2\mathcal{O}_{12;\beta;d;1}$}
& 0 & \footnotesize{$\frac{1}{16}s_{123}^4 A_2(u-v)(u-w)(v-w)$} & $d^{abc}$
\\
\hline
\end{tabular}
\end{table}


\begin{table}[!t]
\caption{Final basis operators at  dimension 16}
\label{tab:dim16}
\vskip 0.4 cm
\hskip -0.7cm
\begin{tabular}{|c|c|c|c|}
\hline
basis operator &  ${\cal F}^{(0)}(-,-,+)$ & ${\cal F}^{(0)}(-,-,-)$  &
  \begin{tabular}{c}  color \\ factor  \\ \end{tabular}
\\
\hline
\footnotesize{$\mathcal{O}_{16;\alpha;f;1}=\frac{1}{16}\partial^8 \mathcal{O}_{8;\alpha;f;1}$}
& \footnotesize{$\frac{1}{16}s_{123}^4 A_1$} & 0 & $f^{abc}$

\\
\footnotesize{$\mathcal{O}_{16;\alpha;f;2}=\frac{1}{8}\partial^6 \mathcal{O}_{10;\alpha;f;2}$}
& \footnotesize{$\frac{1}{16}s_{123}^4 A_1\ u$} & 0 & $f^{abc}$

\\
\footnotesize{$\mathcal{O}_{16;\alpha;f;3}=\frac{1}{4}\partial^4 \mathcal{O}_{12;\alpha;f;3}$}
& \footnotesize{$\frac{1}{16}s_{123}^4 A_1\ u^2$ } & 0 & $f^{abc}$

\\
\footnotesize{$\mathcal{O}_{16;\alpha;f;4}=\frac{1}{4}\partial^4 \mathcal{O}_{12;\alpha;f;4} $}
& \footnotesize{$\frac{1}{16}s_{123}^4 A_1\ (u^2+v^2+w^2)$ }
& 0 & $f^{abc}$

\\
\footnotesize{$\mathcal{O}_{16;\alpha;f;5}=\frac{1}{2}\partial^2 \mathcal{O}_{14;\alpha;f;5}$}
& \footnotesize{$\frac{1}{16}s_{123}^4 A_1\ u^3$ }
& 0 & $f^{abc}$

\\
\footnotesize{$\mathcal{O}_{16;\alpha;f;6}=\frac{1}{2}\partial^2 \mathcal{O}_{14;\alpha;f;6}$}
& \footnotesize{$\frac{1}{16}s_{123}^4 A_1\ (u^3+v^3+w^3)$ }
& 0 & $f^{abc}$

\\
\footnotesize{$\mathcal{O}_{16;\alpha;f;7}=\mathcal{O}''_{16;1}-\mathcal{O}''_{16;22}$}
& \footnotesize{$\frac{1}{16}s_{123}^4 A_1\ u^4$ }
& 0 & $f^{abc}$

\\
\footnotesize{$\mathcal{O}_{16;\alpha;f;8}=\mathcal{O}''_{16;1}+\mathcal{O}''_{16;7}+\mathcal{O}''_{16;10}$}
& \footnotesize{$\frac{1}{16}s_{123}^4 A_1\ u(u^3+v^3+w^3)$ } & 0 & $f^{abc}$

\\
\footnotesize{\hspace{1.7cm}$-\mathcal{O}''_{16;17}-\mathcal{O}''_{16;19}-\mathcal{O}''_{16;22}$}
& & &

\\
\footnotesize{$\mathcal{O}_{16;\alpha;f;9}=\mathcal{O}''_{16;1}+\mathcal{O}''_{16;11}+\mathcal{O}''_{16;15}$}
& \footnotesize{$\frac{1}{16}s_{123}^4 A_1\ (u^4+v^4+w^4)$ }
& 0 & $f^{abc}$

\\
\footnotesize{\hspace{1.7cm}$-\frac{1}{2}\partial^2\mathcal{O}_{14;\beta;f;4}$}
& & &

\\
\hline
\footnotesize{$\mathcal{O}_{16;\alpha;d;1}=\frac{1}{8}\partial^6 \mathcal{O}_{10;\alpha;d;1}$}
& \footnotesize{ $\frac{1}{16}s_{123}^4 A_1\ (w-v)$ }
& 0 & $d^{abc}$

\\
 \footnotesize{$\mathcal{O}_{16;\alpha;d;2}=\frac{1}{4}\partial^4 \mathcal{O}_{12;\alpha;d;2}$}
& \footnotesize{$\frac{1}{16}s_{123}^4 A_1\ u(w-v)$}
 & 0 & $d^{abc}$

\\
\footnotesize{$\mathcal{O}_{16;\alpha;d;3}=\frac{1}{2}\partial^2 \mathcal{O}_{14;\alpha;d;3}$}
& \footnotesize{$\frac{1}{16}s_{123}^4 A_1\ u^2(w-v)$}
& 0 & $d^{abc}$

\\
\footnotesize{$\mathcal{O}_{16;\alpha;d;4}=\frac{1}{2}\partial^2 \mathcal{O}_{14;\alpha;d;4}$}
& \footnotesize{$\frac{1}{16}s_{123}^3 A_1\ (w^3-v^3)$}
& 0 & $d^{abc}$

\\
\footnotesize{$\mathcal{O}_{16;\alpha;d;5}=\mathcal{O}''_{16;7}-\mathcal{O}''_{16;10}+\mathcal{O}''_{16;20}-\mathcal{O}''_{16;21}$}
& \footnotesize{$\frac{1}{16}s_{123}^4 A_1\ u(w^3-v^3)$}
& 0 & $d^{abc}$

\\
\footnotesize{\hspace{1.7cm}$-\frac{1}{4}\partial^4\mathcal{O}_{12;\beta;d;1}$}
& & &

\\
\footnotesize{$\mathcal{O}_{16;\alpha;d;6}=\mathcal{O}''_{16;11}-\mathcal{O}''_{16;15}-\mathcal{O}''_{16;20}+\mathcal{O}''_{16;21}$}
& \footnotesize{$\frac{1}{16}s_{123}^3 A_1\ (w^4-v^4)$}
& 0 & $d^{abc}$

\\
\hline
\footnotesize{$\mathcal{O}_{16;\beta;f;1}=\frac{1}{32}\partial^{10} \mathcal{O}_{6;\beta;f;1}$}
& 0 & \footnotesize{$\frac{1}{32}s_{123}^5 A_2$} & $f^{abc}$

\\
 \footnotesize{$\mathcal{O}_{16;\beta;f;2}=\frac{1}{8}\partial^6 \mathcal{O}_{10;\beta;f;2}$}
& 0 & \footnotesize{$\frac{1}{32}s_{123}^5 A_2\ (u^2+v^2+w^2)$}  & $f^{abc}$

\\
\footnotesize{$\mathcal{O}_{16;\beta;f;3}=\frac{1}{4}\partial^4 \mathcal{O}_{12;\beta;f;3}$}
& 0 & \footnotesize{ $\frac{1}{32}s_{123}^5 A_2\ (u^3+v^3+w^3)$} & $f^{abc}$

\\
\footnotesize{$\mathcal{O}_{16;\beta;f;4}=\frac{1}{2}\partial^2 \mathcal{O}_{14;\beta;f;4}$}
& 0 & \footnotesize{ $\frac{1}{32}s_{123}^5 A_2\ (u^4+v^4+w^4)$} & $f^{abc}$

\\
\footnotesize{$\mathcal{O}_{16;\beta;f;5}=\mathcal{O}''_{16;22}$}
& 0 & \footnotesize{ $\frac{1}{32}s_{123}^5 A_2\ (u^5+v^5+w^5)$} & $f^{abc}$

\\
\hline
\footnotesize{$\mathcal{O}_{16;\beta;d;1}=\frac{1}{4}\partial^4 \mathcal{O}_{12;\beta;d;1}$}
& 0 & \footnotesize{$\frac{1}{32}s_{123}^5 A_2(u-v)(u-w)(v-w)$} & $d^{abc}$

\\
\footnotesize{$\mathcal{O}_{16;\beta;d;2}=\mathcal{O}''_{16;17}-\mathcal{O}''_{16;19}-\mathcal{O}''_{16;20}+\mathcal{O}''_{16;21}$}
& 0 & \footnotesize{$\frac{1}{32}s_{123}^5 A_2(u-v)(u-w)(v-w)$} & $d^{abc}$
\\
& & \footnotesize{$\times(u^2+v^2+w^2)$}   &

\\
\hline
\end{tabular}
\end{table}

\newpage
{\color{white}blank for table}
\newpage
{\color{white}blank for table}
\newpage
{\color{white}blank for table}

In our ancillary file which gives explicit expressions of two-loop finite remainders,
the operators are arranged in following orders: $(f,\alpha)$-sector, $(f,\beta)$-sector,
$(d,\alpha)$-sector, $(d,\beta)$-sector.
\begin{itemize}

\item dimension 8: 1+1+0+0

$\{\mathcal{O}_{8;\alpha;f;1},\mathcal{O}_{8;\beta;f;1}\}$

\item dimension 10: 2+2+1+0

$\{\mathcal{O}_{10;\alpha;f;1},\mathcal{O}_{10;\alpha;f;2},
\mathcal{O}_{10;\beta;f;1},\mathcal{O}_{10;\beta;f;2},
\mathcal{O}_{10;\alpha;d;1} \}$

\item dimension 12: 4+3+2+1

$\{\mathcal{O}_{12;\alpha;f;1},...,\mathcal{O}_{12;\alpha;f;4},
\mathcal{O}_{12;\beta;f;1},...,\mathcal{O}_{12;\beta;f;3},
\mathcal{O}_{12;\alpha;d;1},\mathcal{O}_{12;\alpha;d;2},
\mathcal{O}_{12;\beta;f;1}\}$

\item dimension 14: 6+4+4+1

$\{\mathcal{O}_{14;\alpha;f;1},...,\mathcal{O}_{14;\alpha;f;6},
\mathcal{O}_{14;\beta;f;1},...,\mathcal{O}_{14;\beta;f;4},
\mathcal{O}_{14;\alpha;d;1},...,\mathcal{O}_{14;\alpha;d;4},
\mathcal{O}_{14;\beta;f;1}\}$

\item dimension 16: 9+5+6+2

$\{\mathcal{O}_{16;\alpha;f;1},...,\mathcal{O}_{16;\alpha;f;9},
\mathcal{O}_{16;\beta;f;1},...,\mathcal{O}_{16;\beta;f;5},
\mathcal{O}_{16;\alpha;d;1},...,\mathcal{O}_{16;\alpha;d;6},
\mathcal{O}_{16;\beta;f;1},\mathcal{O}_{16;\beta;f;2}\}$
\end{itemize}
Since descendants are always kept into basis choice,
the remainder list of basis operators at a certain dimension cover all the remainders
of lower dimensional basis.

\section{Dilatation operator under new operator definition}
\label{app:absorb-g}

In this appendix, we provide some details regarding the length-4 operators in Section~\ref{sec:length4}.
\subsubsection*{Dilatation operator with length-4}

Back to perturbative expansion of dilatation operator,
by adding length-4 operators using (\ref{Doperator1})-(\ref{Doperator2}),
we modify the length-3 formula \eqref{Dforlen3-2}  as
\begin{equation}
\begin{aligned}
\mathds{D}^{(1)}&=2\epsilon \Big( Z^{(1)}_{2\rightarrow 2}
+ Z^{(1)}_{3\rightarrow 3}
+ Z^{(1)}_{4\rightarrow 4} \Big),
\\
\label{Doperator2-mid}
\mathds{D}^{(\frac{3}{2})}&=3\epsilon\Big(
Z^{(1)}_{3\rightarrow 4}+Z^{(2)}_{3\rightarrow 2}
+Z^{(2)}_{4\rightarrow 3}
\Big),
\\
\mathds{D}^{(2)}
&=4\epsilon\Big(
Z^{(2)}_{2\rightarrow 2}\big|_{\frac{1}{\epsilon}-\mathrm{part}}
+Z^{(2)}_{3\rightarrow 3}\big|_{\frac{1}{\epsilon}-\mathrm{part}}
+Z^{(2)}_{4\rightarrow 4}\big|_{\frac{1}{\epsilon}-\mathrm{part}}
+Z^{(3)}_{4\rightarrow 2}
\Big)\,.
\end{aligned}
\end{equation}
Above formulae correspond to operator definition which does not
include gauge coupling as mentioned in Footnote \ref{footnote:lengthchanging}.
Again notice there is no mixing from length-2 operator to higher length ones
as mentioned in (\ref{eq:Z2toLasZero}).
For dimension-8 case, there is no operator with length higher than 4,
so including length-4 operators makes the operator basis complete and
(\ref{Doperator2-mid}) provide the full result.

Renormalization matrix is obtained
from UV subtraction of loop form factors.
Using (\ref{formfactor-RG-1})-(\ref{formfactor-RG-2}),
we consider following concrete relations. For the minimal form factors with $E=L$, one has
\begin{align}
\label{FF-pertur-1a}
\mathcal{F}_{{\cal O}_i;R}^{(1)}&=
\mathcal{F}_{{\cal O}_i;B}^{(1)}
+(Z^{(1)}_{L\rightarrow L})_i^{\ j}\mathcal{F}_{{\cal O}_j;B}^{(0)} \,,
\\
\label{FF-pertur-2a}
\mathcal{F}_{{\cal O}_i;R}^{(2)}
&=
\mathcal{F}_{{\cal O}_i;B}^{(2)}
-\frac{\beta_0}{\epsilon}\mathcal{F}_{{\cal O}_i;B}^{(1)}
+(Z^{(1)}_{L\rightarrow L})_i^{\ j}\mathcal{F}_{{\cal O}_j;B}^{(1)}
+(Z^{(2)}_{L\rightarrow L})_i^{\ j}\mathcal{F}_{{\cal O}_j;B}^{(0)}
+(Z^{(2)}_{L\rightarrow L-1})_i^{\ j}\mathcal{F}_{{\cal O}_j;B}^{(0)} \,,
\end{align}
which will be used to determine $Z^{(1)}_{3\rightarrow 3}$, $Z^{(1)}_{4\rightarrow 4}$ $Z^{(2)}_{3\rightarrow 3}$,
$Z^{(2)}_{3\rightarrow 2}$.

We also need to consider non-minimal form factors:
\begin{align}
\label{FF-pertur-3a}
E=L+1:\quad
\mathcal{F}_{{\cal O}_i;R}^{(1)}&=
\mathcal{F}_{{\cal O}_i;B}^{(1)}
-\frac{1}{2}\frac{\beta_0}{\epsilon}
\mathcal{F}_{{\cal O}_i;B}^{(0)}
+(Z^{(1)}_{L\rightarrow L})_i^{\ j}\mathcal{F}_{{\cal O}_j;B}^{(0)}
+(Z^{(1)}_{L\rightarrow L+1})_i^{\ j}\mathcal{F}_{{\cal O}_j;B}^{(0)} \,,
\\
\label{FF-pertur-4a}
E=L-1:\quad
\mathcal{F}_{{\cal O}_i;R}^{(2)}&=
\mathcal{F}_{{\cal O}_i;B}^{(2)}
+(Z^{(2)}_{L\rightarrow L-1})_i^{\ j}\mathcal{F}_{{\cal O}_j;B}^{(0)} \,.
\end{align}
Setting $E=4$, one can use (\ref{FF-pertur-3a}) to
get $Z^{(1)}_{3\rightarrow 4}$ from one-loop 4-gluon form factors length-3 operators,
and use (\ref{FF-pertur-4a}) to get $Z^{(2)}_{4\rightarrow 3}$
from two-loop 3-gluon form factors of length-4.
These provide almost all information to determine the dilatation operator in \eqref{Doperator2-mid},
except matrix elements of $Z^{(2)}_{4\rightarrow 4}$ and $Z^{(3)}_{4\rightarrow 2}$.
To determine $Z^{(2)}_{4\rightarrow 4}$ one needs to compute two-loop 4-gluon form factors of length-4 operators,
while to determine $Z^{(2)}_{4\rightarrow 4}$ one needs to take three-loop calculation,
and these two types of data
will not be discussed in this paper.
Nevertheless, to determine the two-loop anomalous dimension of length-3 operators, the information of $Z^{(2)}_{4\rightarrow 4}$ and $Z^{(2)}_{3\rightarrow 4}$ are not needed as shown in (\ref{eigenvalue-dim8-define2}).

\subsubsection*{Operators with coupling in the definition}

As we mentioned in Footnote \ref{footnote:lengthchanging},
if we multiply all the basis length-4 operators by an overall coupling $g$,
the perturbative order of length-changed matrix elements will change
and running effect of gauge coupling will enter in, and as a result
perturbative expansion (\ref{Doperator2-mid}) will change to:
\begin{align}
\mathds{D}^{(1)}&=2\epsilon
\big(Z^{(1)}_{2\rightarrow 2}
+Z^{(1)}_{3\rightarrow 3}
+Z^{(1)}_{4\rightarrow 4}
+Z^{(1)}_{3\rightarrow 4}
+\frac{\beta_0}{2\epsilon} Z^{(0)}_{4\rightarrow 4}
\big),
\nonumber\\
\label{Doperator2-new}
\mathds{D}^{(\frac{3}{2})}&=3\epsilon\Big(
Z^{(2)}_{3\rightarrow 2}
\Big),&
\\
\mathds{D}^{(2)}
&=4\epsilon\Big(
Z^{(2)}_{2\rightarrow 2}\big|_{\frac{1}{\epsilon}-\mathrm{part}}
+Z^{(2)}_{3\rightarrow 3}\big|_{\frac{1}{\epsilon}-\mathrm{part}}
+Z^{(2)}_{4\rightarrow 4}\big|_{\frac{1}{\epsilon}-\mathrm{part}}
+Z^{(2)}_{3\rightarrow 4}\big|_{\frac{1}{\epsilon}-\mathrm{part}}
+Z^{(2)}_{4\rightarrow 3}
+\frac{\beta_1}{4\epsilon}Z^{(0)}_{4\rightarrow 4}
\Big)\,.\nonumber
\end{align}
The basis (\ref{eq:dim8len234}) we choose in Section \ref{sec:length4} correspond to
this operator redefinition.
Notice perturbative expansion (\ref{FF-pertur-1a})-(\ref{FF-pertur-4a})
will not change because the order change of $Z$-matrix elements will
be compensated for the order change of form factors.

\subsection*{Choice of other basis}

An alternative basis we choose are
\begin{equation}
\begin{aligned}
\label{dim8-basis3}
L=2: &\quad \mathcal{O}_{8;0} \,,
\\
L=3: &\quad \mathcal{O}_{8;\alpha;f;1},\
\mathcal{O}_{8;\beta;f;1} \,,
\\
L=4: &\quad
\Xi_1,\
\Xi_2,\
\Xi_3,\
\Xi_4\,.
\end{aligned}
\end{equation}
Different from (\ref{eq:dim8len234}), the length-4 operators in (\ref{dim8-basis3})
do not contain any gauge coupling. 
As a result, operators basis and $Z$-matrix from  (\ref{eq:dim8len234}) and (\ref{dim8-basis3})
are related by a $g$-dependent transformation:
\begin{align}
\mathcal{O}'_i=\sum_a
M(g)_{\mathcal{O}'_i}^{\ \mathcal{O}_a}\mathcal{O}_a\,,\qquad
\tilde{Z}_{\mathcal{O}'_i}^{\ \mathcal{O}'_j}
=\sum_{a,b}
M(g)_{\mathcal{O}_i}^{\ \mathcal{O}_a}
Z_{O_a}^{\ O_b} (M^{-1}(g_0))_{\mathcal{O}_b}^{\ \mathcal{O}'_j}\,.
\end{align}
Therefore eigenvalues of the two
dilatation operators are not expected to be the same.
There are still three out of seven eigenvalues calculated from (\ref{dim8-basis3})
that are independent of  the unfixed two-loop matrix elements, with values
\begin{align}
\label{eigenvalue-dim8-define2}
\hat\gamma^{(1)}_{\mathcal{O}_8} = \left\{-\frac{22}{3}; 1; \frac{7}{3}\right\}  ,\qquad
\hat{\gamma}^{(2)}_{\mathcal{O}_8}=\left\{
-\frac{136}{3};\frac{25}{3};\frac{75793}{576}
\right\}\,.
\end{align}
The difference between (\ref{eigenvalue-dim8-define1})
and (\ref{eigenvalue-dim8-define2}) lies in the third eigenvalue at
 $\mathcal{O}(\alpha_s^2)$ order.
This difference purely originates from the $g$-dependent basis transformation.

\section{Two-loop remainder of $\mathcal{O}_{8;\alpha;f;1}$ }
\label{app:O81remainder}

In this appendix we provide explicitly the finite remainder of two-loop form factor for $\mathcal{O}_{8;\alpha;f;1}$
under helicity setting $(-,-,+)$, which is used as an example for illustrating analytical structures
in various aspects in Section \ref{sec:finite}.
As shown in (\ref{eq:normalization1}),
the remainder has been normalized by tree-level form factor
$\la12\ra^3[13][23] f^{(0),+}_{\mathcal{O}_{8;\alpha;f;1}}$,
where $f^{(0),+}_{\mathcal{O}_{8;\alpha;f;1}}=1$.
The machine readable form, as well as results for other operators, can be found in the ancillary file submitted together with this paper.

\begin{footnotesize}
\begin{equation}
 \begin{aligned}
 \label{eq:O81deg3}
 &\mathcal{R}^{(2),+}_{\mathcal{O}_{8;\alpha;f;1}}\Big|_{\mathrm{deg} 3}=
\Big[\frac{(8 u^3+9 u^2 w+3 u w^2+2 w^3) T_3(v,w,u)}{6
   u^3}+\frac{(v+w) T_3(u,v,w)}{w}+\frac{(u+w) T_3(v,u,w)}{2 w}
   &\\&
   -\frac{5}{24} \pi ^2\log (v)+v\leftrightarrow w\Big]
   -\frac{1}{24} \pi ^2 \log (u)
   +\frac{143 }{12}\zeta_3\,,&
 \end{aligned}
\end{equation}
\end{footnotesize}
\begin{footnotesize}
\begin{equation}
 \begin{aligned}
 \label{eq:O81deg2}
 &\mathcal{R}^{(2),+}_{\mathcal{O}_{8;\alpha;f;1}}\Big|_{\mathrm{deg} 2}=
\Big[\frac{1}{36 u^2 w^5} T_2(u,v)(108 u^4 v^3+252 u^4 v^2 w+162 u^4 v w^2+36 u^4 w^3+33
   u^3 v^3 w+111 u^3 v^2 w^2+72 u^3 v w^3
   &\\&
   -6 u^3 w^4-20 u^2 v^3 w^2-33 u^2 v^2 w^3-60
   u^2 v w^4-47 u^2 w^5-6 u v^3 w^3-18 u v^2 w^4-12 u v w^5-12 u w^6+6 v^2 w^5-6
   w^7)  +v\leftrightarrow w\Big]
   &\\&
   +\frac{1}{36 u^6}
   T_2(v,w) \big(-60 u^6+54 u^4 v w+18
   u^2 v^2 w^2+(-69 u^5 v-51 u^4 v^2-41 u^3 v^3+9 u^3 v^2 w-60 u^2
   v^3 w+66 u v^3 w^2
   &\\&
    +v\leftrightarrow w)  +180 v^3 w^3 \big)
   +\Big[\frac{\log
   (u) \log (v) (25 u^2-2 u v+2 u w-2 v^2+2 w^2)}{12 u^2}+\frac{(-197
   u^2-12 u w-12 w^2) \log ^2(v)}{72 u^2} +v\leftrightarrow w\Big]
   &\\&
   +\frac{\log (v) \log (w) \big(71 u^2+6 u
   +(6 v^2 +v\leftrightarrow w)\big)}{36 u^2}+\frac{\pi^2\big(143  u^2-u
   +(- v^2 +v\leftrightarrow w) \big)}{36 u^2}-\frac{55}{24} \log ^2(u)\,,
  \end{aligned}
\end{equation}
\end{footnotesize}
\begin{footnotesize}
\begin{equation}
 \begin{aligned}
 \label{eq:O81deg1}
 &\mathcal{R}^{(2),+}_{\mathcal{O}_{8;\alpha;f;1}}\Big|_{\mathrm{deg} 1}=
 \frac{1}{216 u v^4 w^4} \log (u) \big(72 u^3 v^3 w^3+(648 u^3 v^6+1188 u^3 v^5 w+432
   u^3 v^4 w^2+198 u^2 v^6 w+567 u^2 v^5 w^2
   &\\&
   +165 u^2 v^4 w^3-120 u v^6 w^2-138 u v^5 w^3-36 v^6 w^3-282 v^5 w^4 +v\leftrightarrow w)+2366 u v^4 w^4 \big)
   &\\&
   +\Big[\frac{1}{216 u^5 w^4 (u+w)^2}  \log (v) (648 u^8 v^3+1512 u^8 v^2 w+972 u^8 v
   w^2+216 u^8 w^3+1170 u^7 v^3 w+2934 u^7 v^2 w^2+1890 u^7 v w^3
   &\\&
   +3467 u^7 w^4+393 u^6
   v^3 w^2+1305 u^6 v^2 w^3+843 u^6 v w^4+6163 u^6 w^5-204 u^5 v^3 w^3-516 u^5 v^2
   w^4+138 u^5 v w^5+2615 u^5 w^6
   &\\&
   -465 u^4 v^3 w^4-357 u^4 v^2 w^5+87 u^4 v w^6-471 u^4
   w^7-672 u^3 v^3 w^5+348 u^3 v^2 w^6-486 u^3 v w^7-174 u^3 w^8+594 u^2 v^3 w^6
   &\\&
   +702 u^2 v^2 w^7-360 u^2 v w^8+2016 u v^3 w^7+396 u v^2 w^8+1080 v^3 w^8)
   +v\leftrightarrow w\Big]\,,
 \end{aligned}
\end{equation}
\end{footnotesize}
\begin{footnotesize}
\begin{equation}
 \begin{aligned}
 \label{eq:O81deg0}
 &\mathcal{R}^{(2),+}_{\mathcal{O}_{8;\alpha;f;1}}\Big|_{\mathrm{deg} 0}=
  \frac{1}{1728 u^5 v^4 (u+v)^2 w^4 (u+w)^2}
  \big(-2634 v^3 w^3 u^{11}+7902 v^3 w^3 u^{10}-181786 v^4 w^4 u^9+61092 v^4 w^4 u^8
   &\\&
   -347854 v^5 w^5 u^7-34860 v^5 w^5 u^6+68762 v^6 w^6 u^5-35348 v^6 w^6 u^4-68256
   v^7 w^7 u^3+88248 v^7 w^7 u^2-131112 v^8 w^8 u
   &\\&
   +29592 v^8 w^8+(-17064 v^6 u^{11}-14370 v^4 w^2 u^{11}-36072 v^5 w
   u^{11}-51192 v^7 u^{10}+17064 v^6 u^{10}-45680 v^4 w^3 u^{10}
   &\\&
   -144081 v^5 w^2
   u^{10}+14370 v^4 w^2 u^{10}-152370 v^6 w u^{10}+36072 v^5 w u^{10}-51192 v^8
   u^9+34128 v^7 u^9-198560 v^5 w^3 u^9
   &\\&\hspace{-0.3cm}
   +46784 v^4 w^3 u^9-346630 v^6 w^2 u^9+98823 v^5
   w^2 u^9-226206 v^7 w u^9+101826 v^6 w u^9-17064 v^9 u^8+17064 v^8 u^8-294920 v^5 w^4
   u^8
   &\\&
   -297420 v^6 w^3 u^8+100137 v^5 w^3 u^8-325422 v^7 w^2 u^8+161461 v^6 w^2
   u^8-139590 v^8 w u^8+95436 v^7 w u^8-119166 v^6 w^4 u^7
   &\\&
   +26171 v^5 w^4 u^7-160450 v^7
   w^3 u^7+83270 v^6 w^3 u^7-115428 v^8 w^2 u^7+83933 v^7 w^2 u^7-29682 v^9 w u^7+29682
   v^8 w u^7
   &\\&
   -44861 v^6 w^5 u^6+78673 v^7 w^4 u^6-32857 v^6 w^4 u^6-10284 v^8 w^3
   u^6+13755 v^7 w^3 u^6-6925 v^9 w^2 u^6+6925 v^8 w^2 u^6
   &\\&
   +185306 v^7 w^5 u^5-65934 v^6
   w^5 u^5+70604 v^8 w^4 u^5-47559 v^7 w^4 u^5+8260 v^9 w^3 u^5-8260 v^8 w^3 u^5+115441
   v^7 w^6 u^4
   &\\&
   +117010 v^8 w^5 u^4-72617 v^7 w^5 u^4+17209 v^9 w^4 u^4-17209 v^8 w^4
   u^4+28546 v^8 w^6 u^3-1200 v^7 w^6 u^3+30328 v^9 w^5 u^3
   &\\&
   -30328 v^8 w^5 u^3-138726
   v^8 w^7 u^2-6342 v^9 w^6 u^2+6342 v^8 w^6 u^2-52920 v^9 w^7 u+52920 v^8 w^7 u-29592
   v^9 w^8 +v\leftrightarrow w)\big)\,.
 \end{aligned}
\end{equation}
\end{footnotesize}

\providecommand{\href}[2]{#2}\begingroup\raggedright\endgroup

\end{document}